\documentclass[structabstract]{aa}  
\usepackage{graphicx}
\usepackage{txfonts}
\usepackage{natbib}
\usepackage{breqn}
\usepackage{longtable,lscape}
\usepackage{hyperref}
\newcommand{\GG}{\mbox{$G$}}
\newcommand{\GBP}{\mbox{$G_{\rm BP}$}}
\newcommand{\GRP}{\mbox{$G_{\rm RP}$}}
\newcommand{\GGc}{\mbox{$G^\prime$}}
\newcommand{\GGs}{\mbox{$G_{\rm synth}$}}
\newcommand{\GBPs}{\mbox{$G_{\rm BP,synth}$}}

\newcommand{\GGp}{\mbox{$G_{\rm phot}$}}
\newcommand{\GBPp}{\mbox{$G_{\rm BP,phot}$}}
\newcommand{\GRPp}{\mbox{$G_{\rm RP,phot}$}}
\newcommand{\GGcp}{\mbox{$G_{\rm phot}^\prime$}}
\begin{document}

   \title{A reanalysis of the {\em Gaia} Data Release 2 photometric \\ sensitivity curves using HST/STIS spectrophotometry\thanks{Table~\ref{senscurves} 
          is only available at the CDS via anonymous ftp to \url{ftp://cdsarc.u-strasbg.fr}~{\tt (130.79.128.5)} or via 
            http at \url{http://cdsarc.u-strasbg.fr/viz-bin/qcat?J/A+A/vol/page}.}}


   \titlerunning{{\em Gaia} Data Release 2 photometric sensitivity curves from STIS spectra}

   \author{J. Ma{\'\i}z Apell{\'a}niz\inst{1}
           \and
           M. Weiler\inst{2}
          }

   \authorrunning{Ma{\'\i}z Apell{\'a}niz \& Weiler}

   \institute{Centro de Astrobiolog{\'\i}a, CSIC-INTA. Campus ESAC. Camino bajo del castillo s/n. E-28\,692 Villanueva de la Ca\~nada, Spain. \\
              \email{jmaiz@cab.inta-csic.es} \\
         \and
              Departament de F{\'\i}sica Qu\`antica i Astrof{\'\i}sica. Institut de Ci\`encies del Cosmos (ICCUB). Universitat de Barcelona (IEEC-UB). Mart{\'\i} i Franqu\`es 1, 08028. Barcelona, Spain. \\
             }

   \date{Received 8 August 2018; accepted 4 Sep 2018}

 
  \abstract
  {The second data release (DR2) from the European Space Agency mission \mbox{{\it Gaia}} took place on April 2018. DR2 included photometry for more than 
   $1.3 \cdot 10^9$~sources in the three bands \GG, \GBP, and \GRP. Even though the \mbox{{\it Gaia}} DR2 photometry is very precise, there are 
   currently three alternative definitions of the sensitivity curves that show significative differences.}
  {The aim of this paper is to improve the quality of the input calibration data to produce new 
   compatible
   definitions of the \GG, \GBP, and \GRP\ bands and to identify the reasons for the discrepancies between previous definitions.}
  {We have searched the HST archive for STIS spectra with G430L+G750L data obtained with wide apertures and combined them with the CALSPEC library 
   to produce a high quality SED library of 122 stars with a broad range of colors, including three very red stars. 
   This library defines
   new sensitivity curves for \GG, \GBP, and \GRP\ using a functional analytical formalism.}
  {The new sensitivity curves are significantly better than the two previous attempts we use as a reference, REV and WEI. For \GG\ we confirm the 
   existence of a systematic bias in magnitude and correct a color term present in REV. For \GBP\ we confirm the need to define two magnitude
   ranges with different sensitivity curves and measure the cut between them at $\GGp = 10.87$~mag with a significant increase in precision. The 
   new curves also fit the data better than either REV or WEI. 
   For \GRP, our new sensitivity curve fits the STIS spectra better and the differences with previous attempts reside in a systematic effect between 
   ground-based and HST spectral libraries.
   Additional evidence from color-color diagrams indicate that the new 
   sensitivity curve is more accurate. Nevertheless, there is still room for improvement in the accuracy of the sensitivity curves because of the
   current dearth of good-quality red calibrators: adding more to the sample should be a priority before \mbox{{\it Gaia}} data release 3 takes place.}
  {}


   \maketitle
%

\section{Introduction}

$\,\!$\indent The second data release (DR2) of the {\it Gaia} mission \citep{Prusetal16} took place in April 2018 \citep{Browetal18}. {\it Gaia} DR2 
includes photometry for over $1.3 \cdot 10^9$~sources in the three bands \GG, \GBP, and \GRP. The \GG\ photometry was extracted using PSF fitting and 
has formal uncertainties under 1 mmag for most stars brighter than $\GG = 16$. The \GBP\ and \GRP\ magnitudes were obtained through aperture photometry 
and have larger formal uncertainties, of the order of a few mmag for stars brighter than $\GG = 15$, larger than those for \GG\ because they are 
measured just once per transit as opposed to the nine measurements per transit for \GG. {\it Gaia} DR2 constitutes the first all-sky 
multiband high-precision deep optical photometric survey and as such is likely to be considered an astronomical milestone that will be used as a
reference and a calibration source for many studies. However, a high formal precision does not necessarily imply a high accuracy, as one needs to 
read the ``fine print'' of how the photometry was obtained to determine the applicability of the published magnitudes and uncertainties. For 
example, the different nature of the photometry (PSF vs. aperture) leads to \GG\ being more accurate than \GBP\ and \GRP\ in crowded (where 
multiple sources can be included more easily) or nebular (where the background model can be biased) regions \citep{Evanetal18}.

Another accuracy issue, which is the main subject of this paper, is the comparison between the observed magnitudes ($m_{\rm phot}$) and the 
synthetic ones ($m_{\rm synth}$) derived from the spectral energy distributions (SEDs) of the sources. In this paper we use Vega magnitudes, as 
customary for Gaia photometry, and the reader is referred to the Appendix to see how we define the relevant quantities, including the zero points
(ZPs) that are one of our results.  An accurate definition of the sensitivity curves is especially important for the {\it Gaia} photometric system 
because the three passbands are very broad: \GG\ has an effective width\footnote{There are different ways to measure the center and width of a 
passband (see e.g. section 5.1 in \citealt{synphot}) but that does not affect
the argument here.} around 2900~\AA\ (centered around 6400~\AA) while those of \GBP\ and \GRP\ are close to 1900~\AA\ (with that of \GRP\ slightly
larger) and centered around 5100~\AA\ and 7800~\AA, respectively. For comparison, the widths of the Johnson $UBV$ system are 500-700~\AA. When 
doing broad-band photometry of sources with very different intrinsic SEDs and degrees of extinction one needs to integrate each SED to calculate the 
magnitudes, as a simple evaluation of the flux at a central wavelength does not work. Already for the Johnson $UBV$ system the classical $Q$
approximation to calculate extinction \citep{JohnMorg53} breaks down in many practical situations (see Appendix~B in \citealt{MaizBarb18}) due to 
the non-linearity of the extinction trajectories in the $U-B$ + $B-V$ plane induced by this effect. For {\it Gaia} photometry such extinction 
non-linearities in a color-color plane are even larger and more dependent on the precise definition of the passbands, as we will show later on in this 
paper.

The first sensitivity curves for the three {\it Gaia} passbands were published by \citet{Jordetal10} but those were based on pre-launch data that 
were later modified. In one of the {\it Gaia} DR1 calibration papers, \citet{Carretal16} noted that if one used those curves a color term was 
present in the \GG\ photometry and \citet{Maiz17} published a modified sensitivity curve 
that was able to correct for it. An independent analysis by \citet{Weiletal18} found a very similar sensitivity curve. The {\it Gaia} DR1 photometry
was affected by a contamination effect caused by water freezing in some optical elements \citep{Prusetal16} so the {\it Gaia} DR2 \GG\ data were
expected to be characterized by a different sensitivity curve. Indeed, \citet{Evanetal18} published not a set but two sets of sensitivity
curves for the \GG, \GBP, and \GRP\ photometry in the second data release: one they called DR2 and another one they called REV (for revised, that
set was considered the preferred one by the authors). Later on, \citet{Weil18} provided a third set that differed from the other two, and that we 
refer to as WEI. All three by now published sets of {\it Gaia} DR2 passbands are based on the same set of calibration sources, the 
''Spectrophotometric Standard Stars'' (SPSS, \citealt{Pancetal12, Altaetal15}) with the only exception of the WEI \GBP\ passbands, which 
were derived using CALSPEC 
\citep{Bohletal14,Bohletal17}.
The SPSS set of calibration spectra is being constructed for the calibration of {\it Gaia}, 
and a first subset of 92 stars was made available for deriving {\it Gaia} DR2 passbands. Other spectral libraries, namely CALSPEC, the Next 
Generation Spectral Library (NGSL, \citealt{HeapLind07}), and the library by \citet{Strietal05} have been used for validation purposes by 
\citet{Evanetal18} and \citet{Weil18}.

The DR2, REV, and WEI results are similar (but not identical) for $G$: the three sensitivity curves show few 
differences and agree in their overall shape. They all require a correction for a drift in the zero point of the observed \GG\ photometry, as 
discussed further in section~3 (the corrected \GG\ magnitude is denoted \GGc\ here). The results are more different for 
\GBP, as \citet{Weil18} found that bright and faint stars follow different sensitivity curves and that there is a jump of 
20~mmag in zero point between the two. The WEI sensitivity curves for \GBP\ for the bright and the faint stars both differ 
strongly in their overall shape from the DR2 and REV passbands. For \GRP\ the differences in shape between the DR2 and REV sensitivity 
curves and the WEI curve are large, too, although resulting in a small improvement for the SPSS calibration spectra only. Furthermore, 
\citet{Weil18} noted that, while the WEI sensitivity curve for \GRP\ improves the results for the SPSS, Stritzinger, and NGSL libraries, it yields a 
worse result than the REV passband for the CALSPEC spectra.
\citet{Weil18} also compared synthetic color-color relationships with observed relationships to test the consistency of 
a set of sensitivity curves for the three different {\it Gaia} passbands. This consistency test showed that the REV set of sensitivity curves fails
to reproduce the observed color-color relationships. On the other hand, the WEI sensitivity curves have been designed not only to result in a good 
reproduction of the observed photometry for each passband individually, but also to reproduce the observed color-color relationships even 
outside the range in colors covered by the calibration spectra.

In this work, we first compile a new set of calibration spectra based on high-quality HST/STIS optical observations. This set of calibration 
spectra extends the CALSPEC set significantly, both in number and coverage of different spectral types. In section~2 we describe this set 
of calibration spectra in detail. We then use the new set of calibration spectra to derive refined sensitivity curves (that will be referred to as 
MAW from our last names) for all {\it Gaia} passbands in section~3. Finally, in section~4 we demonstrate that the calibration data of this work is 
superior in quality to existing sets of calibration spectra. We also compare synthetic color-color relationships with observed ones, both for main 
sequence stars and for highly reddened stars, demonstrating that the sensitivity curves derived in this work with a new set of calibration spectra are 
the most accurate available to date.



\section{A new compilation of HST/STIS optical spectra}

$\,\!$ \indent One of us (J.M.A.) has performed several analyses of the validity of sensitivity curves for different photometric systems
\citep{Maiz05b,Maiz06a,Maiz07a,Maiz17}. In those works the main source of spectrophotometric data was NGSL,
a spectrophotometric library built from HST/STIS data obtained with the three gratings G230LB+G430L+G750L that covers the
1700-10\,200~\AA\ for several hundreds of stars of diverse spectral type. Our original idea for this work was to base it also on NGSL but after
several tests we discovered that the quality of their absolute and relative calibrations was not good enough for the purposes of calibrating
{\it Gaia} DR2 photometry. This issue can be seen, for example, in Figs.~3~and~8 of \citet{Weil18}, where the dispersion for NGSL stars is higher
than for the other three libraries. The likely reason for this problem is that the NGSL data were obtained with a narrow STIS slit, 52x0.2, for 
which absolute flux calibration is difficult to attain. 

Not being able to use NGSL, we looked for spectrophotometric substitutes in the HST archive subject to the following criteria:

\begin{itemize}
 \item Existence of data for at least the G430L+G750L grating to allow for coverage of the 2900-10\,200~\AA\ range. Note that \GG\ and \GRP\ have
       some sensitivity at longer wavelengths but that is small enough that the SED can be interpolated between the STIS data and NIR photometry 
       without a significant bias in the analysis.
 \item Use of a wide STIS slit (52x0.5 or wider) to avoid flux calibration issues. Note that in a single case below we relax this criterion.
 \item S/N large enough in the STIS data for the CTI correction not to introduce large uncertainties.
 \item Existence of good-quality {\it Gaia} DR2 \GG, \GBP, and \GRP\ photometry.
 \item Lack of extended nebulosity around the object and of notorious variability.
\end{itemize}

\begin{table*}
\caption{Sample used in this paper sorted by $\GBPp-\GRPp$.}
\label{sample}
\centerline{
\begin{tabular}{lcrrrr}
\hline
Name & Type & \multicolumn{1}{c}{\GGp} & \multicolumn{1}{c}{\GGcp} & \multicolumn{1}{c}{\GBPp} & \multicolumn{1}{c}{\GRPp} \\
\hline
Tyc 4547-01009-1          & C & 11.8797 & 11.8609 & 11.6383 & 12.2257 \\
BD $+$52 913              & C & 11.7379 & 11.7195 & 11.4865 & 12.0665 \\
PG 1657$+$344             & C & 16.4476 & 16.4142 & 16.1955 & 16.7473 \\
2MASS J12570233$+$2201526 & C & 13.3224 & 13.2990 & 13.0813 & 13.6290 \\
2MASS J13233526$+$3607595 & C & 11.6350 & 11.6170 & 11.3913 & 11.9377 \\
Feige 110                 & C & 11.7924 & 11.7739 & 11.5571 & 12.1033 \\
Feige 34                  & C & 11.1072 & 11.0909 & 10.8753 & 11.4203 \\
Tyc 6429-00796-1          & H & 11.6968 & 11.6786 & 11.4641 & 12.0045 \\
2MASS J05522761$+$1553137 & C & 13.0255 & 13.0030 & 12.7700 & 13.2991 \\
HS 2027$+$0651            & C & 16.6542 & 16.6201 & 16.4011 & 16.8739 \\
$\mu$ Col                 & C &  5.1009 &  5.1253 &  4.9745 &  5.4056 \\
$\lambda$ Lep             & C &  4.1705 &  4.2201 &  4.1165 &  4.5018 \\
HD 205\,805               & H & 10.1483 & 10.1350 & 10.0158 & 10.3958 \\
2MASS J16293576$+$5255532 & C & 15.6821 & 15.6511 & 15.5033 & 15.8646 \\
10 Lac                    & C &  4.7935 &  4.8262 &  4.7230 &  5.0390 \\
2MASS J03094790$-$5623494 & C & 14.1201 & 14.0941 & 13.9764 & 14.2793 \\
2MASS J13385054$+$7017077 & C & 12.7910 & 12.7693 & 12.6678 & 12.9688 \\
Tyc 5818-00926-1          & H & 11.3146 & 11.2976 & 11.1987 & 11.4633 \\
2MASS J03552198$+$0947180 & C & 14.5644 & 14.5370 & 14.5107 & 14.6442 \\
BD $-$00 4234 B           & C & 14.6630 & 14.6353 & 14.5682 & 14.6960 \\
HD 172\,140               & M &  9.9273 &  9.9147 &  9.8847 &  9.9872 \\
HD 60\,753                & C &  6.6425 &  6.6404 &  6.6213 &  6.7079 \\
HD 93\,028                & M &  8.3368 &  8.3293 &  8.3067 &  8.3930 \\
HD 116\,405               & C &  8.3129 &  8.3055 &  8.2967 &  8.3716 \\
$\xi^{2}$ Cet             & C &  4.1797 &  4.2290 &  4.2475 &  4.3091 \\
CPD $-$57 3507            & M &  9.2364 &  9.2260 &  9.2161 &  9.2685 \\
HD 46\,966 AaAb           & H &  6.8063 &  6.8037 &  6.8000 &  6.8434 \\
HD 142\,165               & M &  5.3251 &  5.3434 &  5.3479 &  5.3406 \\
Tyc 5709-00698-1          & C & 12.2707 & 12.2506 & 12.2277 & 12.2204 \\
$\lambda$ Lib             & M &  4.9466 &  4.9751 &  4.9853 &  4.9731 \\
BD $+$60 1753             & C &  9.6784 &  9.6666 &  9.7016 &  9.6666 \\
HD 220\,057               & M &  6.8945 &  6.8916 &  6.9316 &  6.8647 \\
Tyc 4201-01542-1          & C & 12.0086 & 11.9894 & 12.0278 & 11.9551 \\
HD 198\,781               & M &  6.3899 &  6.3887 &  6.4278 &  6.3308 \\
HD 197\,512               & M &  8.5140 &  8.5060 &  8.5528 &  8.4492 \\
9 Sgr AB                  & H &  5.8728 &  5.8762 &  5.9165 &  5.8052 \\
HD 193\,322 AaAb          & M &  5.8868 &  5.8899 &  5.9454 &  5.8081 \\
HD 164\,073               & M &  7.9627 &  7.9564 &  8.0217 &  7.8762 \\
Tyc 4209-01396-1          & C & 12.2958 & 12.2757 & 12.3446 & 12.1857 \\
Tyc 4424-01286-1          & C & 12.5264 & 12.5055 & 12.5770 & 12.4163 \\
HD 165\,459               & C &  6.8342 &  6.8315 &  6.9153 &  6.7515 \\
BD $-$13 5550             & H & 11.3153 & 11.2983 & 11.3131 & 11.1399 \\
HD 180\,609               & C &  9.4099 &  9.3990 &  9.4823 &  9.3078 \\
HD 158\,485               & C &  6.4601 &  6.4586 &  6.5523 &  6.3571 \\
HD 46\,150                & H &  6.7031 &  6.7009 &  6.7802 &  6.5752 \\
HDE 228\,199              & H &  9.2412 &  9.2308 &  9.3181 &  9.1112 \\
HD 46\,106                & M &  7.8884 &  7.8824 &  7.9676 &  7.7546 \\
HD 37\,725                & C &  8.3003 &  8.2929 &  8.3936 &  8.1679 \\
BD $+$69 1231             & M &  9.2387 &  9.2283 &  9.3312 &  9.0897 \\
HD 210\,121               & M &  7.6199 &  7.6147 &  7.7212 &  7.4706 \\
ALS 8988                  & M &  9.6817 &  9.6699 &  9.7709 &  9.5158 \\
Tyc 4207-00219-1          & C & 12.4698 & 12.4491 & 12.5538 & 12.2945 \\
HD 14\,943                & C &  5.8643 &  5.8680 &  5.9788 &  5.7186 \\
Tyc 4205-01677-1          & C & 11.7214 & 11.7031 & 11.8145 & 11.5535 \\
Tyc 4433-01800-1          & C & 12.0457 & 12.0264 & 12.1386 & 11.8680 \\
HD 163\,466               & C &  6.8136 &  6.8110 &  6.9377 &  6.6646 \\
HD 92\,044                & M &  8.1879 &  8.1809 &  8.2902 &  8.0099 \\
HD 147\,196               & M &  6.9740 &  6.9709 &  7.0897 &  6.7943 \\
HD 38\,087                & M &  8.2053 &  8.1982 &  8.3168 &  7.9949 \\
HD 112\,607               & M &  8.0194 &  8.0129 &  8.1437 &  7.8189 \\
HD 18\,352                & M &  6.7468 &  6.7444 &  6.8756 &  6.5475 \\
\hline
\end{tabular}
}
\end{table*}

\addtocounter{table}{-1}
\begin{table*}
\caption{Continued.}

\centerline{
\begin{tabular}{lcrrrr}
\hline
Name & Type & \multicolumn{1}{c}{\GGp} & \multicolumn{1}{c}{\GGcp} & \multicolumn{1}{c}{\GBPp} & \multicolumn{1}{c}{\GRPp} \\
\hline
Tyc 4212-00455-1          & C & 11.7746 & 11.7561 & 11.9036 & 11.5514 \\
CPD $-$41 7715            & M & 10.1991 & 10.1857 & 10.3231 &  9.9671 \\
$\lambda$ Cep             & H &  4.9432 &  4.9718 &  5.1041 &  4.7427 \\
HD 93\,250 AB             & H &  7.2615 &  7.2575 &  7.3970 &  7.0342 \\
2MASS J17430448$+$6655015 & C & 13.4843 & 13.4604 & 13.6124 & 13.2442 \\
HDE 239\,745              & M &  8.8216 &  8.8126 &  8.9728 &  8.5722 \\
HD 13\,338                & M &  9.0632 &  9.0534 &  9.2155 &  8.8109 \\
HD 146\,285               & M &  7.8435 &  7.8376 &  7.9998 &  7.5935 \\
BD $-$13 4930             & H &  9.3195 &  9.3089 &  9.4721 &  9.0631 \\
HD 207\,198               & H &  5.8222 &  5.8270 &  5.9895 &  5.5661 \\
HD 14\,321                & M &  9.1446 &  9.1345 &  9.3048 &  8.8779 \\
BD $+$56 576              & M &  9.3124 &  9.3018 &  9.4775 &  9.0395 \\
BD $+$56 517              & M & 10.3636 & 10.3496 & 10.5268 & 10.0780 \\
CPD $-$59 2600            & M &  8.4564 &  8.4485 &  8.6206 &  8.1641 \\
HD 14\,250                & M &  8.8985 &  8.8892 &  9.0803 &  8.6025 \\
HD 192\,639               & H &  7.0036 &  7.0004 &  7.2083 &  6.6980 \\
HD 68\,633                & M &  7.8110 &  7.8052 &  8.0126 &  7.4855 \\
HD 110\,336               & M &  8.5490 &  8.5408 &  8.7688 &  8.2145 \\
NU Ori                    & M &  6.7031 &  6.7009 &  6.9293 &  6.3508 \\
HD 74\,000                & C &  9.5268 &  9.5155 &  9.7828 &  9.1408 \\
HDE 284\,248              & C &  9.0847 &  9.0748 &  9.3502 &  8.6841 \\
BD $+$26 2606             & C &  9.5704 &  9.5590 &  9.8411 &  9.1661 \\
HDE 292\,167              & M &  9.0862 &  9.0763 &  9.3557 &  8.6771 \\
BD $+$17 4708             & C &  9.3217 &  9.3111 &  9.5978 &  8.9135 \\
HD 199\,216               & M &  6.8745 &  6.8717 &  7.1566 &  6.4709 \\
HDE 233\,511              & C &  9.5615 &  9.5501 &  9.8363 &  9.1498 \\
HD 160\,617               & C &  8.5746 &  8.5664 &  8.8591 &  8.1584 \\
HD 31\,128                & C &  8.9817 &  8.9722 &  9.2710 &  8.5569 \\
BD $+$02 3375             & C &  9.7858 &  9.7737 & 10.0782 &  9.3565 \\
HD 209\,458               & C &  7.5087 &  7.5039 &  7.8127 &  7.0896 \\
HD 70\,614                & M &  9.1148 &  9.1048 &  9.4194 &  8.6746 \\
HD 38\,949                & C &  7.6617 &  7.6564 &  7.9761 &  7.2280 \\
2MASS J03323287$-$2751483 & C & 16.3779 & 16.3447 & 16.6706 & 15.9210 \\
BD $+$60 513              & H &  9.2502 &  9.2398 &  9.5539 &  8.7987 \\
BD $+$29 2091             & C & 10.0987 & 10.0856 & 10.4061 &  9.6498 \\
HD 168\,075               & H &  8.5644 &  8.5562 &  8.8700 &  8.1122 \\
HD 106\,252               & C &  7.2628 &  7.2588 &  7.5948 &  6.8103 \\
HD 205\,905               & C &  6.5849 &  6.5830 &  6.9230 &  6.1368 \\
Tyc 4413-00304-1          & C & 11.8744 & 11.8556 & 12.1946 & 11.4033 \\
HD 159\,222               & C &  6.3607 &  6.3595 &  6.7067 &  5.9062 \\
HD 37\,962                & C &  7.6767 &  7.6713 &  8.0301 &  7.2069 \\
2MASS J16313382$+$3008465 & C & 12.8618 & 12.8398 & 13.1992 & 12.3679 \\
2MASS J15591357$+$4736419 & C & 13.3330 & 13.3095 & 13.6751 & 12.8342 \\
HD 185\,975               & C &  7.9308 &  7.9246 &  8.3012 &  7.4476 \\
CPD $-$59 2591            & M & 10.6266 & 10.6118 & 10.9678 & 10.0989 \\
2MASS J16194609$+$5534178 & C & 16.0753 & 16.0431 & 16.4239 & 15.5545 \\
2MASS J16181422$+$0000086 & C & 16.5842 & 16.5503 & 16.9507 & 16.0362 \\
HD 200\,654               & C &  8.8749 &  8.8657 &  9.2623 &  8.3426 \\
HD 217\,086               & H &  7.4470 &  7.4424 &  7.8353 &  6.9121 \\
HD 149\,452               & M &  8.8659 &  8.8567 &  9.2581 &  8.3145 \\
HD 15\,570                & H &  7.8787 &  7.8727 &  8.3448 &  7.2653 \\
HD 9051                   & C &  8.6779 &  8.6693 &  9.1521 &  8.0724 \\
2MASS J19031201$-$3729105 & M &  9.8271 &  9.8149 & 10.3070 &  9.1827 \\
2MASS J17551622$+$6610116 & C & 13.0497 & 13.0271 & 13.5668 & 12.4061 \\
2MASS J17583798$+$6646522 & C & 13.6304 & 13.6060 & 14.2579 & 12.8873 \\
HD 164\,865               & M &  7.1962 &  7.1924 &  7.8392 &  6.4310 \\
HD 29\,647                & M &  7.8164 &  7.8106 &  8.4899 &  7.0295 \\
HD 147\,889               & M &  7.4894 &  7.4846 &  8.1656 &  6.6878 \\
BD $-$11 3759             & O &  9.8788 &  9.8664 & 11.5941 &  8.6616 \\
Proxima Cen               & O &  8.9536 &  8.9441 & 11.3829 &  7.5864 \\
2MASS J16553529$-$0823401 & C & 13.8384 & 13.8133 & 17.0754 & 12.3215 \\
\hline
\end{tabular}
}
\end{table*}

With those criteria, we found useful STIS spectrophotometry for 122 stars, listed in Table~\ref{sample}. The Type column there refers to the first
letter of the four sets of data:

\paragraph{1. CALSPEC:} This library was already used as a secondary source in our previous works (e.g. \citealt{Maiz06a}) and has been built over
the years to calibrate STIS (and other HST instruments) in absolute flux (\citealt{Bohletal17} and references therein). It constitutes the most 
reliable source, as some of the sources have been observed repeatedly under different conditions and because it uses the widest STIS slit 52x2. For 
this set we downloaded the reduced spectra directly from the CALSPEC web site\footnote{\url{http://www.stsci.edu/hst/observatory/crds/calspec.html}.}. 
CALSPEC contributes with 63 stars.

\paragraph{2. HOTSTAR:} This set is described in \citet{KhanWort18}. It is a hot star extension to NGSL that used the 52x0.5 slit, allowing it to be 
included in this work.  In this case we downloaded the raw data from the HST archive and processed it ourselves using the STSDAS package in pyraf. The 
only exception to the latter is BD~$-$13~4930, whose data is not yet public at the time of this writing; for that star we used the author's reduction.
HOTSTAR contributes with 17 stars.

\paragraph{3. Massa:} The third set is that of HST program 13\,760 (P.I.: Massa). To our knowledge, no paper has appeared that uses those data.
The Massa set uses the 52x2 slit and contains 40 stars. As for the previous case, we downloaded the raw data from the archive and processed it
ourselves.

\paragraph{4. Other:} One problem that is crucial for an accurate calibration of the {\it Gaia} DR2 photometry is the use of SEDs covering a wide 
range in color. In particular the calibration of very red sources of M type require SEDs of such objects in the set of calibration spectra 
\citep{Weiletal18}. The 120 stars in the three sets described above however only contain stars with a  $\GBPp-\GRPp$ color less than 1.5, with the 
only exception of the M-dwarf 
2MASS J16553529$-$0823401
at $\GBPp-\GRPp = 2.9$. It is therefore desirable to include more very red objects in our 
set of calibrations spectra and to extend it to even redder objects. To address this issue we searched the HST archive for additional red objects 
with little variability. We found two M dwarfs that satisfy those conditions:
BD~$-$11~3759 and Proxima Cen. The first one was observed by HST program 8422 (P.I.: Ferguson) using the 52x2 slit. As we did with the previous two
sets, we downloaded the raw data from the HST archive and processed it ourselves. For Proxima Cen we used the reduced spectrum provided by
\citet{Ribaetal17}. Note that this second star was observed with the 52x0.2 slit but in that paper it was recalibrated in absolute flux using 
external information. 

A fraction of red dwarfs is known to be variable but only 8\% of them show variations above 20~mmag \citep{Hoseetal15}. We therefore performed 
checks on the three M-type targets to ensure they are not too strongly variable. We searched the literature for indications of variability. 
\citet{Hoseetal15} lists an amplitude of 13.8~mmag in the $V$ band for BD~$-$11~3759, which is small enough for our purposes. Proxima Cen 
experiences flux variations due to rotational modulation of surface inhomogenities \citep{Ribaetal17}. However, in the optical those 
have relatively large amplitudes only at short wavelengths. In the $V$ band the amplitude is only of the order of 20~mmag and at longer wavelengths,
where we are more interested, is even smaller, a dependence with wavelength that is typical of variable red dwarfs.
2MASS~J16553529$-$0823401 is the faintest of the three red dwarfs and there is less information
about variability than for the other two. The AAVSO International Variable Star Index in VizieR lists an amplitude of 45~mmag in $V$,
which likely contains a low-S/N component as the object has a magnitude of 16.7 in that band. On the other hand, 2MASS~J16553529$-$0823401 
shows very little variation in the WISE \citep{Cutretal13} bands, where it is an eighth-magnitude star and has a variability flag of 1 (in a scale
of 0-9 with variability starting at 6). Ideally, one would use the variability flag provided by {\it Gaia} itself when selecting calibration sources, 
but this flag will not be available for most of the observed sources until the next data release (DR3). 
A proxy for variability, however, is the uncertainty on the mean fluxes provided in {\it Gaia} DR2. The uncertainty of the mean flux is computed as the 
standard deviation of the mean flux, which is the standard deviation of the sample of all epoch photometry of a particular source that entered into the 
computation of the mean flux, divided by the square root of the number of observations \citep{Carretal16}. We 
therefore multiplied the standard deviation of the mean flux with the square root of the number of observations to obtain the standard deviation of 
the epoch photometry for the three M dwarfs. The standard deviation of the sample of epoch photometry in {\it Gaia} DR2 is a very strong function of
magnitude and color, though. For a meaningful comparison with typical values for the standard deviation of the epoch photometry we therefore 
computed the median standard deviation for stars in a small bins in the magnitude-color plane and compared the standard deviation of the three 
M-dwarfs with the medians of the nearby bins. For all three M dwarfs and for all three {\it Gaia} passbands, the standard deviation is smaller than 
the median standard deviation of stars in the corresponding region of the magnitude-color diagram. The available {\it Gaia} DR2 photometry shows 
thus no indication for significant variability for the three M dwarfs in our set of calibration spectra.

\section{Generating the new sensitivity curves}

\subsection{Theoretical approach}

$\,\!$\indent For the determination of the sensitivity curves we use the formalism developed in \citet{Weiletal18}. This formalism has 
already been applied in the computation of the {\it Gaia} DR2 sensitivity curves by \citet{Weil18}, where it has been described in detail. We 
therefore only provide a sketch of the method here.

The idea of this formalism is to describe a particular sensitivity curve by the sum of two orthogonal functions. One of these functions, denoted 
the parallel component of the passband, is a linear combination of the principal components of the set of calibration spectra. This function is 
uniquely defined by the calibration spectra and their corresponding photometric observations. It can be derived by solving a simple set of linear 
equations for the coefficients of the principal components derived from the calibration spectra. The second function, denoted the orthogonal 
component of the passband, is unconstrained by the calibration data and is derived in such a way that the passband resulting from adding the parallel and 
orthogonal component satisfies the physical requirements for the passband (such as non-negativity, being bound to unity, and not oscillating). When 
deriving several passbands, as is the case here, we can also use the requirement of reproducing observed color-color relationships with the 
synthetic photometry resulting from the set of passbands as an additional constraint on the choice of the orthogonal component. The derivation of the 
orthogonal component is performed using an initial guess for the passband and defining a multiplicative linear model for the deformation of the initial 
guess. For this modification model, we use a linear combination of B-spline basis functions, which is multiplied to the initial guess for the 
passband. We then choose the coefficients of the B-spline basis functions in such a way that the passband resulting from the multiplication is close to 
the initial guess, under the constraint that the parallel component of the resulting passband is is agreement with the formal interval of confidence on
the parallel component.

\begin{table*}
\caption{Sensitivity curves for \GG, \GBP, and \GRP\ derived in this work separated into their parallel and orthogonal components. This table is only 
 available at the CDS via anonymous ftp to \url{ftp://cdsarc.u-strasbg.fr~(130.79.128.5)} or via http at 
 \url{http://cdsarc.u-strasbg.fr/viz-bin/qcat?J/A+A/vol/page}.}
\label{senscurves}
\end{table*}

For a detailed description of this formalism and its practical implementation the reader is referred to \citet{Weil18}. The sensitivity curves are
given in Table~\ref{senscurves}, separated into their parallel and orthogonal components, respectively. This table is available in electronic form only.

\subsection{Sensitivity curve for \GG}

\begin{figure}
\centerline{\includegraphics[width=\linewidth]{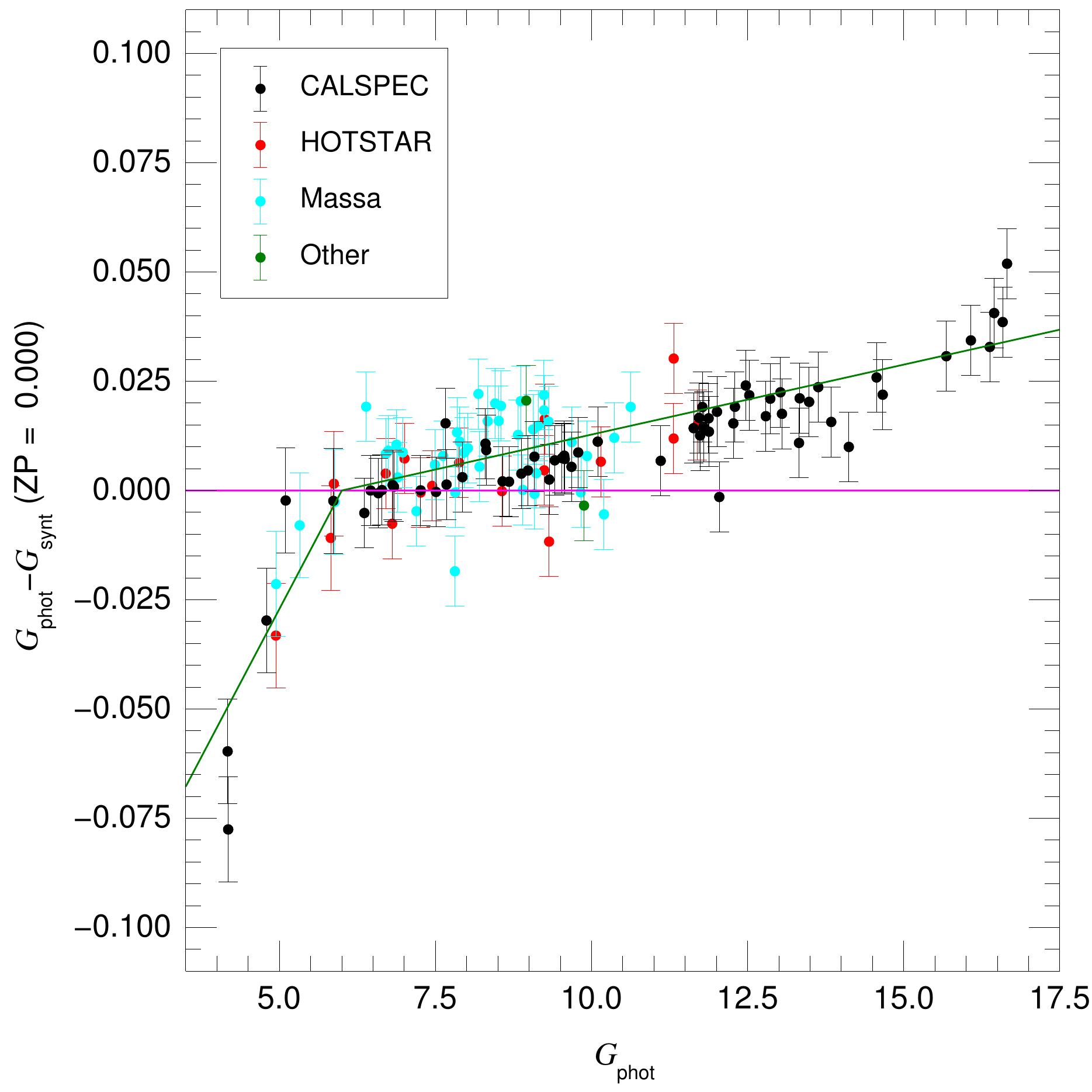}}
\caption{Correction of the systematic errors in \GG. The horizontal axis is the observed (uncorrected) \GG\ magnitude and the vertical axis is
         the difference between that value and the synthetic \GG\ magnitude assuming the \GG\ sensitivity curve in this paper and a ZP of 0. The 
         points with error bars show the data, color-coded according to data set, and the dark green solid line shows the fit used to derive the 
         correction proposed in this paper.  The size of the error bars is explained in section~4.}
\label{Gcor4}
\end{figure}

\begin{figure}
\centerline{\includegraphics[width=\linewidth]{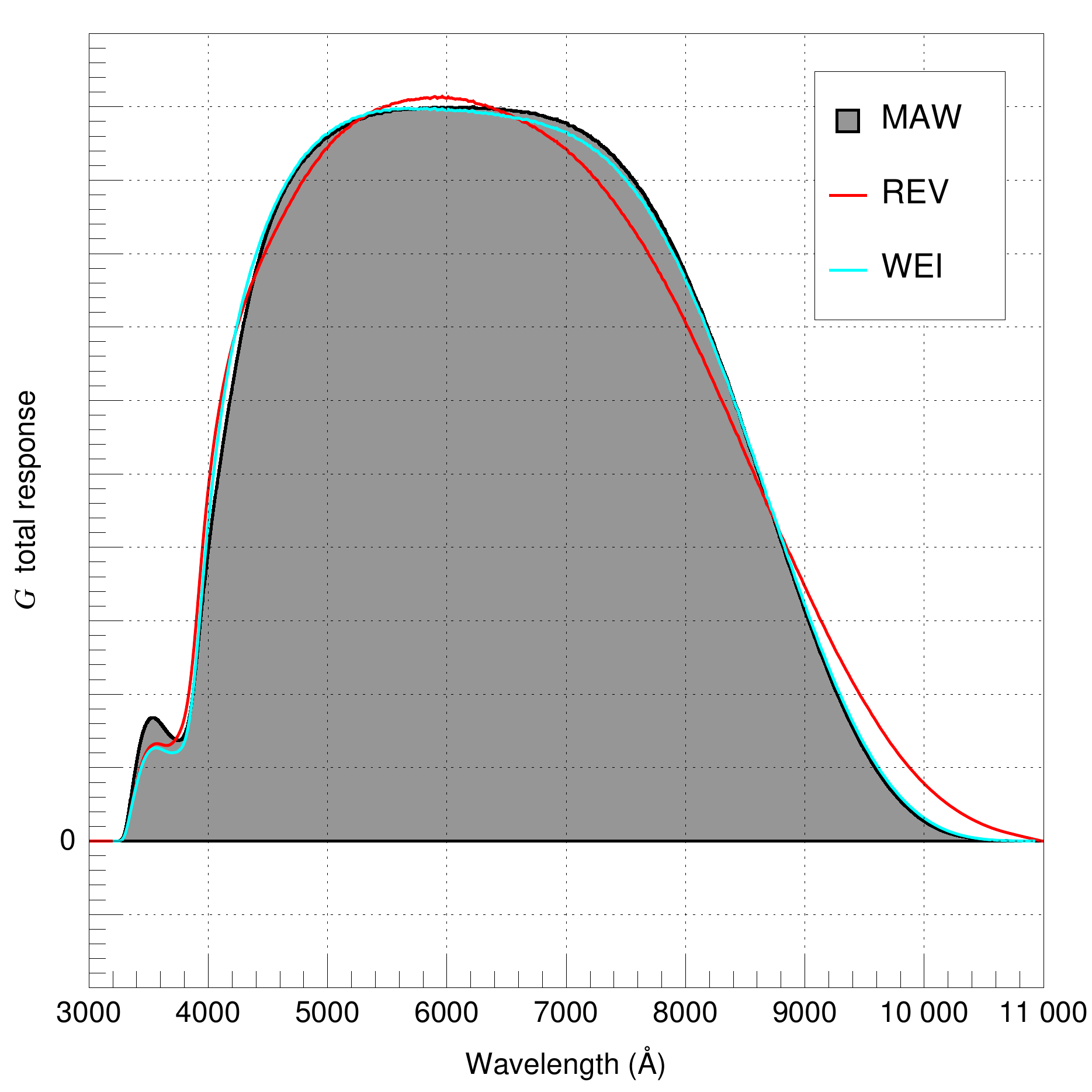}}
\centerline{\includegraphics[width=\linewidth]{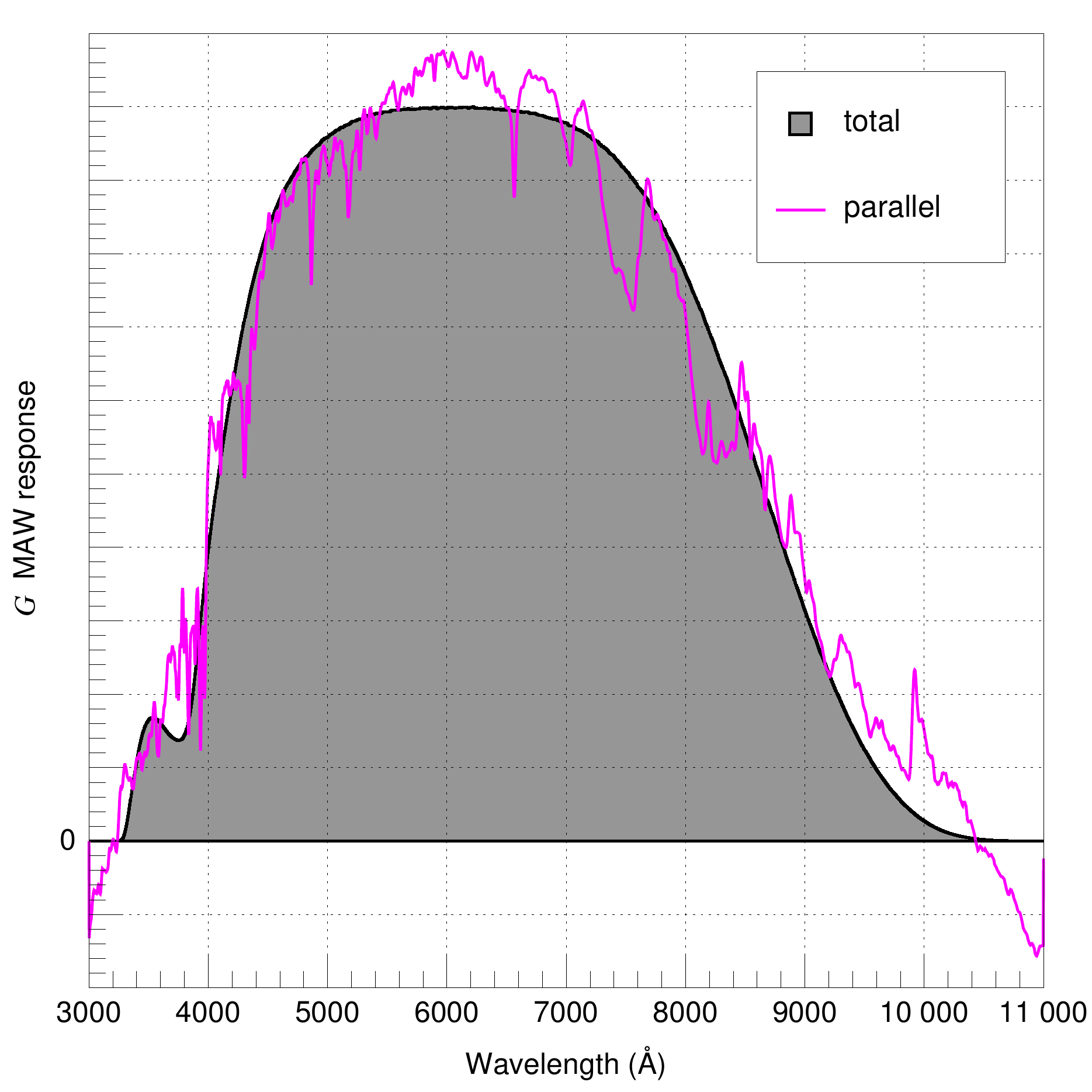}}
\caption{\GG\ sensitivity curves. (top) Total responses for this work (MAW), \citealt{Evanetal18} (REV), and \citealt{Weil18} (WEI) normalized to the
         same area. (bottom) Total response and parallel component for this work with the same scale as on the top panel.}
\label{Gsens}
\end{figure}

$\,\!$\indent  The {\it Gaia} DR2 photometry in the \GG\ band is affected by systematic errors. \citet{Arenetal18} noticed that 
$\GG - \GBP$ magnitude shows a systematic trend with \GG\ magnitude, which is approximately linear between about 6 and 16 in \GG. \citet{Weil18} 
and \citet{CasaVand18} noticed an approximately linear trend in the difference between the observed \GG\ magnitude (\GGp) and the
synthetic \GG\ magnitude (\GGs) computed with the REV passband for the CALSPEC spectra. This trend was estimated to be $\rm 3.5 \pm 0.3$ mmag/mag and
manifests itself in a magnitude dependence of the zero point of the \GG\ passband, which can be removed by introducing a correction factor for 
\GGp.  

Here we estimate the correction again excluding sources with $\GG < 6$ and $\GG > 16$ from our data set. Furthermore, as the Massa data set shows a 
slight systematic deviation from the remaining calibration spectra, we also exclude them from estimating the trend in \GG. We find a value of 
$\rm 3.2 \pm 0.3$ mmag/mag, thus slightly lower than previous works\footnote{Including the Massa stars and those with $\GG > 16$ 
produces a very similar result of $\rm 3.0 \pm 0.3$ mmag/mag.}. This is the value we use for producing \GGcp, the corrected \GGp\ magnitudes, 
before computing the sensitivity curves, i.e. we assume a relationship:

\begin{equation}
\GGcp = -2.5 \cdot 0.9968 \cdot \log_{10}(I_G) + {\rm AZP}
\label{Gcor1}
\end{equation}

\noindent between \GGcp\ and the number of photoelectrons in the \GG\ band, $I_G$, where AZP is the absolute zero point (not to be mistaken with
the Vega zero points used elsewhere in this paper). Alternatively, \GGcp\ can expressed as:

\begin{equation}
\GGcp = \GGp - 0.0032 \cdot (\GGp-6) \;\; {\rm for} \; 6 < \GGp < 16.
\label{Gcor2}
\end{equation}

Eqn.~\ref{Gcor2} can be used to correct the \GG\ magnitudes downloaded from the archive before comparing them with external photometry or synthetic
photometry from SEDs. We have no calibration spectra for sources fainter than 16.7 available. The analysis of \cite{Arenetal18} suggests a more complex 
systematic error for fainter sources, so the linear correction derived in this work may not apply there.
Since our data set also contains eleven stars brighter than $\GG = 6$ we can produce a correction for them, which we find to
be an order of magnitude larger, $\rm 27.1 \pm 5.8$ mmag/mag, i.e.:

\begin{equation}
\GGcp = \GGp + 0.0271 \cdot (6-\GGp) \;\; {\rm for} \; \GGp < 6.
\label{Gcor3}
\end{equation}

The correction for these 
eleven
bright stars is larger because of saturation but is much smaller than the equivalent discussed for Gaia DR1 photometry by
\citet{Maiz17}, indicating that Gaia DR2 did a much better job with them than Gaia DR1. Note that our eleven stars are all fainter than $\GG = 4$, 
so the correction may fail for brighter stars. \citet{Evanetal18} did another saturation analysis and found a correction with the same sign in the
range probed here, though its value was slightly larger. In any case, these bright stars were not included in our calculation of the \GG\ 
sensitivity curve. The fits for the two magnitde ranges are shown in Fig~\ref{Gcor4}.

The \GG\ sensitivity curve is computed using 6 basis functions obtained with the functional principal component analysis of the set of calibration 
spectra and using the \citet{Weil18} \GG\ passband as the initial guess. The resulting sensitivity curve is shown in Fig.~\ref{Gsens} and compared to 
the REV and WEI ones. The new curve is very similar to the WEI one, with the main difference between the two in the secondary peak to the left of the
Balmer jump which is larger in the new curve. On the other hand, there are significant differences with the REV curve, which has a lower sensitivity 
in the 7000-8000~\AA\ region and a higher one beyond 9000~\AA, i.e. it is ``redder'' (in the sense of being more sensitive at longer wavelengths or, in
stellar terms, for cooler temperatures). The consequences of this difference will be explored in the next section.

\subsection{Sensitivity curves for \GBP}

\begin{figure*}
\centerline{\includegraphics[width=0.49\linewidth]{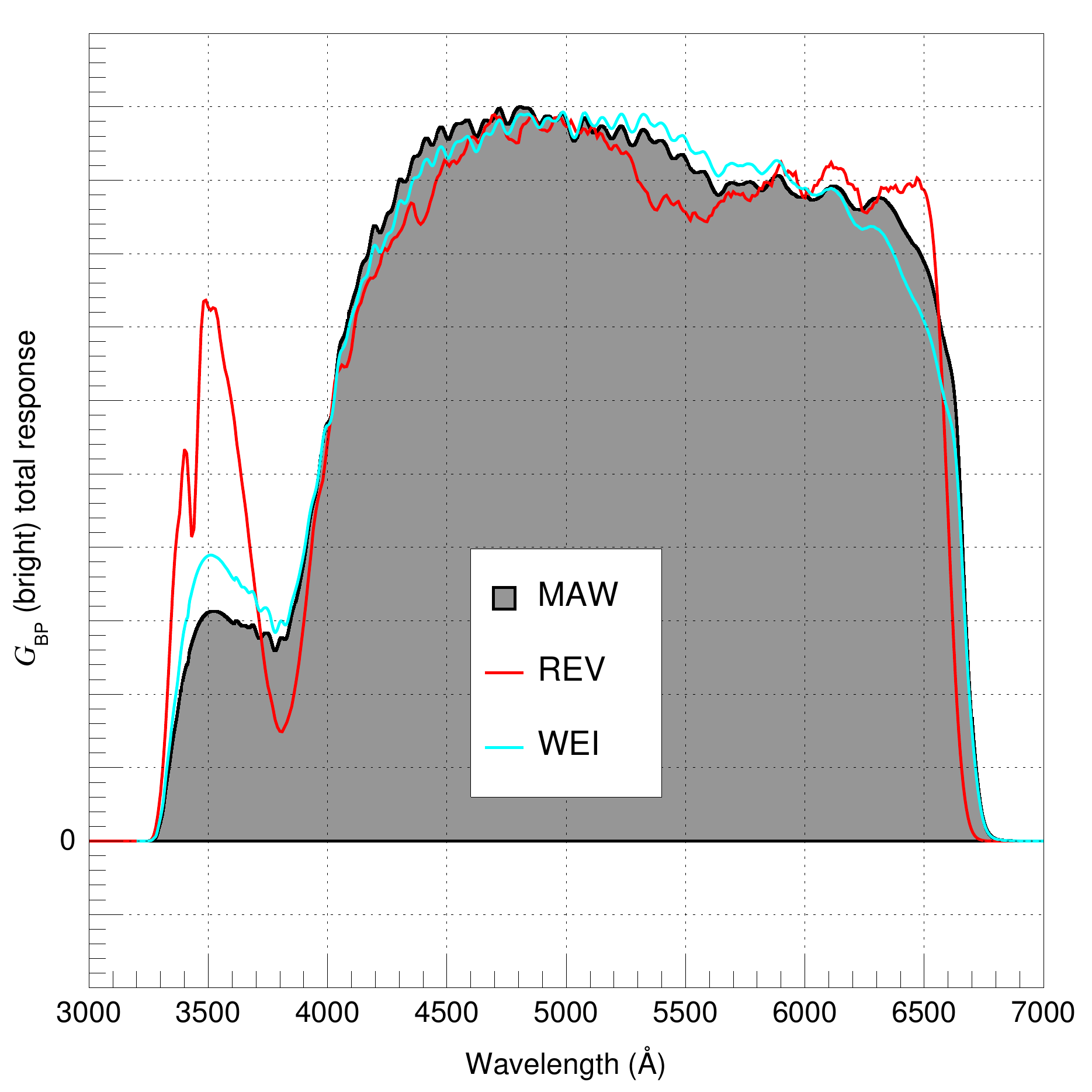} \
            \includegraphics[width=0.49\linewidth]{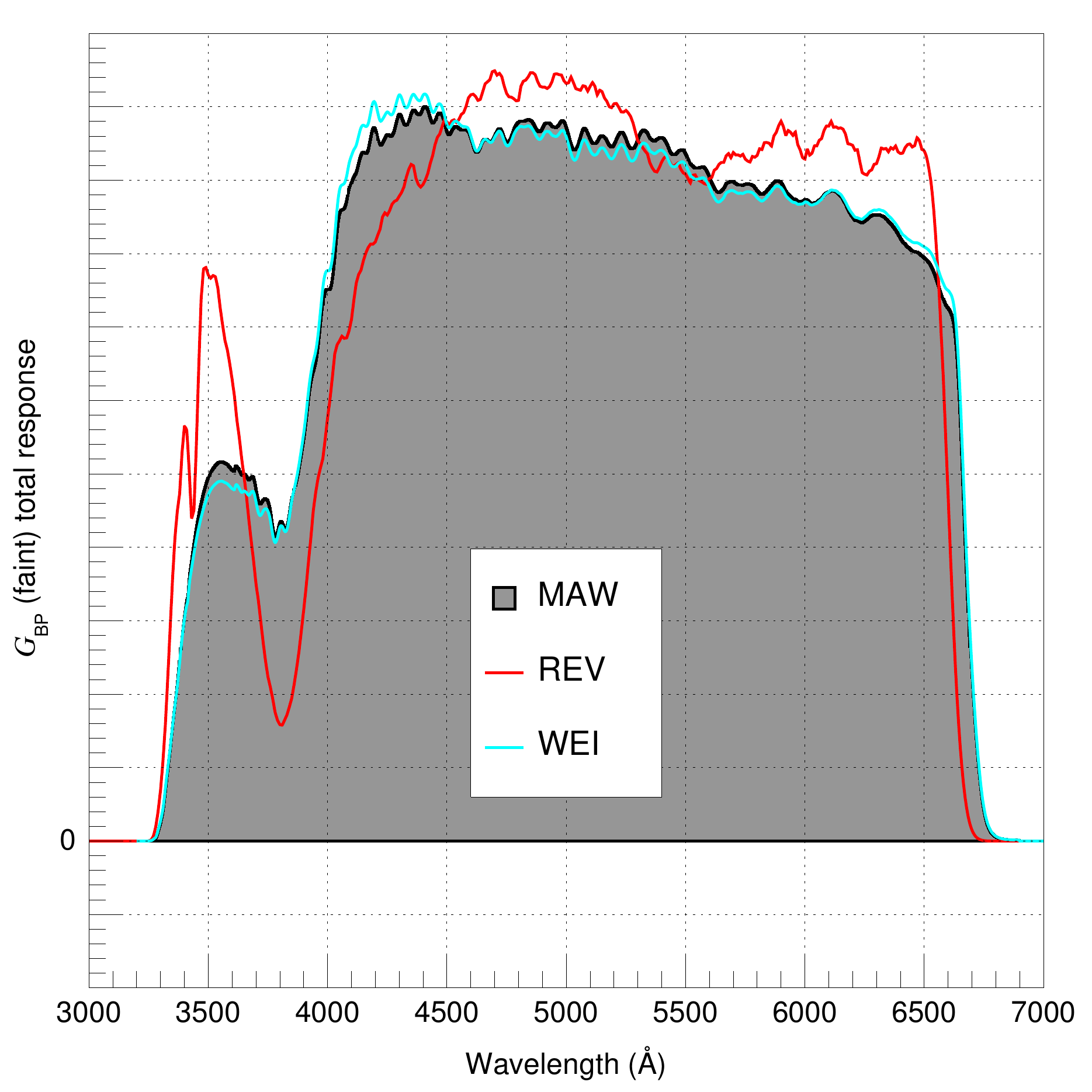}}
\centerline{\includegraphics[width=0.49\linewidth]{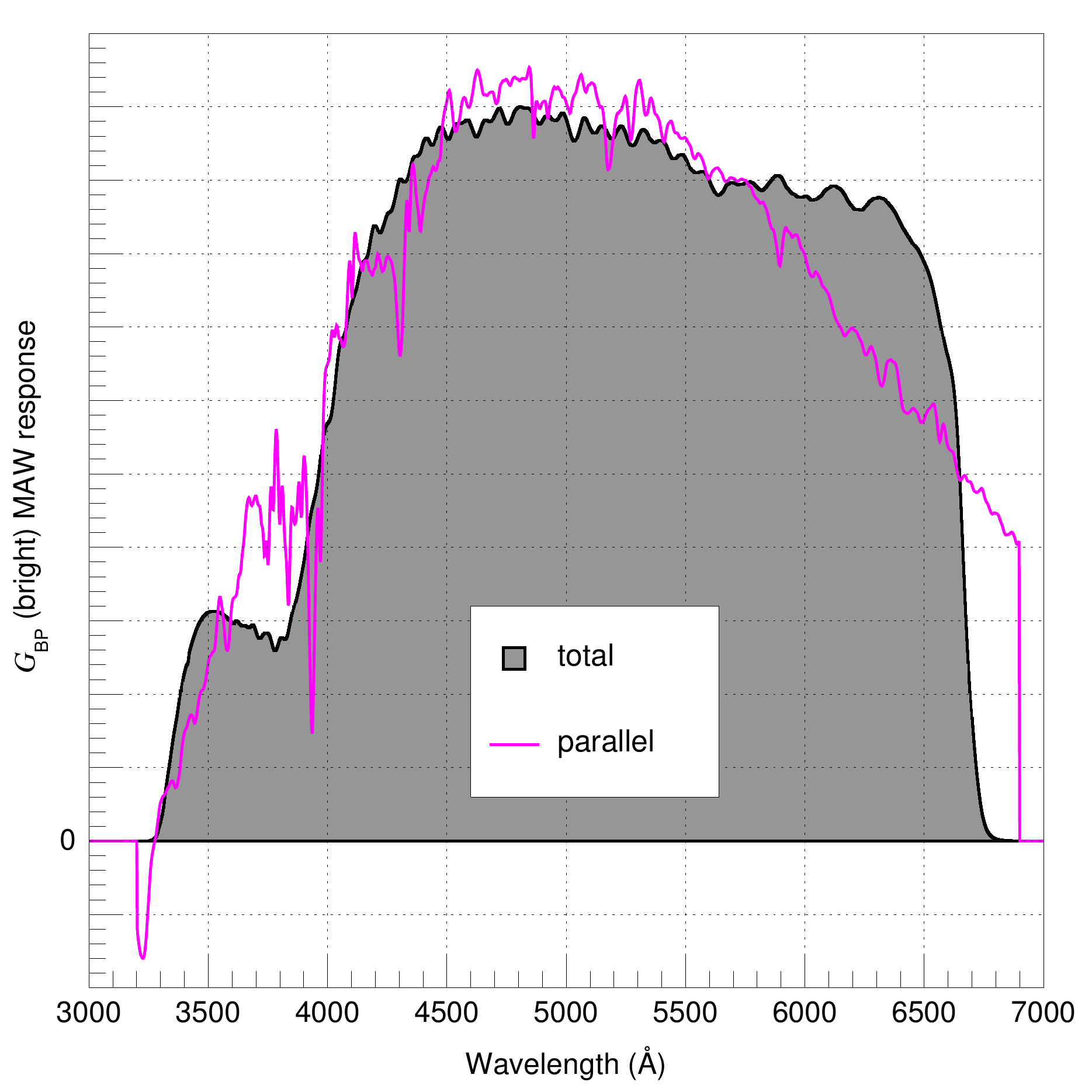} \
            \includegraphics[width=0.49\linewidth]{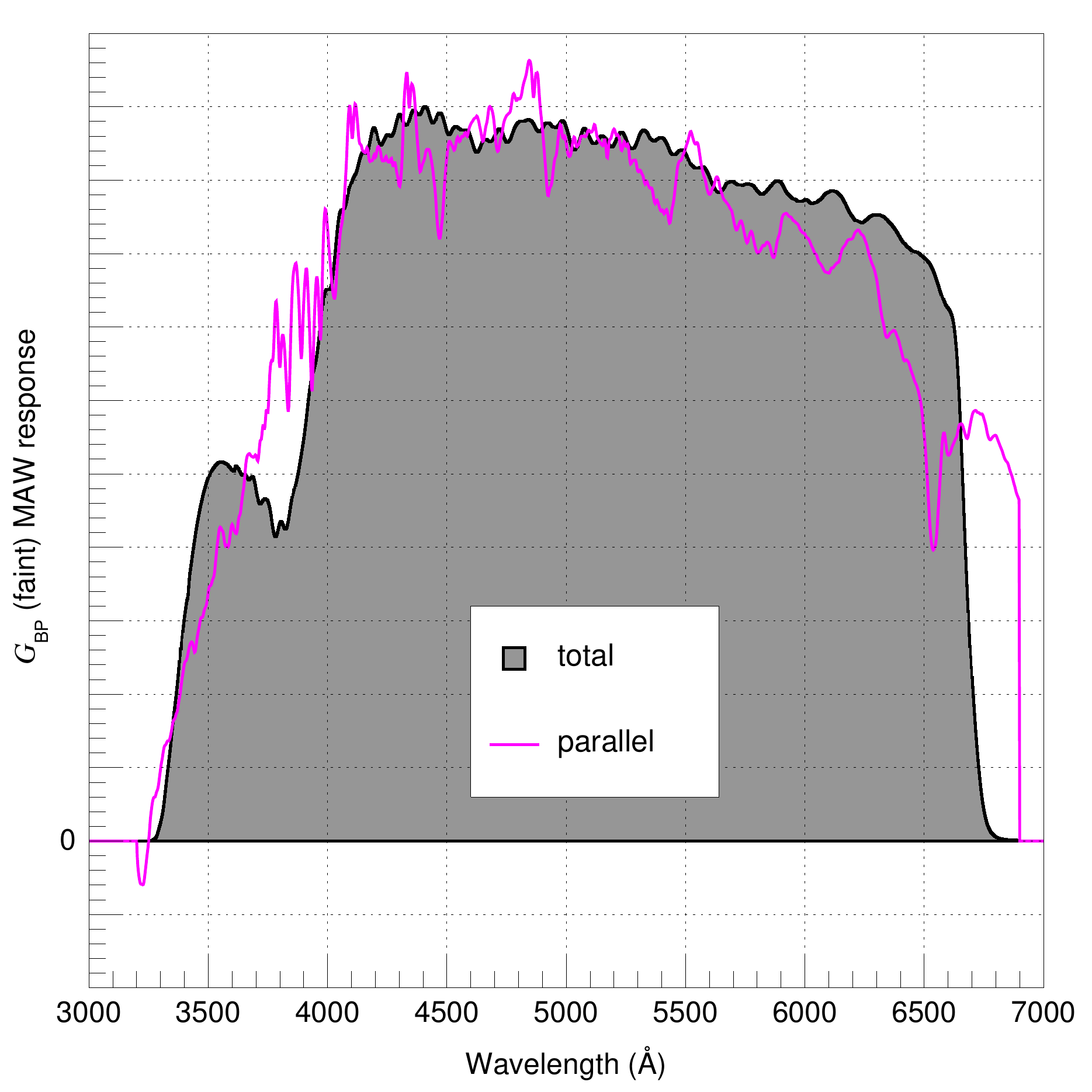}}
\caption{Same as Fig.~\ref{Gsens} for \GBP. (left) Bright magnitude range. (right) Faint magnitude range.}
\label{GBPsens}
\end{figure*}

\begin{figure}
\centerline{\includegraphics[width=\linewidth]{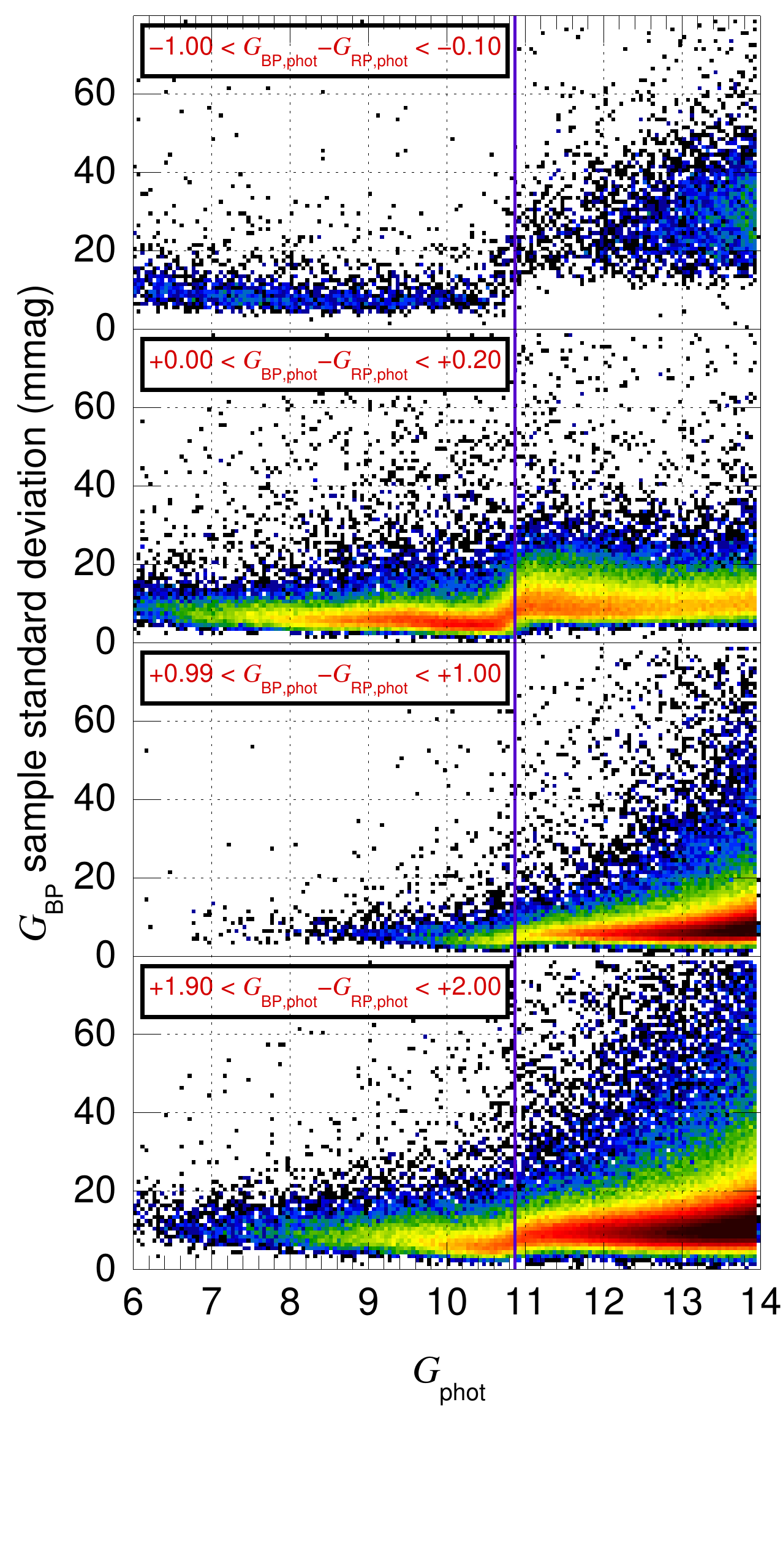}}
\centerline{\vspace{-22mm}}
\caption{Sample standard deviations of the \GBP\ flux as a function of \GGp\ for {\it Gaia} DR2 stars far from the Galactic Plane
         ($|b| > 30\degr$) for four color ranges $\GBPp-\GRPp$. The histogram shows the logarithm of the number of stars per bin on a common scale
         for all four panels. Note that the total number of stars per panel increases from the top one to the bottom one. The location of the break
         is marked with a purple line.}
\label{GBPstdev}
\end{figure}

$\,\!$\indent Systematic errors were also previously detected for {\it Gaia} DR2 \GBP\ photometry. \citet{Arenetal18} describe a branching of the 
$\GBP - \GRP$ 
versus $\GG - \GBP$ color-color relation for very blue sources, occurring at a \GG\ magnitude around 11. This branching results in a ``jump'' of 
approximately 20 mmag in \GBP. \citet{Weil18} confirmed the inconsistency in the \GBP\ photometry by comparing observed and synthetic magnitudes 
resulting from the REV passband for four different spectral libraries, and located the position of the jump in the range between \GG\ magnitudes of 
10.47 and 10.99. As the differences between the \GBP\ magnitudes for sources brighter and fainter than the position of the jump depends on color, 
different sensitivity curves for both sides of the jump are required to describe the \GBP\ photometry. \citet{Weil18} thus presented two different 
passbands for \GBP, valid for sources brighter and fainter than 10.99 in \GG.

The set of calibration spectra used in this work confirms the existence of the inconsistency of the \GBP\ photometry for bright and faint blue 
sources. In order to better constrain the position of the jump, we used the errors on the mean \GBP\ fluxes. Multiplying these values, 
provided with {\it Gaia} DR2, with the square root of the number of observations results in the standard deviation of the epoch fluxes that entered 
into the computation of the mean fluxes \citep{Carretal16}, which we refer to as the sample standard deviation to clearly distinguish it from the 
standard deviation of the mean flux. Figure~\ref{GBPstdev} shows this sample standard deviation for all sources with Galactic latitude 
$\rm |b| > 30^\circ$ for small $\GBPp-\GRPp$ color intervals. An abrupt break occurs at a \GGp\ magnitude of about 10.87, with a clearly increased 
standard deviation for blue sources fainter than 10.87~mag than for sources brighter than that limit. This jump in the sample standard deviations 
decreases strongly for redder sources. Assuming that the jump in the sample standard deviations in {\it Gaia's} \GBP\ photometry has the same origin 
as
the jump in the mean magnitudes, we can thus locate the position of the jump more accurately than was done by \citet{Weil18} and we obtain a value 
of 10.87~mag in \GGp.

Although the jump in the sample standard deviations for \GBP\ decreases strongly with increasing color index, it remains detectable up to a 
$\GBPp-\GRPp$ color of about two. When comparing the position of the jump as a function of color in \GG\ and \GBP, we find that the position in \GG\
is far less dependent on color than it is in \GBP. We can therefore confirm that the position of the jump in \GBP\ photometry is determined by the 
\GG\ magnitude rather than the \GBP\ magnitude of a source. As the choice of the instrumental configuration (gate and window class) under which a 
star is observed by {\it Gaia} is chosen according to an estimate of the \GG\ magnitude, the abrupt jump in \GBP\ at 10.87~mag points 
to a problem with the calibration of the \GBP\ photometry at different instrumental configurations. The increased sample standard deviation for blue 
sources fainter than 10.87~mag suggests that the calibration of faint blue sources is less accurate than for the brighter sources, resulting also in the 
observed systematic color-dependent differences in the mean \GBP\ fluxes and magnitudes.

To take this effect properly into account, we derive two different \GBP\ passbands in this work as was already done in \citet{Weil18}, valid for 
sources brighter and fainter than 10.87~mag in \GGp, respectively. The two sensitivity curves were computed using the \citet{Weil18} \GBP\ passbands as 
the initial guess, and using 6 and 5 basis functions for the bright and faint magnitude range, respectively. The resulting bright and faint 
passbands are shown in Fig.~\ref{GBPsens} and compared to the REV and WEI ones, noting that the REV result is the same for both magnitude ranges. The 
same pattern as with \GG\ takes place here: the new curves are more similar to the WEI result 
than
to the REV result. The main difference of either MAW
or WEI with REV is that REV shows a much more prominat peak for $\lambda < 3800$~\AA, i.e. to the left of the Balmer jump. The peak is weak in the WEI 
curves and in the faint MAW curve but has almost disappeared from the bright MAW curve. The WEI and MAW faint curves are very similar while the WEI 
bright curve is slightly ``bluer'' (in the sense of being more sensitive at shorter wavelengths) than the MAW equivalent.

\subsection{Sensitivity curve for \GRP}

\begin{figure}
\centerline{\includegraphics[width=\linewidth]{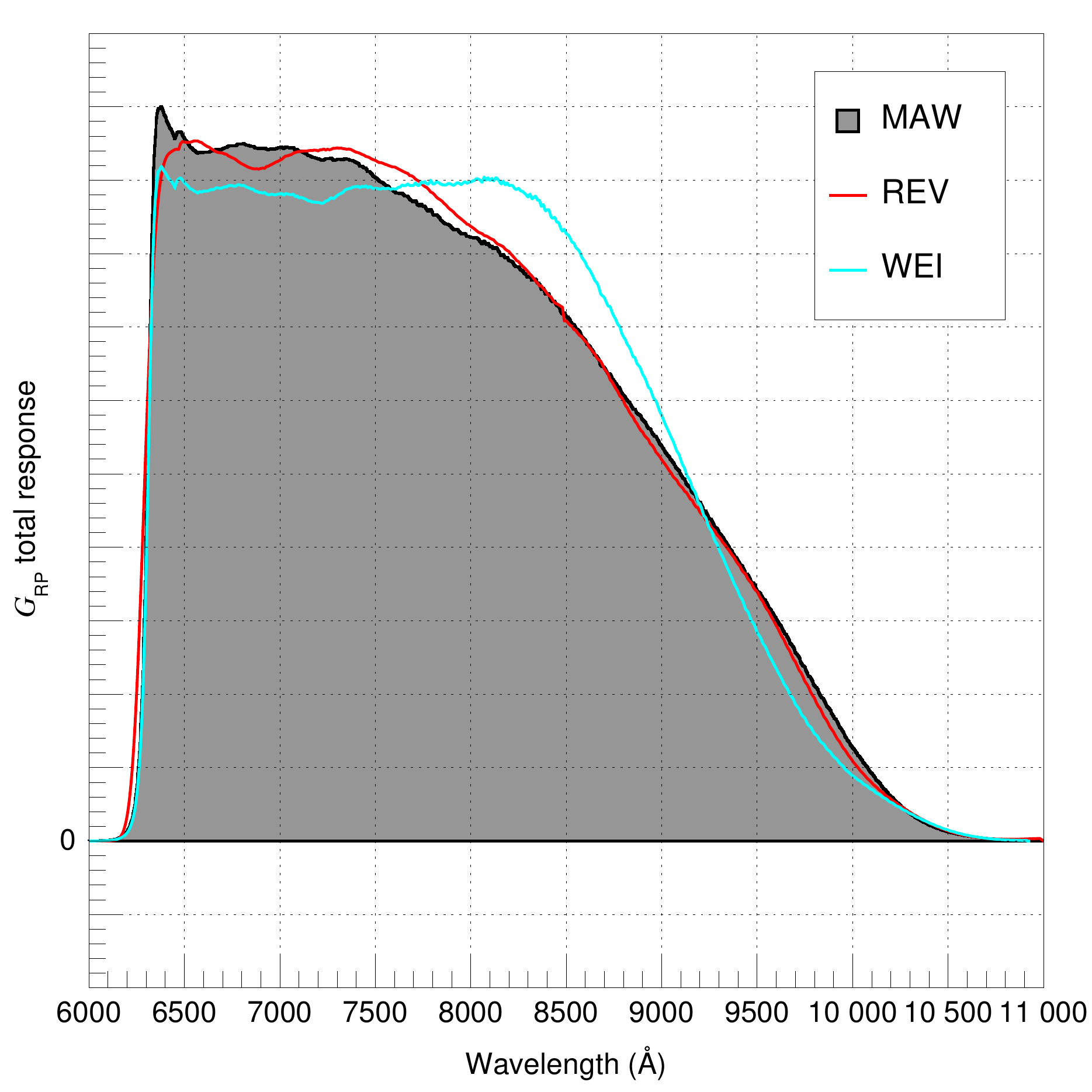}}
\centerline{\includegraphics[width=\linewidth]{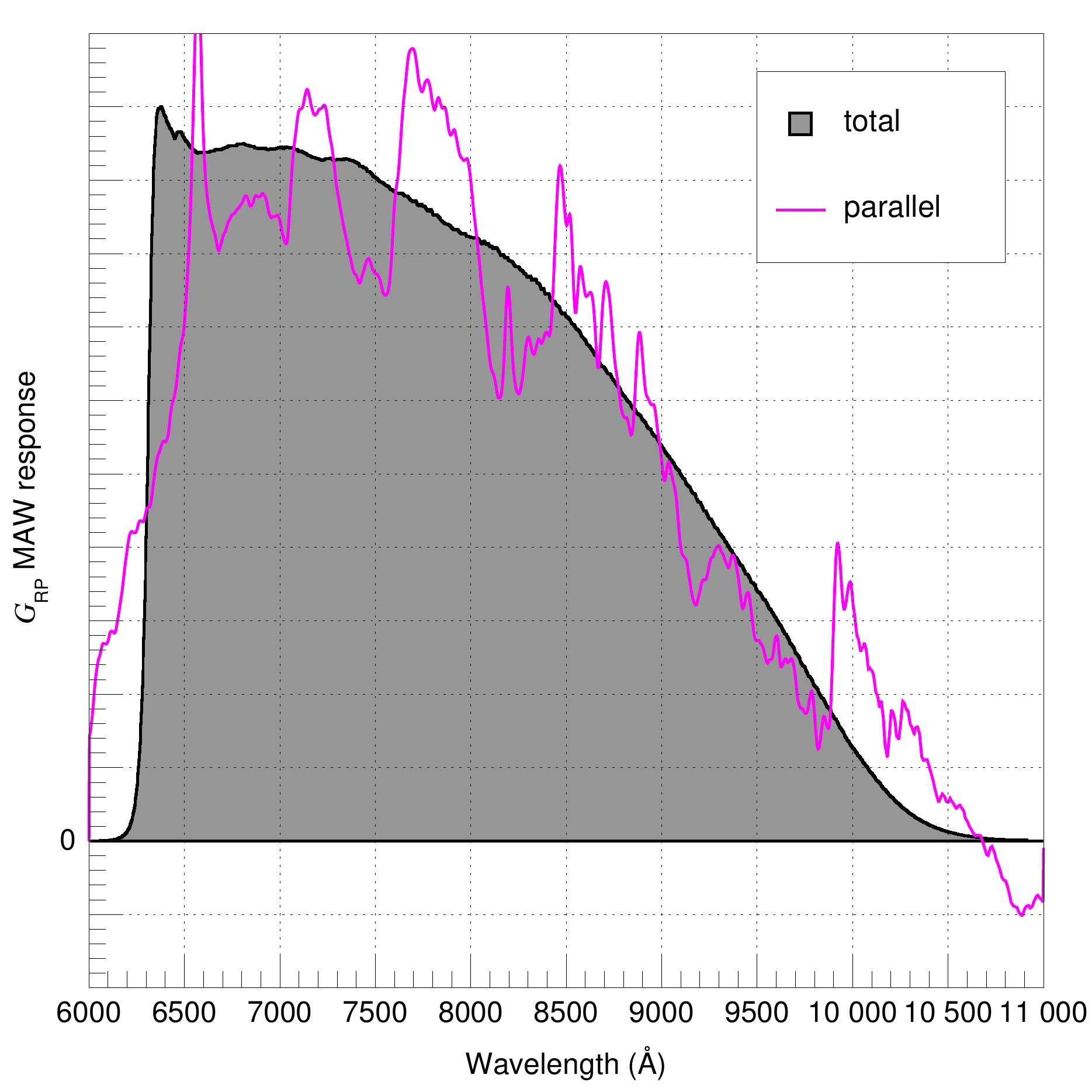}}
\caption{Same as Fig.~\ref{Gsens} for \GRP.}
\label{GRPsens}
\end{figure}

The \GRP\ photometry is less affected by systematic errors than the \GG\ and \GBP\ photometry. \citet{Weil18} derived a \GRP\ passband that differs 
clearly from the REV passband. The main effect of the strong change in the RP passband was the removal of a small color tendency in the residuals 
for the SPSS and Stritzinger spectra. \citet{Weil18} however noted that while improving the \GRP\ residuals for the SPSS and Stritzinger spectra, 
the residuals for the CALSPEC spectra are worse with the WEI passband. Having more calibration spectra available for this work, we can assess the 
differences between different sets of spectra in more detail in this work.

The \GRP\ passband in this work is computed again with the WEI passband as an initial guess and with 6 basis functions. The solution found in this 
work is presented in Fig.~\ref{GRPsens} and compared to the REV and WEI ones. The situation for \GRP\ is different than for \GG\ or \GBP\ in the sense
that MAW is more similar to REV than to WEI. WEI is more sensitive in the 8000-9000~\AA\ region and less sensitive in the 9500-10\,000~\AA\ region, which
is the opposite situation to what we find for \GG, where REV was the one that had those differences in similar wavelength ranges. MAW and REV are not
identical and their most importance difference is found below 8000~\AA, where REV is redder. In the next section we will explore the consequences of these
effects.

\section{Testing the new sensitivity curves}

$\,\!$\indent We test the MAW sensitivity curves we have generated for the three {\it Gaia} DR2 passbands using a similar strategy to the one we
recently employed for {\it Gaia} DR1 \GG\ photometry \citep{Maiz17} and for the three 2MASS bands \citep{MaizPant18}. We check that there are no
magnitude or color terms when plotting the difference between observed and synthetic magnitudes and we determine the Vega ZPs (ZP$_{{\rm Vega},p}$, see 
the Appendix for definitions and for how to use the alternative AB or ST systems) for each of the three filters $p$. We do the same comparison for the 
REV and WEI sensitivity curves.

For REV and WEI we start by calculating and applying corrections to \GG\ in the same way we did for MAW with 
Eqns.~\ref{Gcor2}~and~\ref{Gcor3}. We then calculate a minimum uncertainty in each band $\sigma_{{\rm min},p}$ (divided by magnitude or color ranges, 
as appropriate, see below) from the dispersion of the data, as we did in \citet{Maiz06a} for Johnson $UBV$ and Str\"omgren $uvby$ photometry and in 
\citet{Maiz17} for {\it Gaia} DR1 photometry. The minimum uncertainty is the threshold value that should be applied when comparing observed and synthetic
magnitudes and depends on the accuracies of the absolute calibration of the spectrophotometric library and the passband characterization as well as on 
the possible existence of variabilty in the sample. The individual uncertainty for the magnitude of a given star used in this paper is the larger one of 
(a) the minimum uncertainty for that filter and (b) the published one in each case. We then fit a restricted (slope forced to zero) and an unrestricted
linear fit to the difference between (corrected if needed) observed magnitudes and synthetic ones as a function of the $\GBPp-\GRPp$ color. The value of 
the restricted fit yields the ZP$_{{\rm Vega},p}$ and the slope of the unrestricted fit $b_p$ indicates the possible existence of a color term. Our 
proposed values for ZP$_{{\rm Vega},p}$, $\sigma_{{\rm min},p}$, and $b_p$ are given in Table~\ref{results}.

\begin{table*}
\caption{Results for the three {\it Gaia} bands using the MAW, REV, and WEI sensitivity curves.}
\label{results}
\centerline{
\renewcommand{\arraystretch}{1.3}
\begin{tabular}{lcr@{.}lr@{.}lr@{.}l}
\hline
\multicolumn{1}{c}{Property} & \multicolumn{1}{c}{Range} &  \multicolumn{2}{c}{MAW} &  \multicolumn{2}{c}{REV} &  \multicolumn{2}{c}{WEI} \\
\hline
$\;\;$ ZP$_{{\rm Vega},G}$             &                   & $+0$&$000\pm 0.001$ mag  & $-0$&$004\pm 0.001$ mag  & $-0$&$005\pm 0.001$ mag  \\
$\;\;$ $\sigma_{{\rm min},G}$          & $\GGp > 6 $       &  $0$&$008$ mag           &  $0$&$013$ mag           &  $0$&$008$ mag           \\
                                       & $\GGp < 6 $       &  $0$&$012$ mag           &  $0$&$012$ mag           &  $0$&$012$ mag           \\
$\;\;$ $b_G$                           &                   & $-0$&$8\pm 1.0$ mmag/mag & $+9$&$1\pm 1.6$ mmag/mag & $-0$&$4\pm 1.0$ mmag/mag \\ 
\hline
$\;\;$ ZP$_{{\rm Vega},G_{\rm BP}}$    & $\GGp > 10.87 $   & $+0$&$005\pm 0.002$ mag  & $+0$&$023\pm 0.002$ mag  & $+0$&$003\pm 0.002$ mag  \\
                                       & $\GGp < 10.87 $   & $+0$&$026\pm 0.001$ mag  & $+0$&$042\pm 0.002$ mag  &  $0$&$029\pm 0.001$ mag  \\
$\;\;$ $\sigma_{{\rm min},G_{\rm BP}}$ & $\GGp > 10.87 $   &  $0$&$009$ mag           &  $0$&$012$ mag           &  $0$&$010$ mag           \\
                                       & $\GGp < 10.87 $   &  $0$&$009$ mag           &  $0$&$020$ mag           &  $0$&$011$ mag           \\
$\;\;$ $b_{G_{\rm BP}}$                &                   & $+0$&$0\pm 1.1$ mmag/mag & $-3$&$3\pm 1.8$ mmag/mag & $-3$&$8\pm 1.3$ mmag/mag \\ 
\hline
$\;\;$ ZP$_{{\rm Vega},G_{\rm RP}}$    &                   & $+0$&$012\pm 0.001$ mag  & $+0$&$011\pm 0.001$ mag  & $+0$&$016\pm 0.001$ mag  \\
$\;\;$ $\sigma_{{\rm min},G_{\rm RP}}$ & $\GBPp-\GRPp < 2$ &  $0$&$010$ mag           &  $0$&$011$ mag           &  $0$&$011$ mag           \\
                                       & $\GBPp-\GRPp > 2$ &  $0$&$022$ mag           &  $0$&$030$ mag           &  $0$&$025$ mag           \\
$\;\;$ $b_{G_{\rm RP}}$                &                   & $-1$&$6\pm 1.7$ mmag/mag & $-3$&$9\pm 1.9$ mmag/mag & $+5$&$8\pm 1.9$ mmag/mag \\ 
\hline
\end{tabular}
\renewcommand{\arraystretch}{1.0}
}
\end{table*}

\subsection{\GG}

\begin{figure*}
\centerline{\includegraphics[width=0.49\linewidth, bb=28 90 566 512]{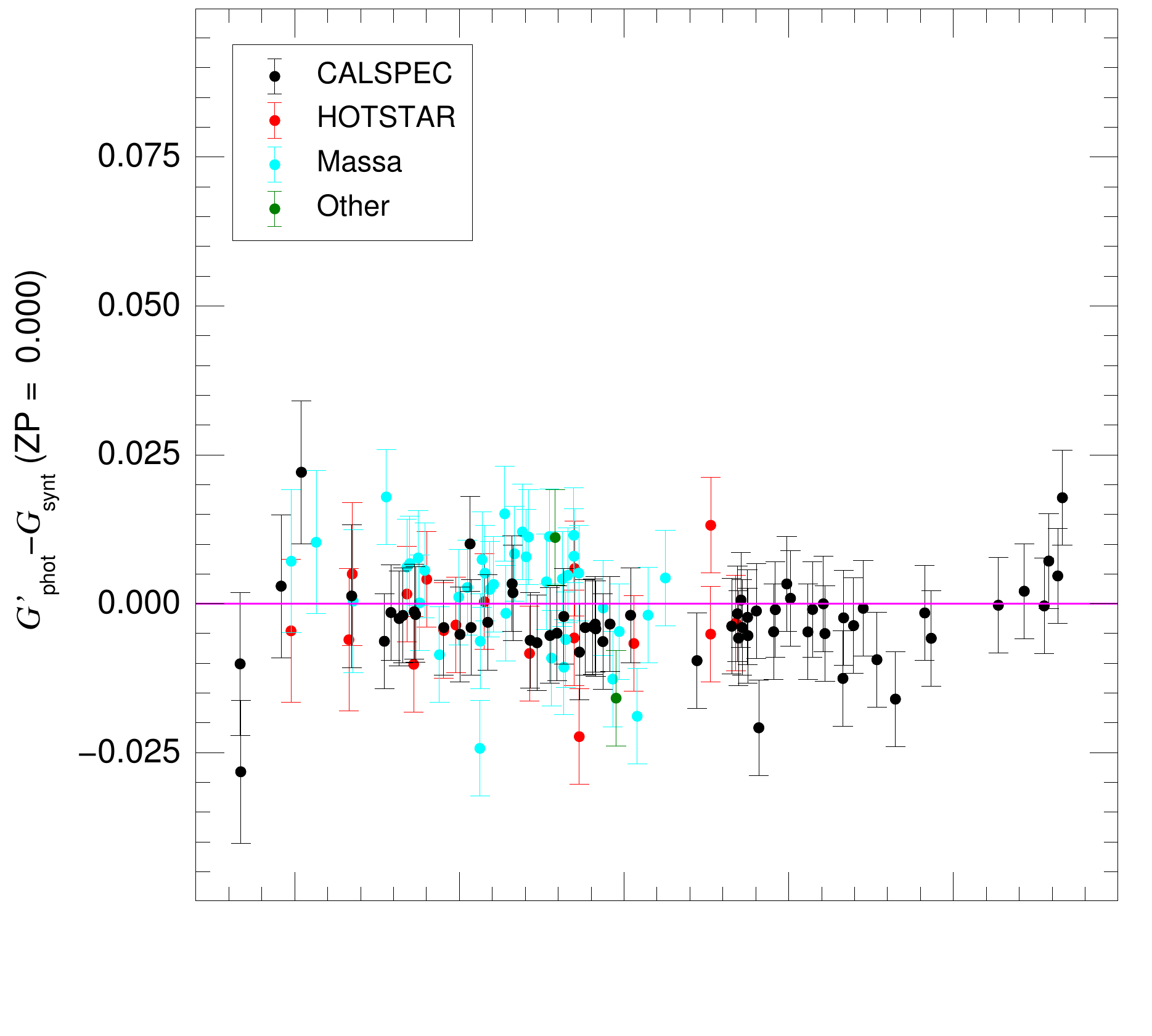} \
            \includegraphics[width=0.49\linewidth, bb=28 90 566 512]{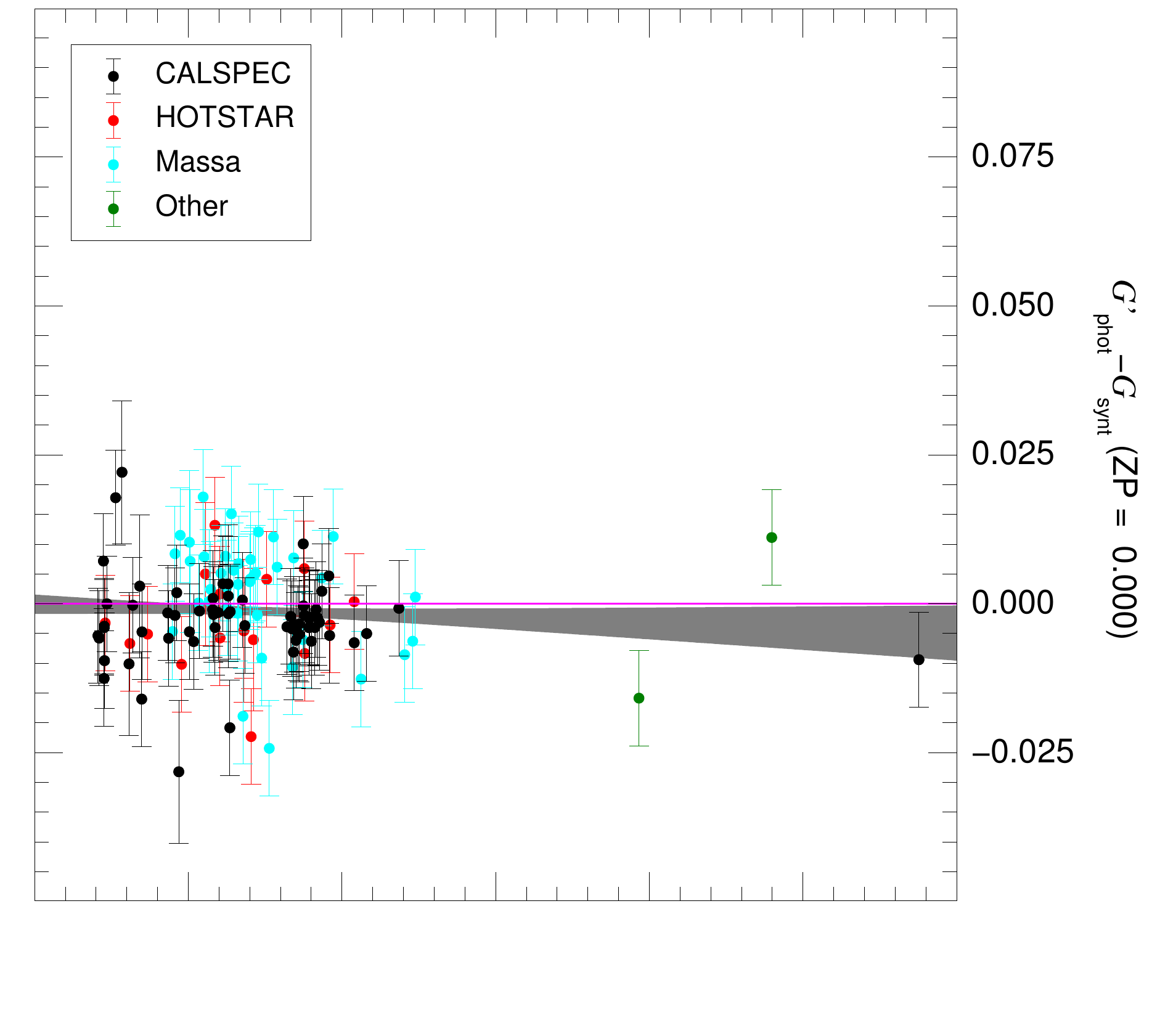}}
\centerline{\includegraphics[width=0.49\linewidth, bb=28 90 566 512]{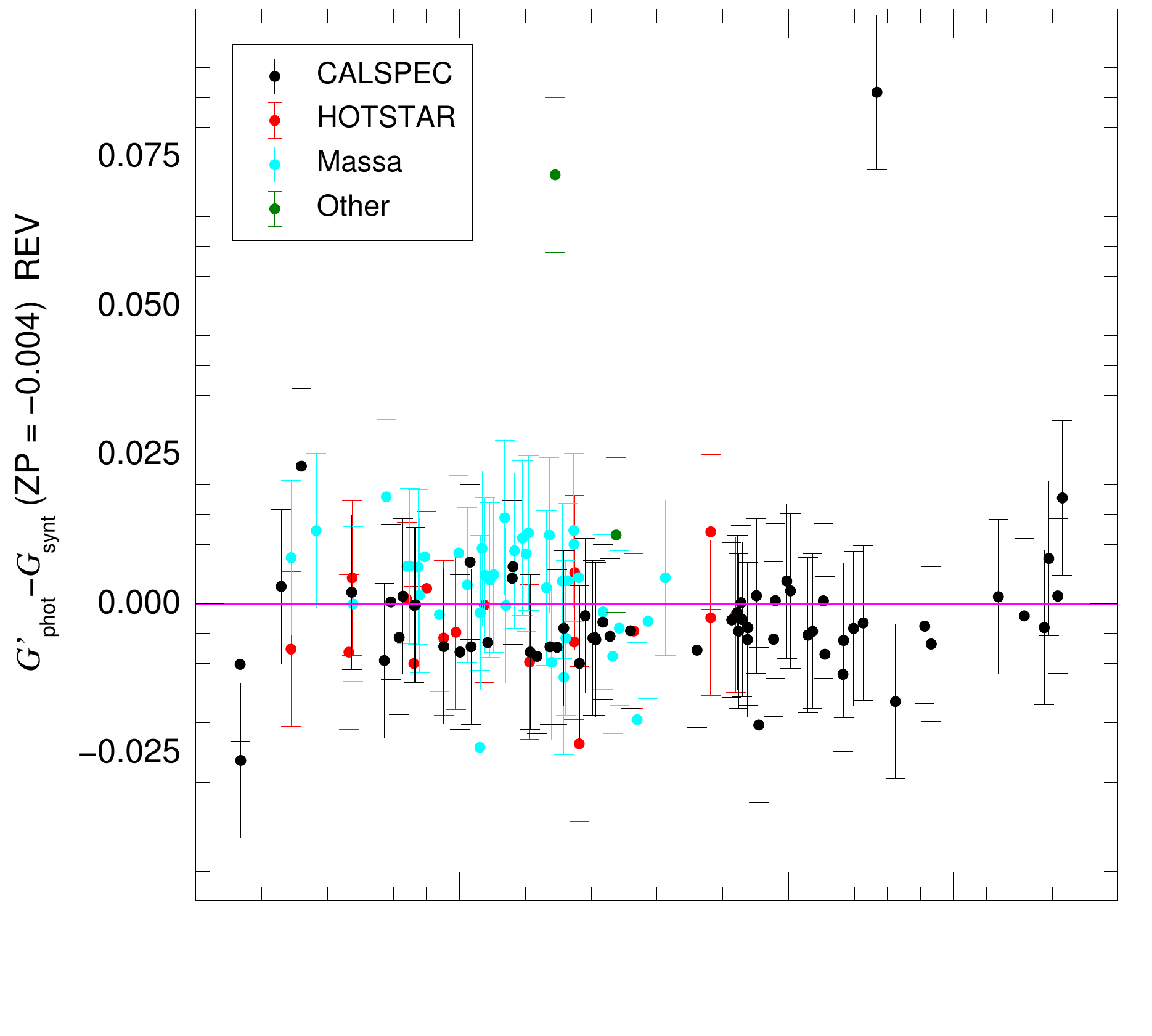} \
            \includegraphics[width=0.49\linewidth, bb=28 90 566 512]{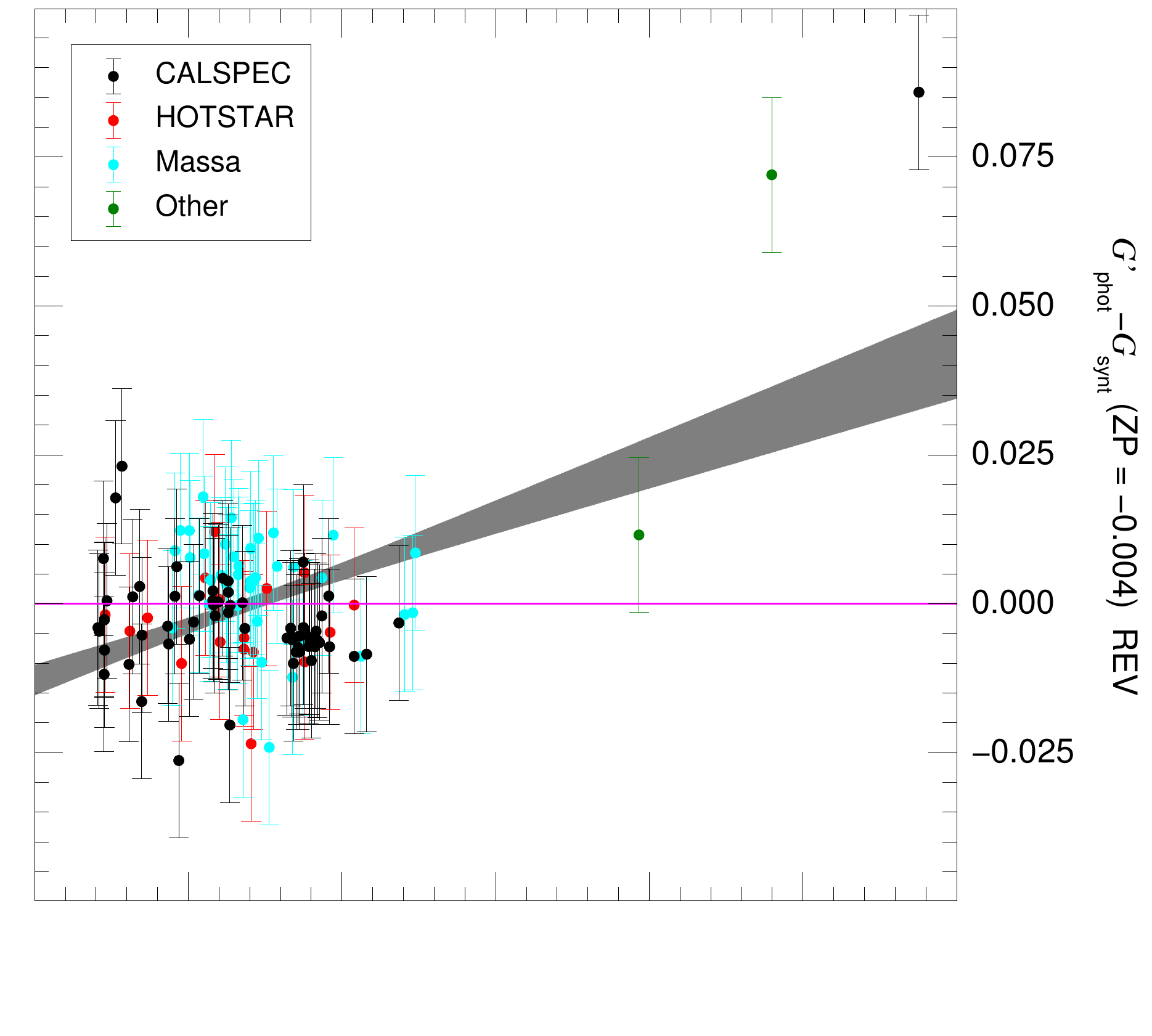}}
\centerline{\includegraphics[width=0.49\linewidth, bb=28 90 566 512]{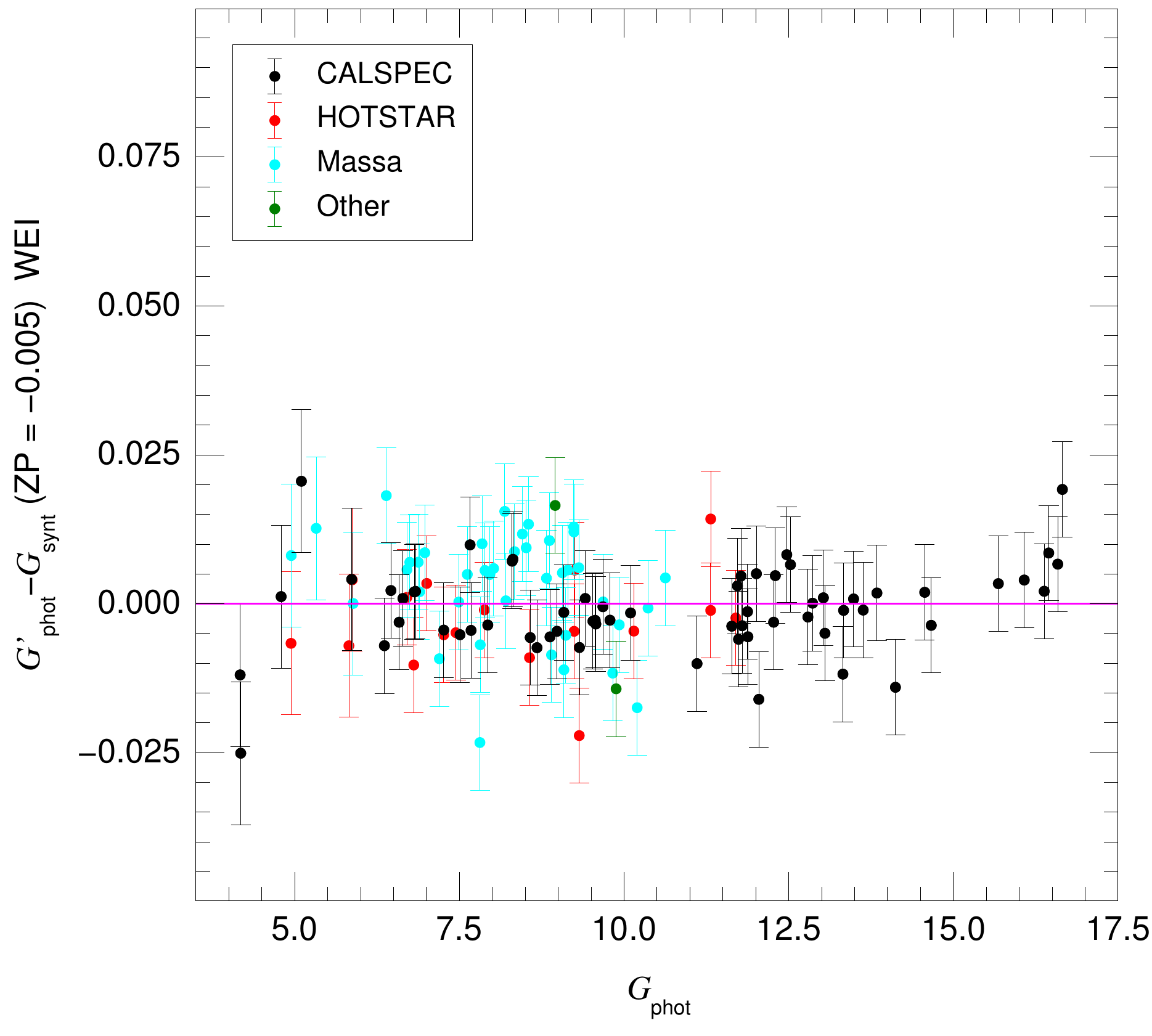} \
            \includegraphics[width=0.49\linewidth, bb=28 90 566 512]{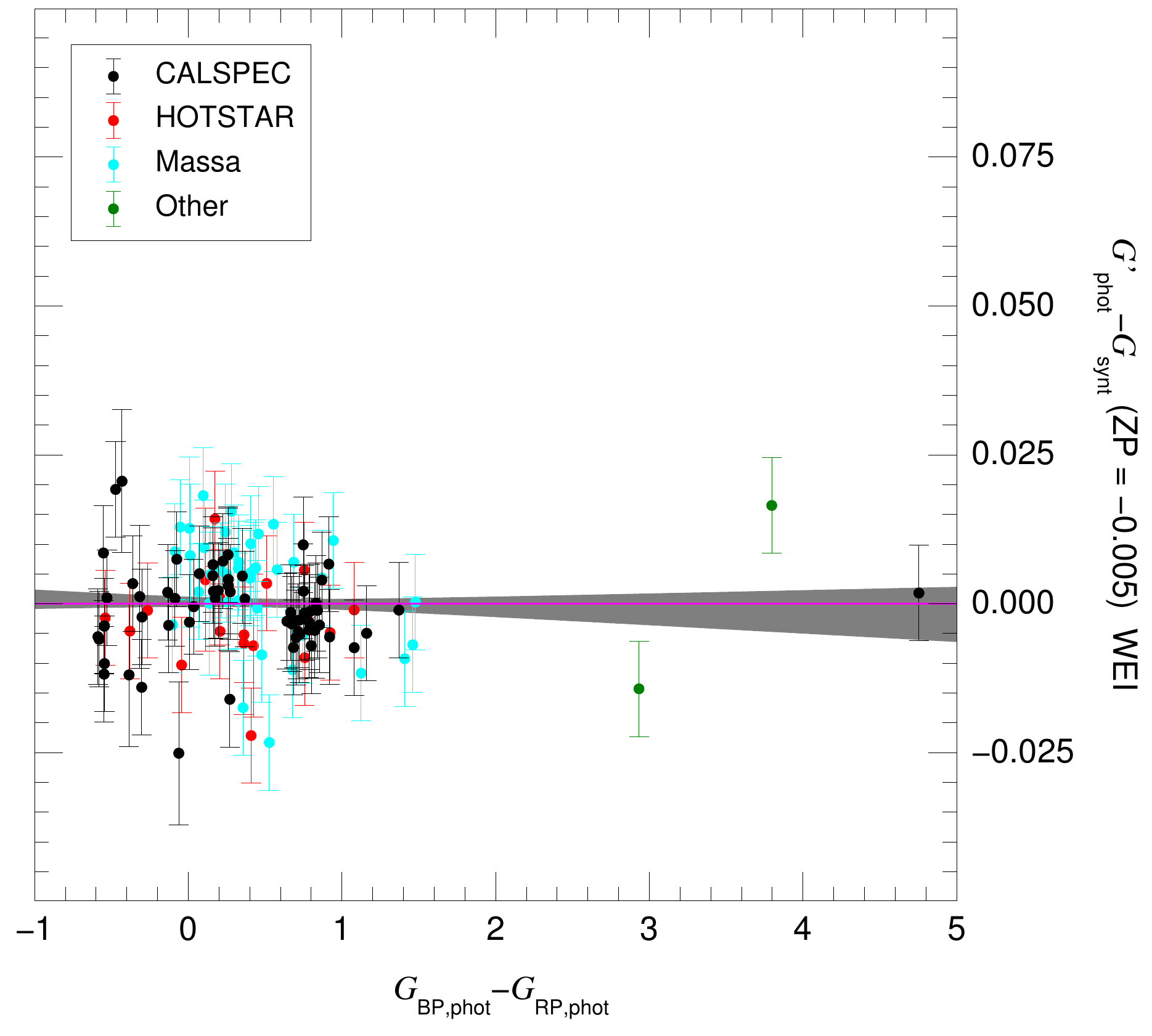}}
\centerline{\vspace{12mm}}
\caption{Comparison between the corrected observed \GG\ magnitudes and the synthetic \GG\ magnitudes as a function of \GGp\ (left column) and as
         a function of $\GBPp-\GRPp$ (right column). The first, second, and third row show the result for MAW, REV, and WEI, 
         respectively. Data points and error bars are color-coded by data set. 
         The region shaded in gray in the 
         right
         column shows the 1~$\sigma$ confidence range for the unrestricted fit.}
\label{Gplots}
\end{figure*}

\begin{figure*}
\centerline{\includegraphics[width=0.49\linewidth, bb=28 90 566 512]{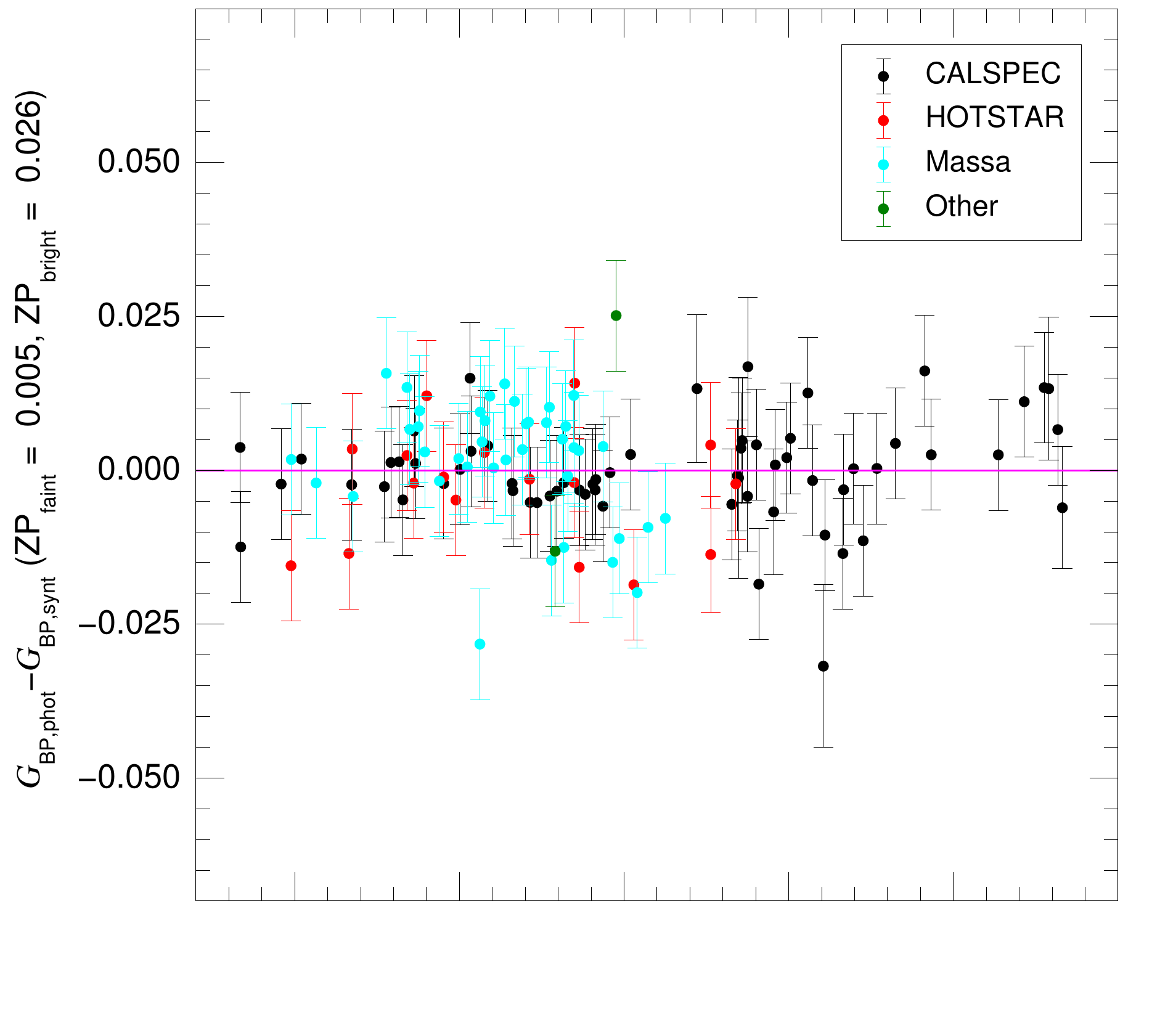} \
            \includegraphics[width=0.49\linewidth, bb=28 90 566 512]{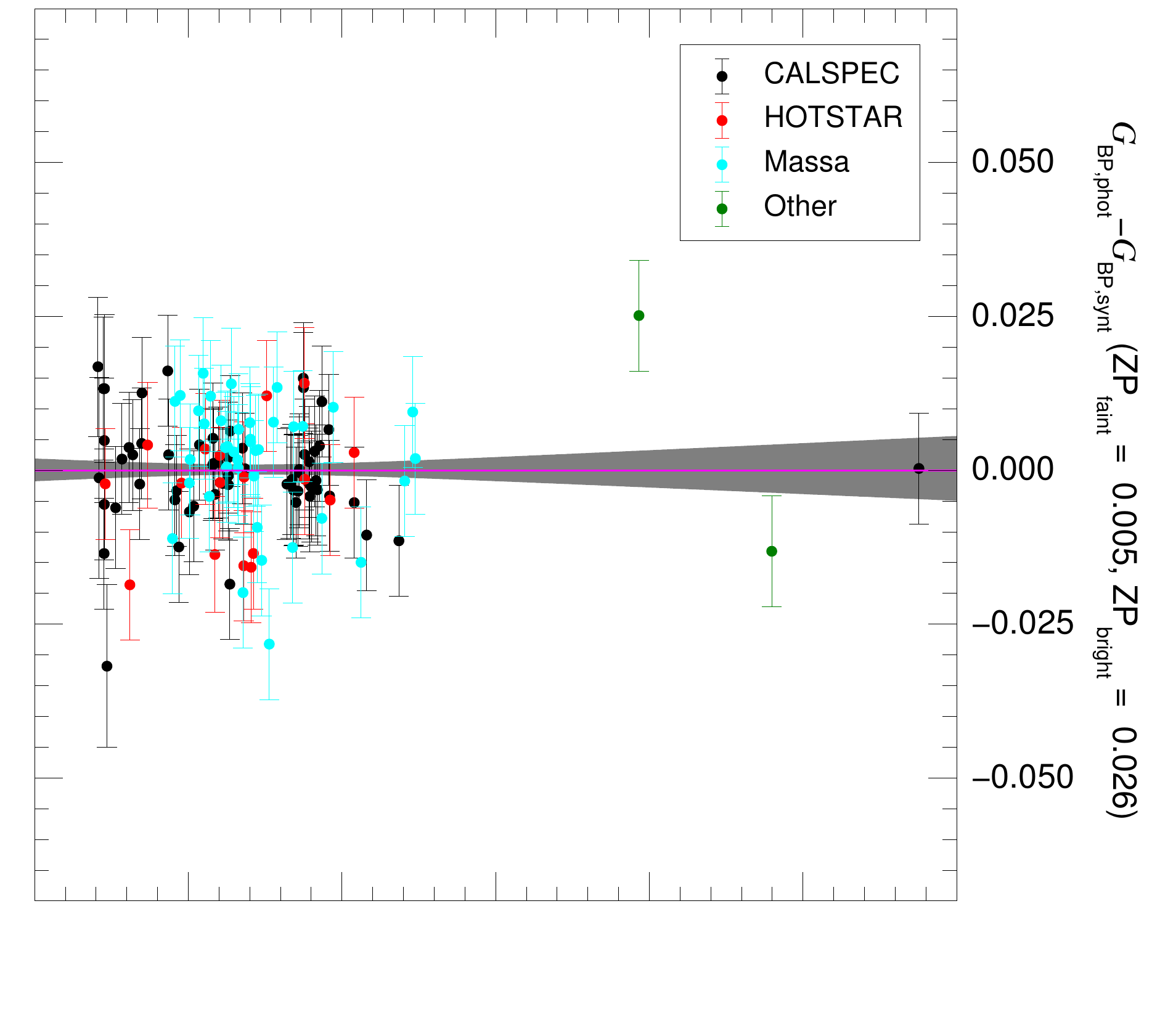}}
\centerline{\includegraphics[width=0.49\linewidth, bb=28 90 566 512]{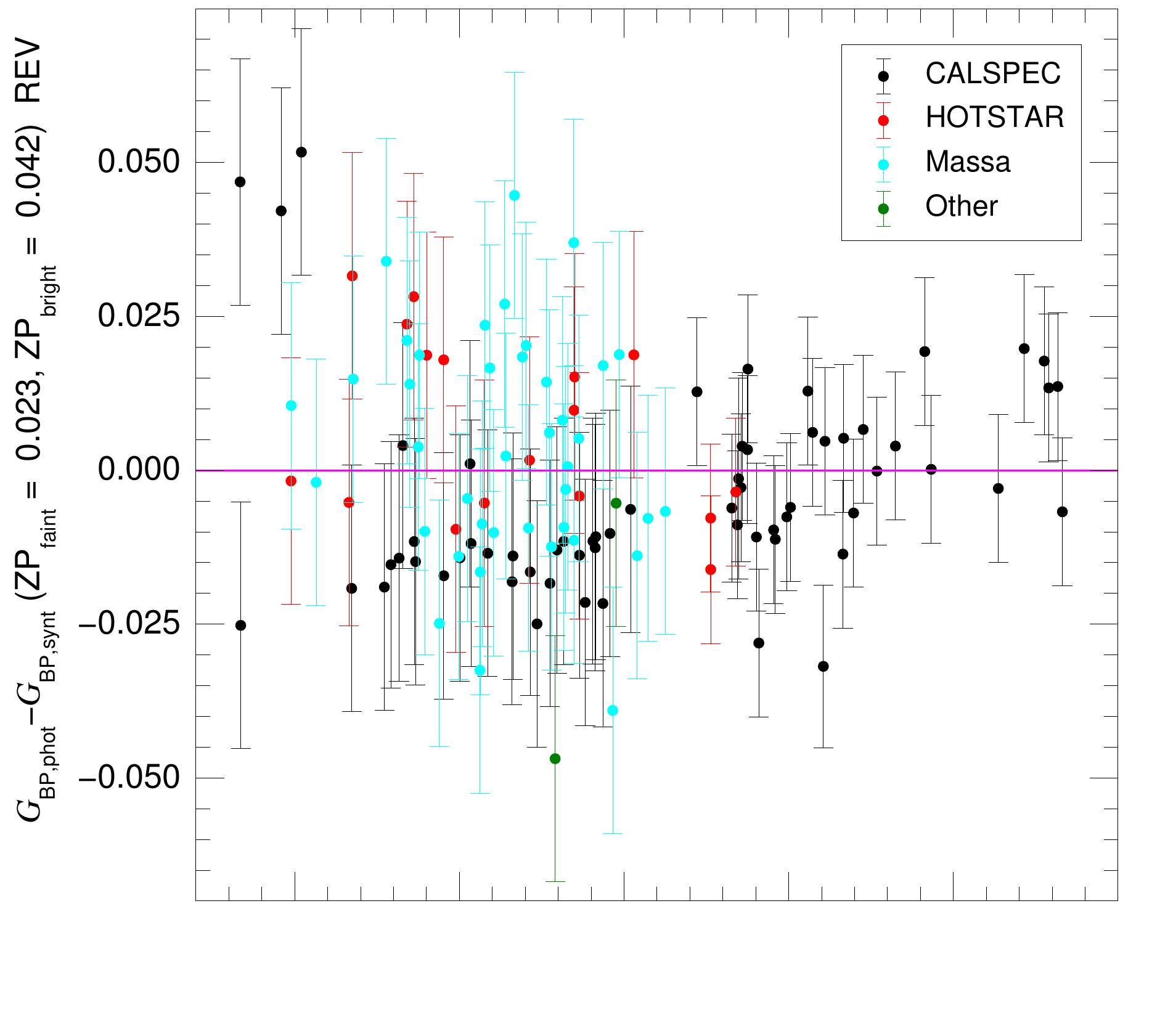} \
            \includegraphics[width=0.49\linewidth, bb=28 90 566 512]{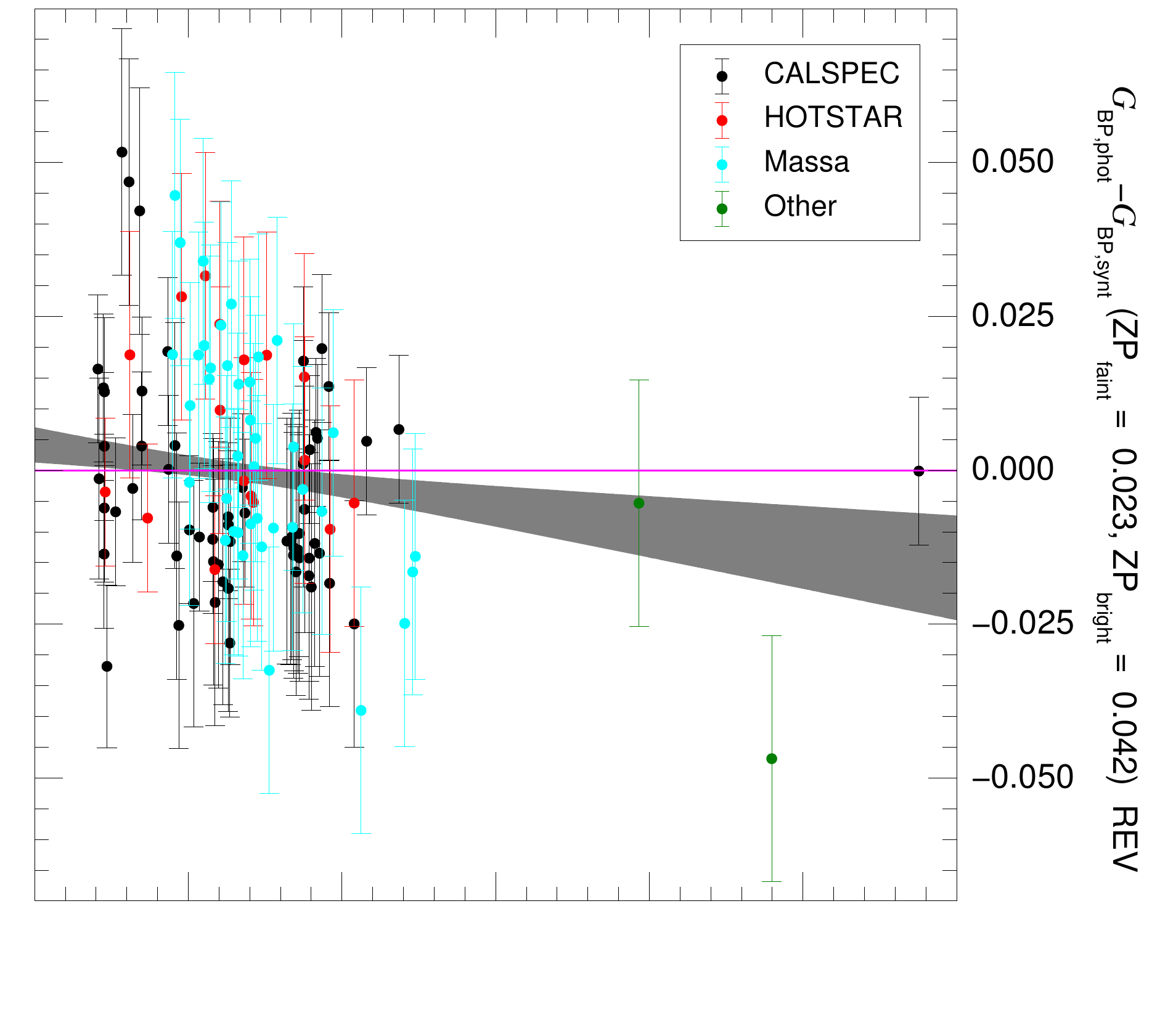}}
\centerline{\includegraphics[width=0.49\linewidth, bb=28 90 566 512]{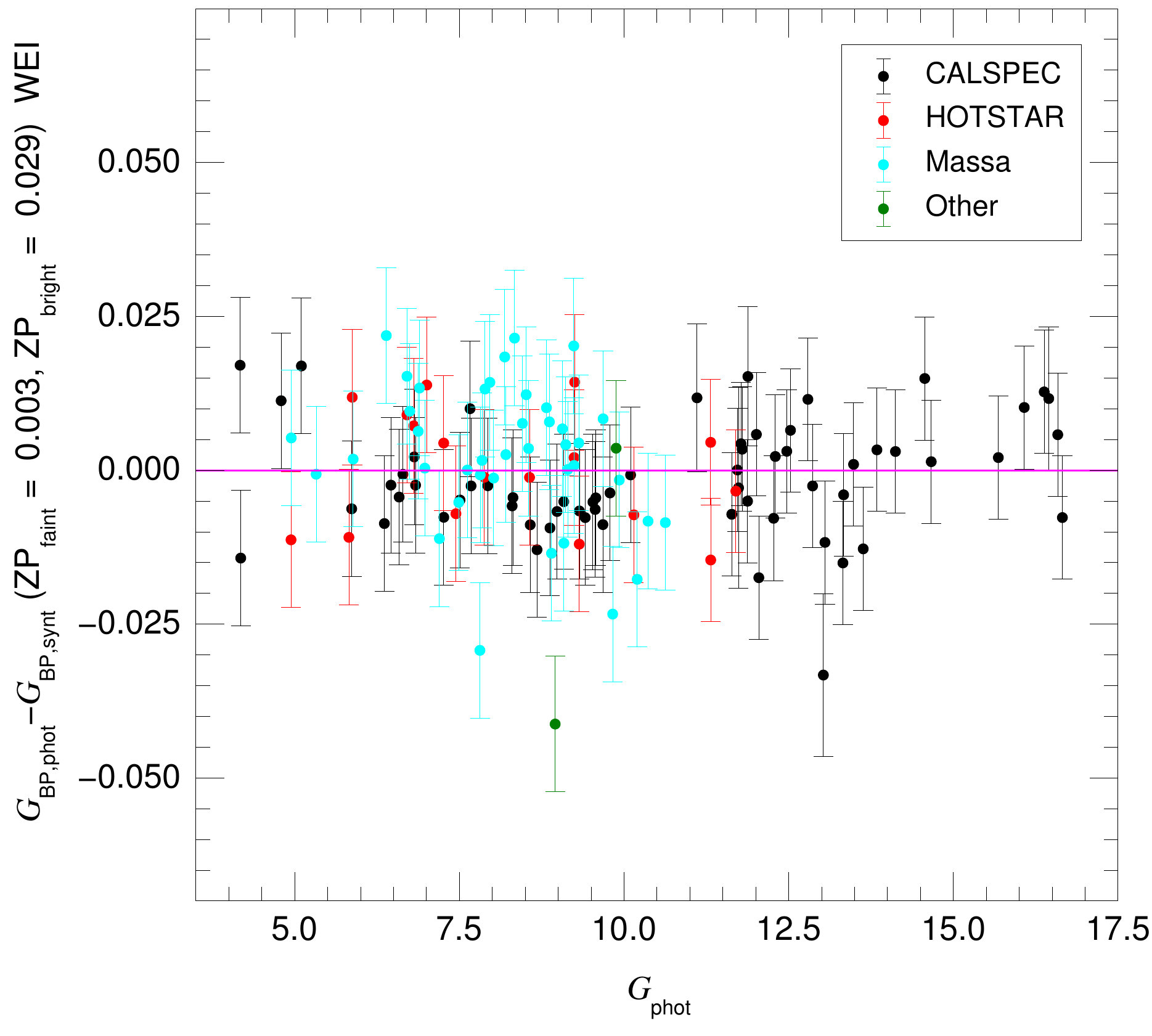} \
            \includegraphics[width=0.49\linewidth, bb=28 90 566 512]{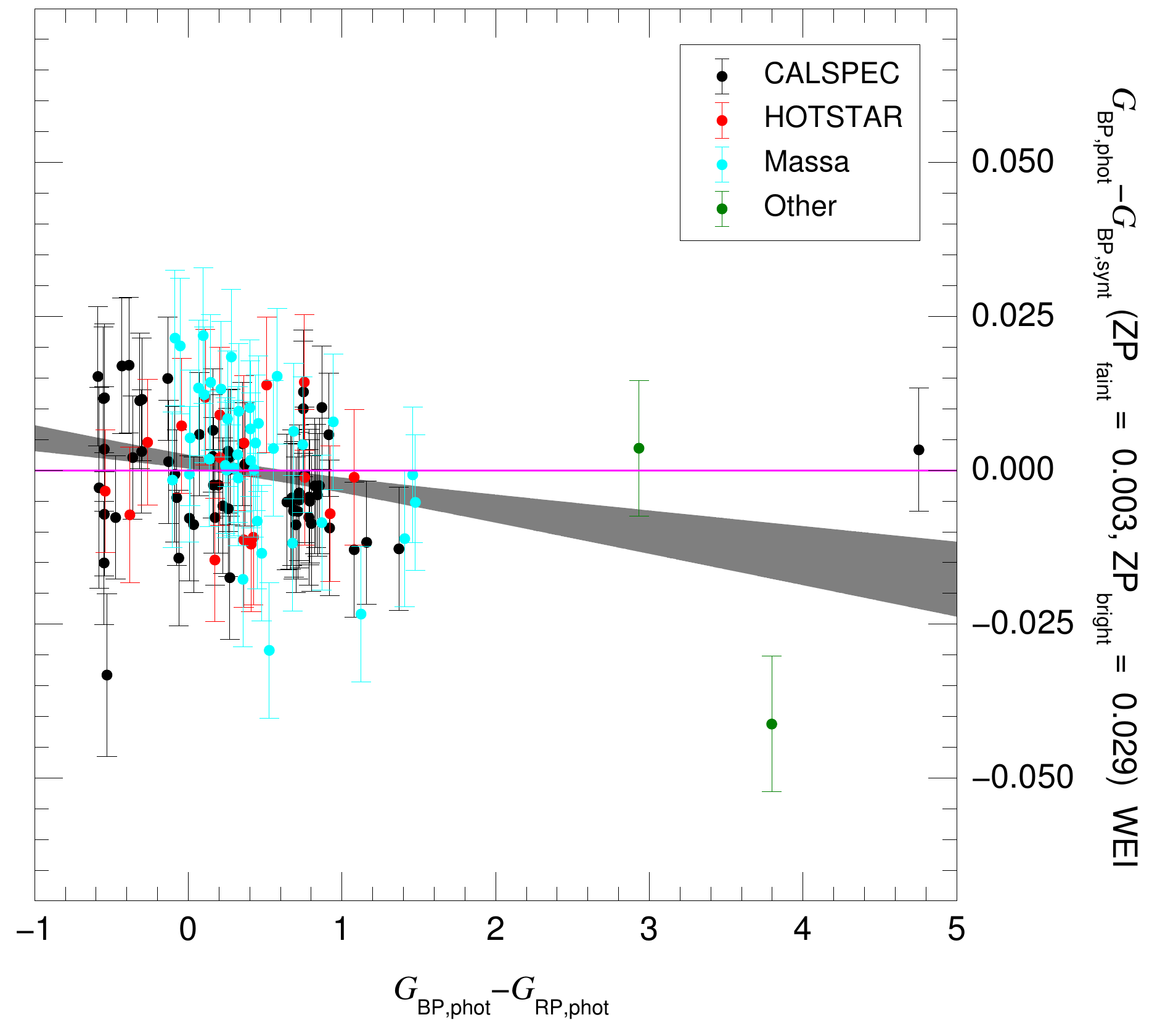}}
\centerline{\vspace{12mm}}
\caption{Same as Fig.~\ref{Gplots} for \GBP.}
\label{GBPplots}
\end{figure*}

\begin{figure*}
\centerline{\includegraphics[width=0.49\linewidth, bb=28 90 566 512]{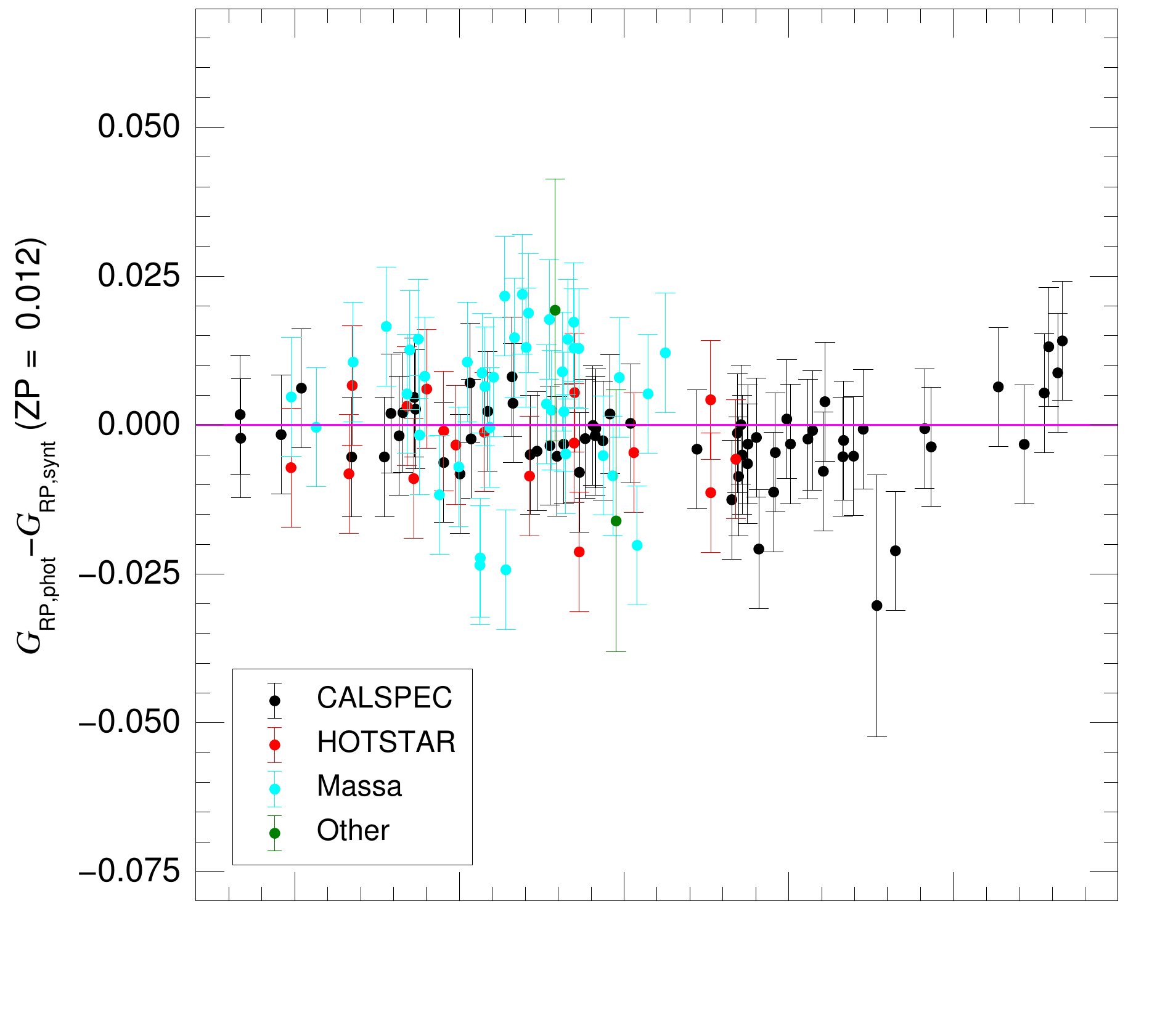} \
            \includegraphics[width=0.49\linewidth, bb=28 90 566 512]{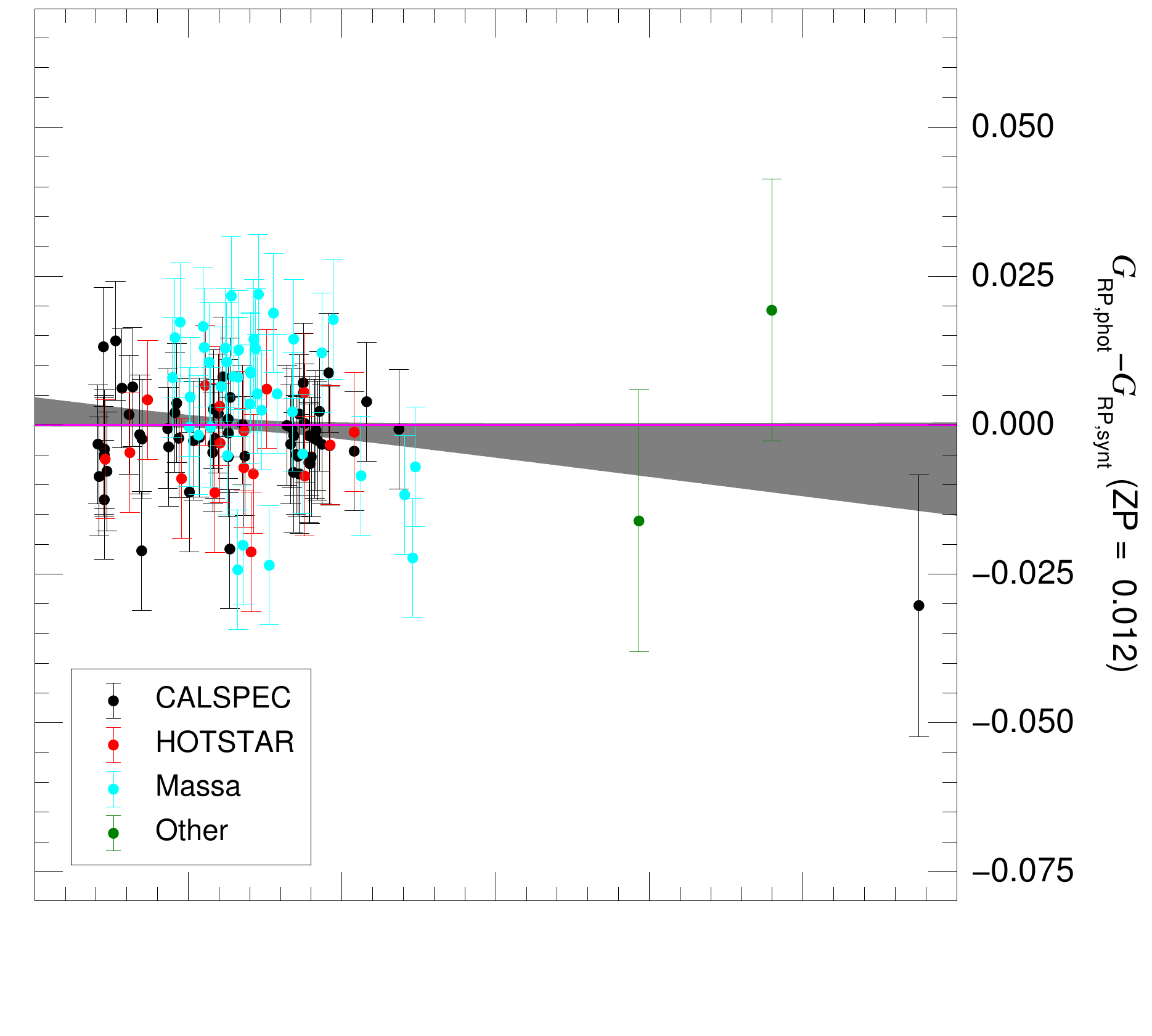}}
\centerline{\includegraphics[width=0.49\linewidth, bb=28 90 566 512]{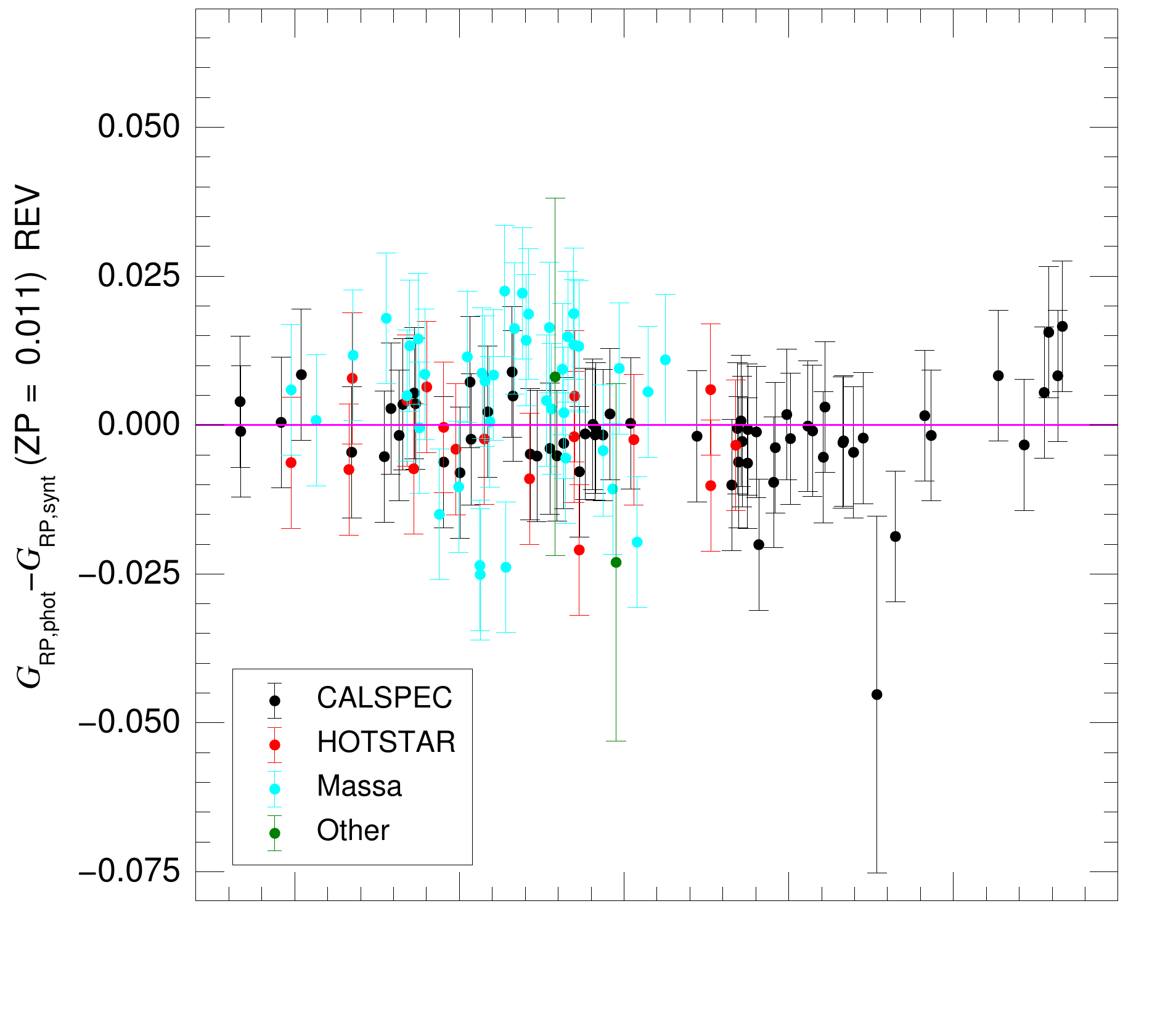} \
            \includegraphics[width=0.49\linewidth, bb=28 90 566 512]{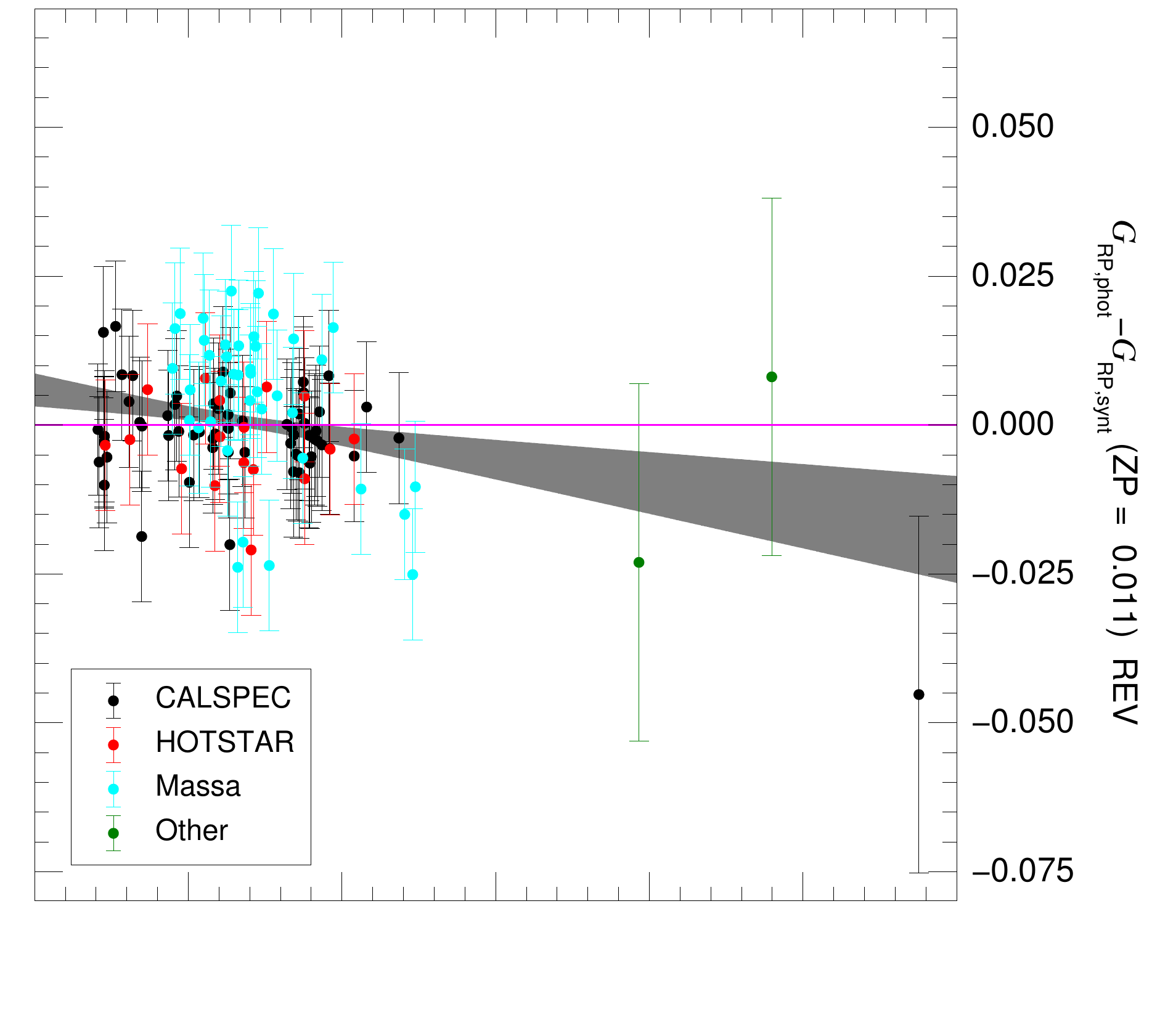}}
\centerline{\includegraphics[width=0.49\linewidth, bb=28 90 566 512]{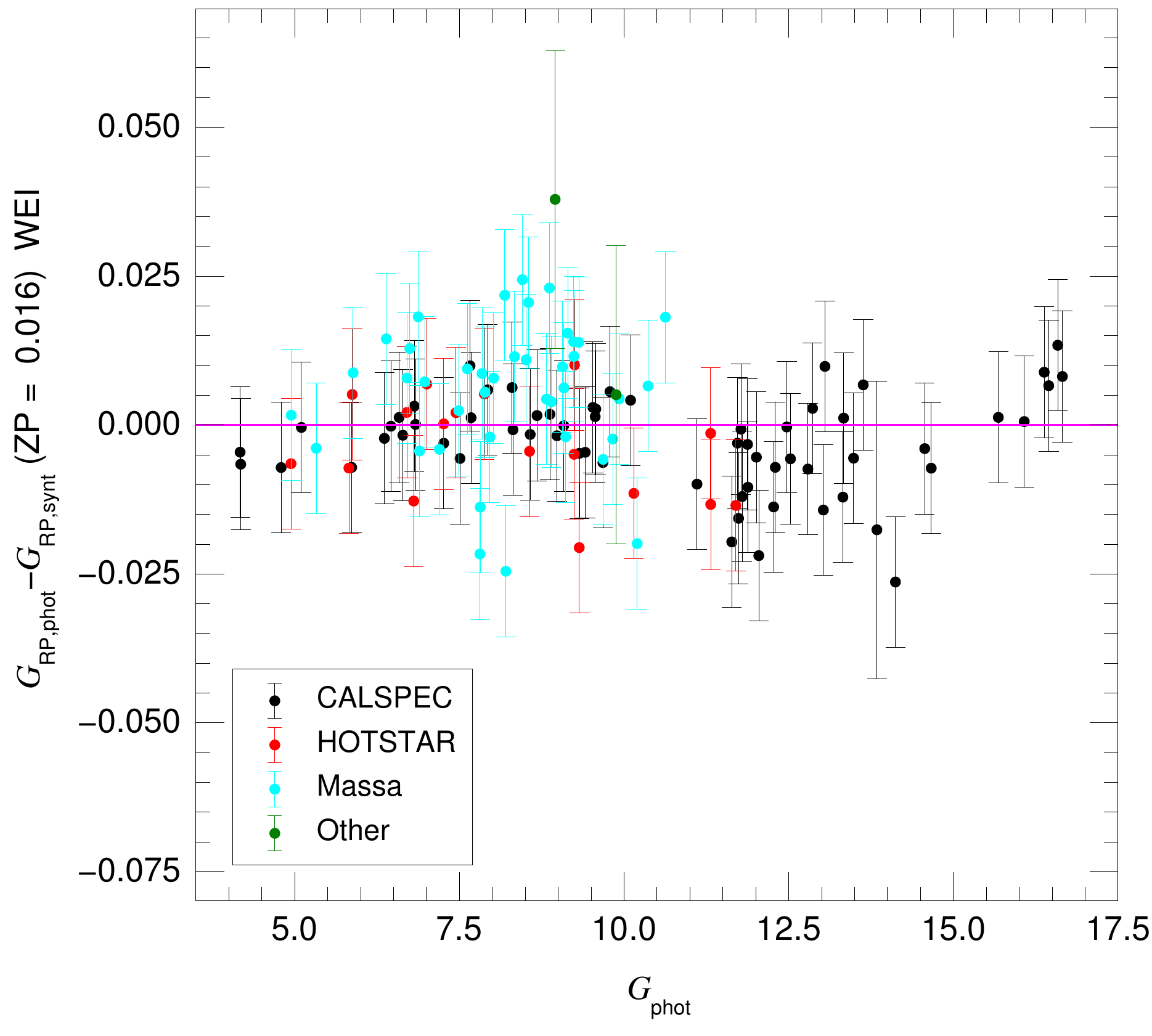} \
            \includegraphics[width=0.49\linewidth, bb=28 90 566 512]{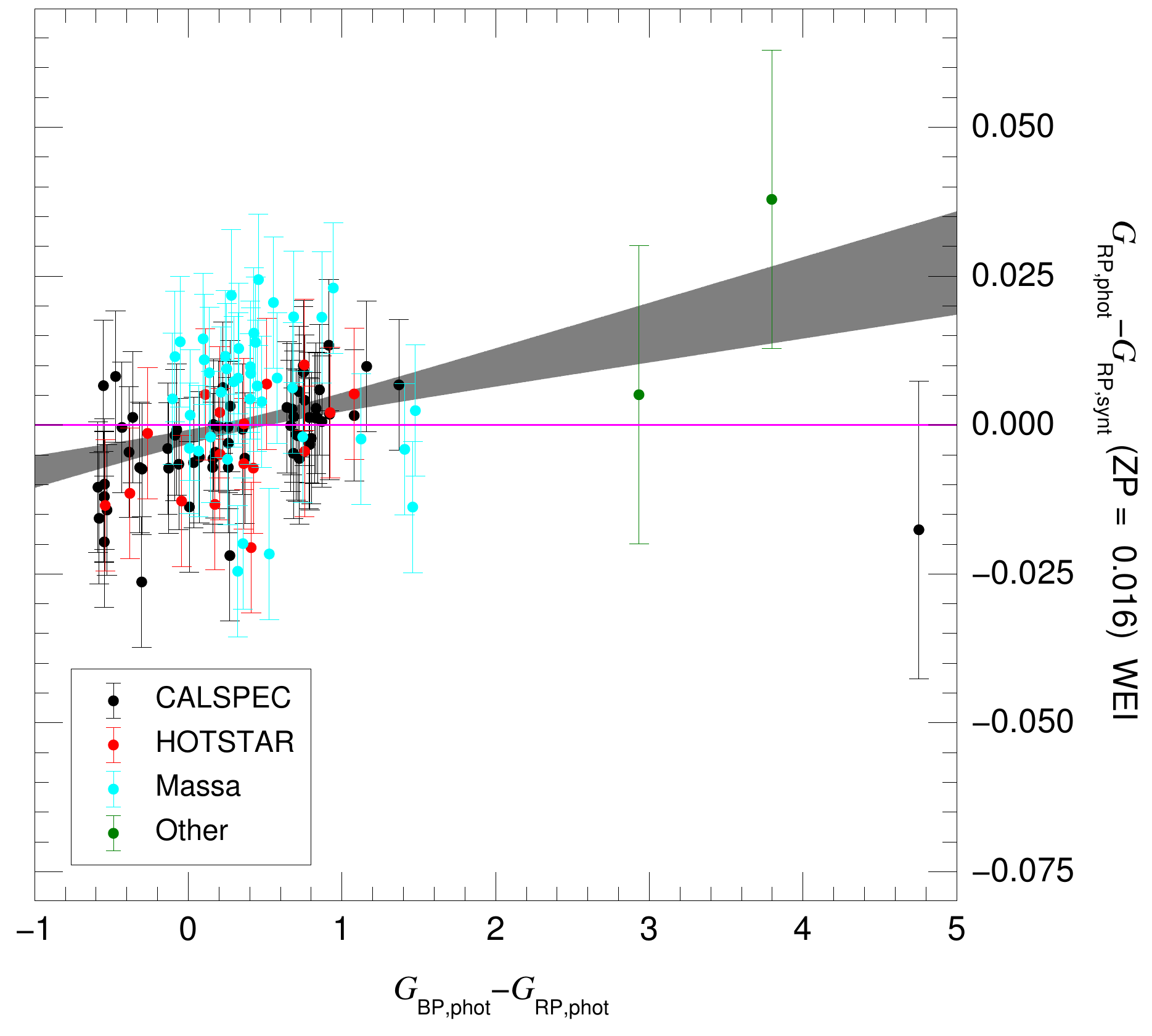}}
\centerline{\vspace{12mm}}
\caption{Same as Fig.~\ref{Gplots} for \GRP.}
\label{GRPplots}
\end{figure*}

$\,\!$\indent Our results for \GG\ are given in Table~\ref{results} and Fig.~\ref{Gplots}. The left panels of Fig.~\ref{Gplots} show that the magnitude
dependence is well removed for the three tested sensitivity curves. The right panels of Fig.~\ref{Gplots} show that there is no significant color term
for either MAW or WEI ($b_G$ is less than one sigma from zero) and both solutions provide very similar results, as expected from the comparison of the
sensitivity curves (Fig.~\ref{Gsens}). On the other hand, REV has a strong color term indicated not only by a large value of $b_G$ but also by the
increasing deviation in the residual $\GGcp-\GGs$ for the three M stars of increasing color. This is a consequence of the redder nature of the REV
sensitivity curve previously mentioned. This improvement in the \GG\ passband (already present in WEI) remained undetected before, as no sufficiently 
red calibration sources were available before this work. In fact, the modification of the \GG\ passband is highly relevant for getting a good 
representation of the color-color relationships, as we will see below. 

We derive values of $\sigma_{{\rm min},G}$ for the three \GG\ sensitivity curves by measuring the dispersion with respect to the ZP in the restricted
fit. As saturation effects set in at $\GGp = 6$, we use that value as the breaking point to define two magnitude ranges. Results are the same 
for MAW and WEI, 8~mmag for faint stars and 12~mmag for bright ones, with the value for faint stars in REV significantly worse (13 mmag). Those values
are much smaller than the 30~mmag (faint) and 74~mmag (bright) derived by \citet{Maiz17} but it is clear now that the reason for such high values was the
use of the NGSL spectral library (see above).
The $\sigma_{{\rm min},G}$ values are also
significantly lower than their equivalents for literature Johnson photometry (20 mmag for $B-V$ and 28 mmag for $U-B$, \citealt{Maiz06a}) and, 
furthermore, they refer to the absolute (magnitude) calibration and not to the relative (color) one. Ground-based optical surveys also have higher
values of $\sigma_{\rm min}$ than Gaia: see \citet{Padmetal08} for SDSS and \citet{Drewetal14} for VPHAS+.
The slightly larger, though still insignificant color term for MAW as compared to WEI results from the Massa spectra, which tend to cluster in a small 
color range, have a small but systematic offset with respect to the other calibration spectra of this work, as already mentioned in section 3.2. 
Excluding Massa spectra from the computation, the color term actually becomes zero. The improvement of the MAW as compared to WEI is essentially the 
removal of a small systematic deviation of the CALSPEC spectra at colors around $\GBP-\GRP$ of 0.3.
Considering all of the above and that the formal uncertainty on ZP$_{{\rm Vega},G}$ is just 1~mmag, we can conclude that the 
current calibration of the Gaia~DR2 \GG\ magnitude has an unparalleled quality among deep all-sky photometric survey. 

\subsection{\GBP}

$\,\!$\indent Our results for \GBP\ are given in Table~\ref{results} and Fig.~\ref{GBPplots}. As previously discussed, we have divided our sample into
two with a break at $\GGp = 10.87$~mag, which we also use to divide the calculation of $\sigma_{{\rm min},G_{\rm BP}}$. As a result, the two subsamples
are clearly divided in the left panels of Fig.~\ref{GBPplots} but are mixed in the right panels.

The jump at $\GGp = 10.87$~mag manifests itself in the ZP$_{{\rm Vega},G_{\rm BP}}$ for the two ranges, with differences of 21, 19, and 26~mmag for
MAW, REV, and WEI, respectively. There is a large difference in $\sigma_{{\rm min},G_{\rm BP}}$ for bright stars between REV (20~mmag) and either MAW 
(9~mmag) or WEI (11~mmag). This is a sign of the reality of different \GBP\ sensitivity curves for the two ranges, a factor included in MAW and WEI but 
not in REV. The difference is smaller for faint stars, indicating that the REV calibration is not as bad there. This is in agreement with the SPSS 
calibration spectra used to derive the REV passband being dominated by sources in the faint magnitude regime. The small difference between MAW and 
WEI points towards an improvement of the results in this paper.

Looking at the right panels in Fig.~\ref{GBPplots} we see a difference between MAW and either REV or WEI. The latter two have a significant negative
value of $b_{G_{\rm BP}}$ while that of the former is zero. This indicates that, as it happened with \GG, the addition of very red sources introduces an
improvement in the sensitivity curves that was previously undetected. Furthermore, in this case the improvement takes place in the transition from WEI
to MAW while for \GG\ it was in the transition from REV to WEI. Considering also that the formal uncertainties on ZP$_{{\rm Vega},G}$ for the MAW
calibration are just 1~mmag (faint) and 2~mmag (bright), we conclude that the current calibration of the Gaia~DR2 \GBP\ magnitude has a similar quality 
to that of \GG.

We also mention that we attempted an alternative procedure by using a single sensitivity curve for \GBP\ (the faint one) and correcting the bright
\GBPp\ values into the faint system using a second degree polynomial in $\GBPp - \GRPp$. When doing so, we derived a reasonable transformation but the
values of $\sigma_{{\rm min},G_{\rm BP}}$ were higher than the ones described above. Looking into the detailed behavior of the sample, we
realized that the reason resided in the different behavior of late-B and early-A stars i.e. those with a large Balmer jump, which (for a given
extinction) have intermediate $\GBPp - \GRPp$ values between those of O/early-B stars and late-type stars. As the most important differences between the
bright and faint \GBP\ sensitivity curves are to the left of the Balmer jump, those stars deviate from a correction defined mostly from other types
of stars\footnote{Note that it is not possible to unequivocally identify stars of a given spectral type by a single color due to extinction.}. Therefore, 
we decided that alternative procedure (correcting \GBPp), though attractive due to its simplicity, should be discarded in favor of using different 
definitions of \GBPs\ for different ranges of \GGp.

\subsection{\GRP}

$\,\!$\indent Our results for \GRP\ are given in Table~\ref{results} and Fig.~\ref{GRPplots}. In this case we do not need to divide our
sample in magnitude ranges, as there is no magnitude-dependent correction (as for \GG) or need for two different sensitivity curves (as for \GBP).
However, we divided the sample by color in order to increase the value of $\sigma_{{\rm min},G_{\rm RP}}$ for the three red dwarfs, which play a large 
role in the calibration of the passband.

The left panels of Fig.~\ref{GRPplots} do not show trends in magnitude or large differences among the three sensitivity curves and the derived values 
of $\sigma_{{\rm min},G_{\rm RP}}$ (excluding the three red dwarfs) are also similar (10~mmag for MAW, 11~mmag for the other two). On the other hand, 
the right panels show significant differences: REV yields a negative value of $b_{G_{\rm RP}}$, WEI a positive one, and only MAW yields one that is
within one sigma of zero. Therefore, the new calibration is an improvement over the previous two but, in this case, the result at this point is more 
uncertain as it depends mostly on the three red dwarfs. For that reason, we explore the issue in more detail in the next two subsections, where we
discuss the effect of using different spectrophotometric libraries and employ additional information from color-color diagrams.

\subsection{Comparing different spectrophotometric libraries}

\begin{figure*}
\centerline{\includegraphics[width=0.33\linewidth,  bb=28 90 448 512]{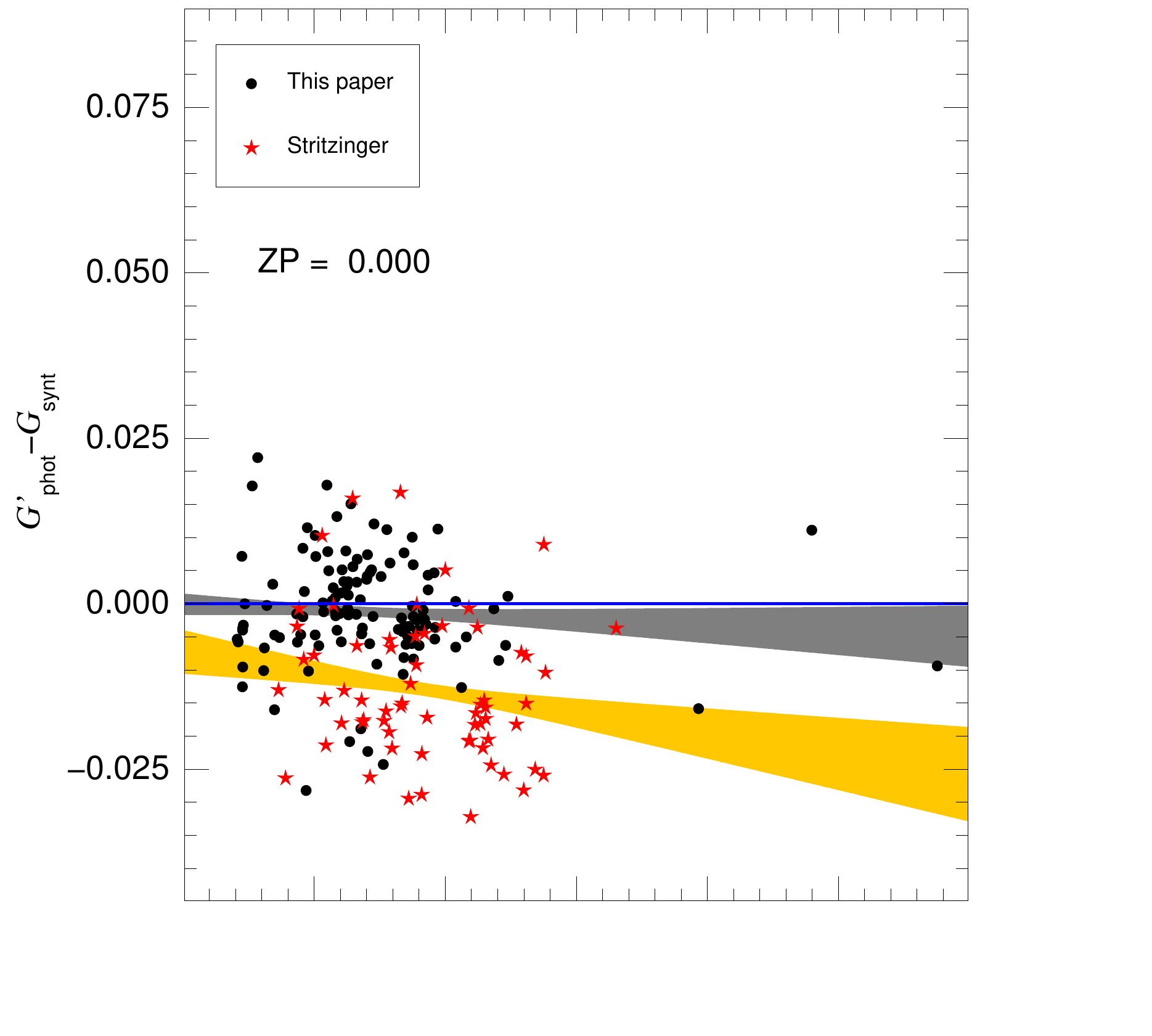}  \
            \includegraphics[width=0.33\linewidth,  bb=87 90 507 512]{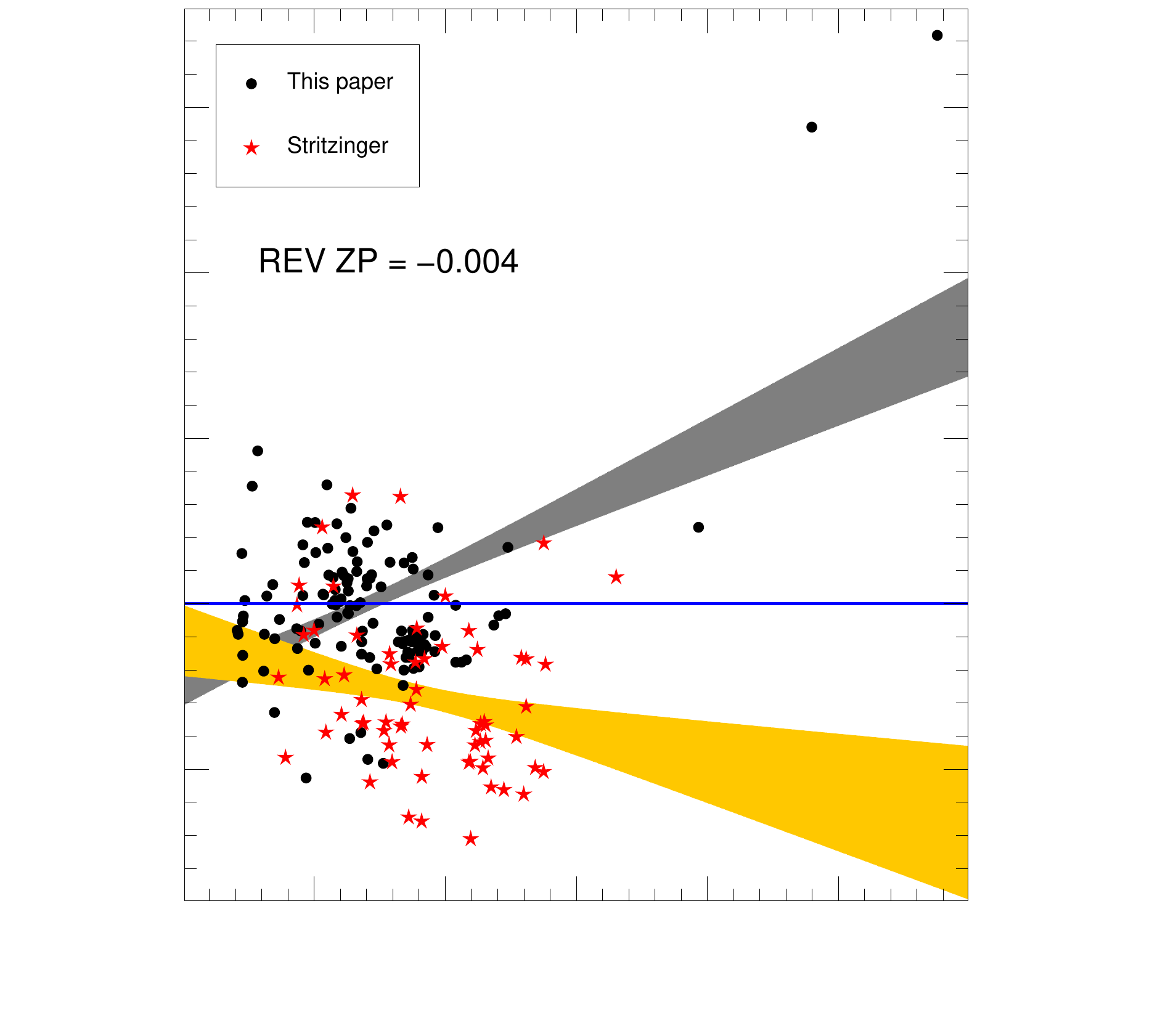} \
            \includegraphics[width=0.33\linewidth, bb=146 90 566 512]{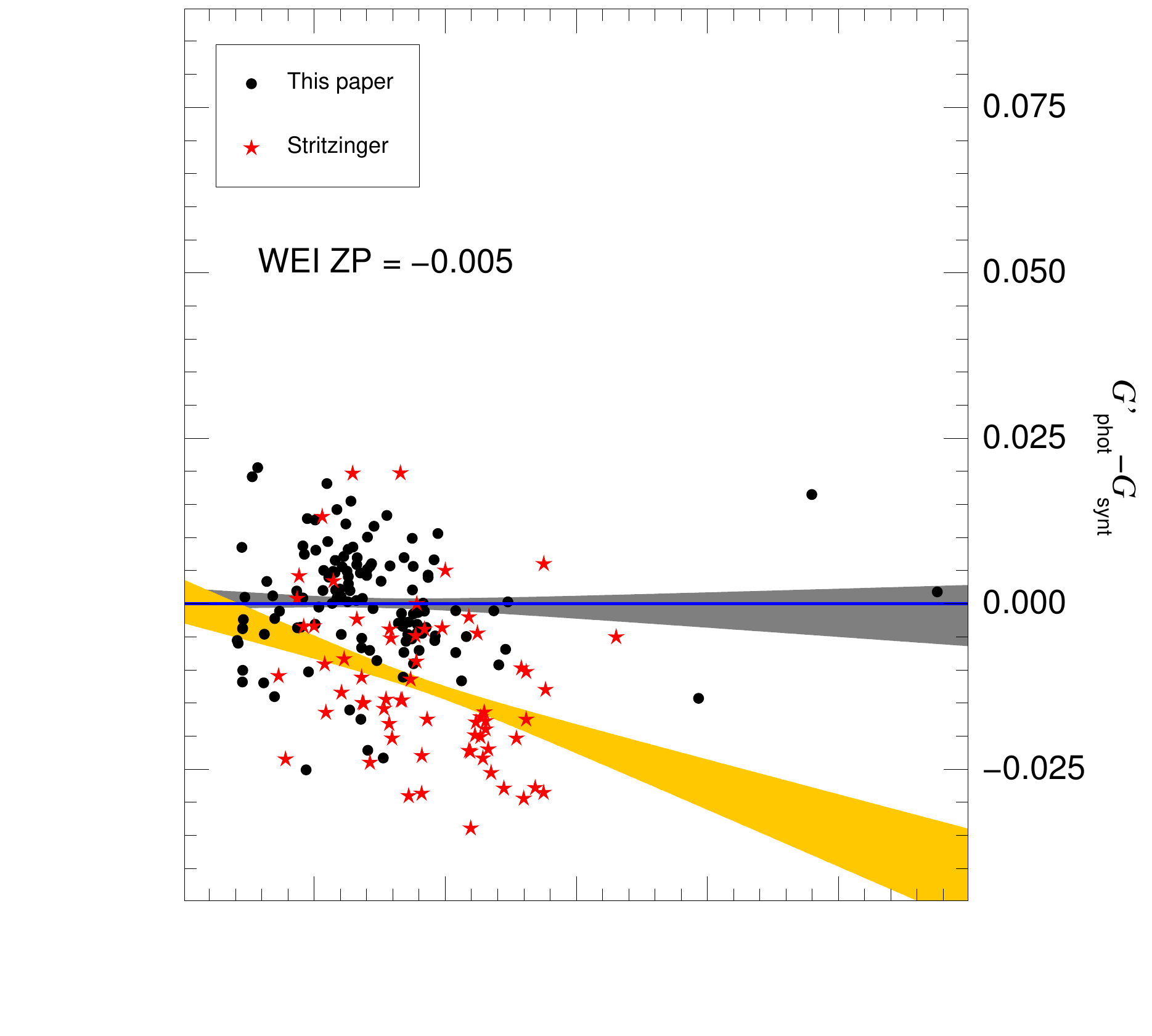}}
\centerline{\includegraphics[width=0.33\linewidth,  bb=28 90 448 512]{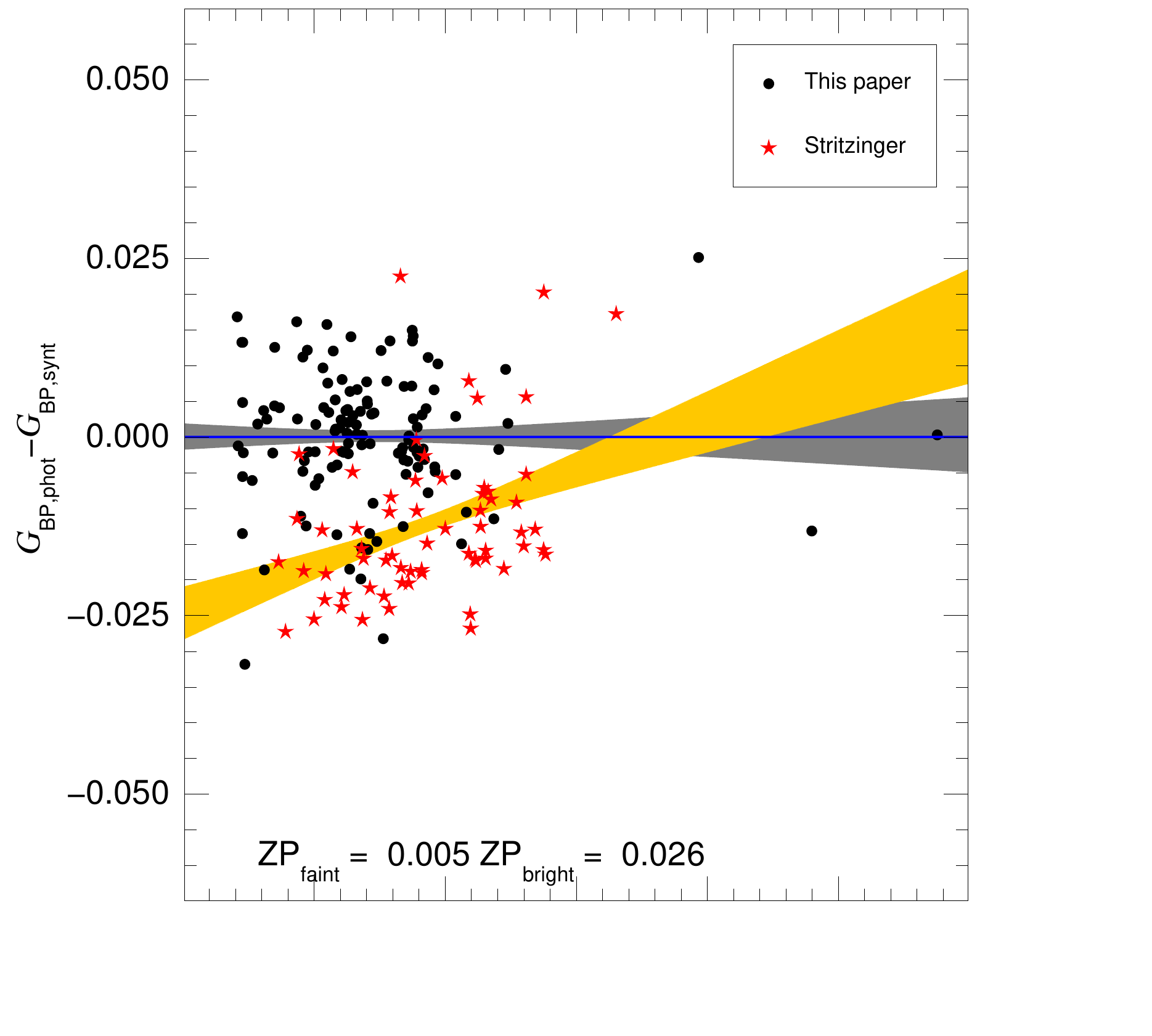} \
            \includegraphics[width=0.33\linewidth,  bb=87 90 507 512]{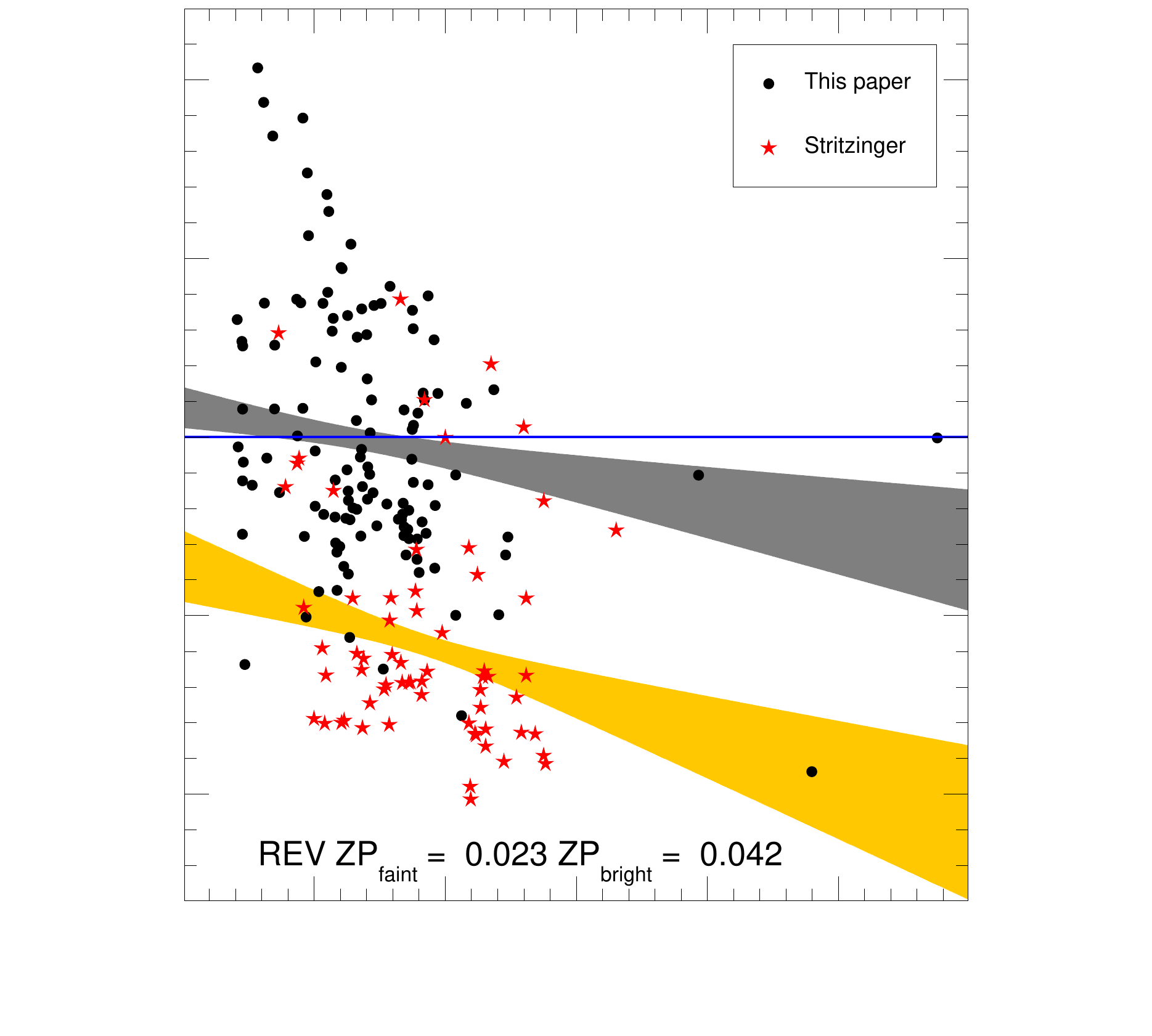} \
            \includegraphics[width=0.33\linewidth, bb=146 90 566 512]{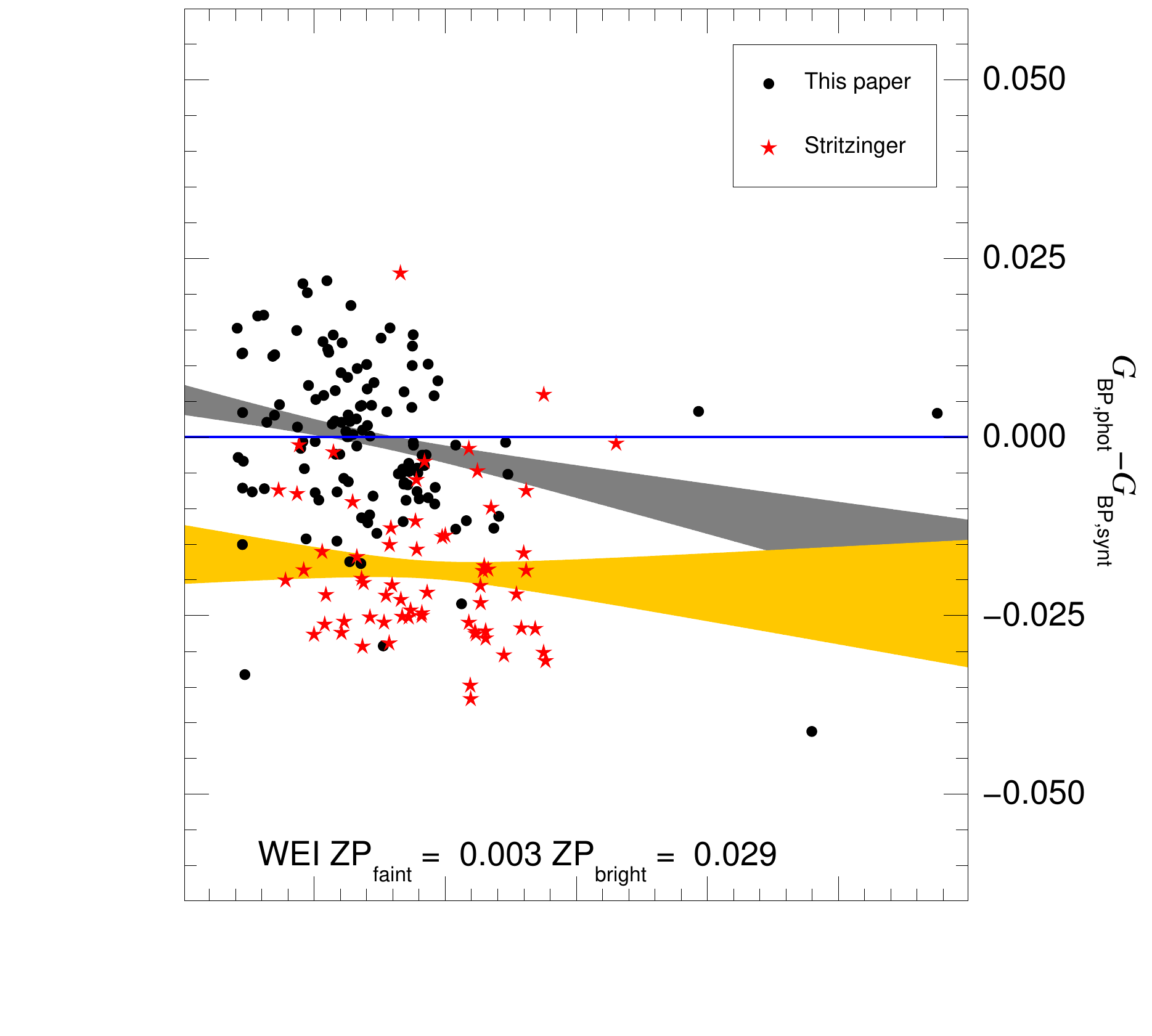}}
\centerline{\includegraphics[width=0.33\linewidth,  bb=28 90 448 512]{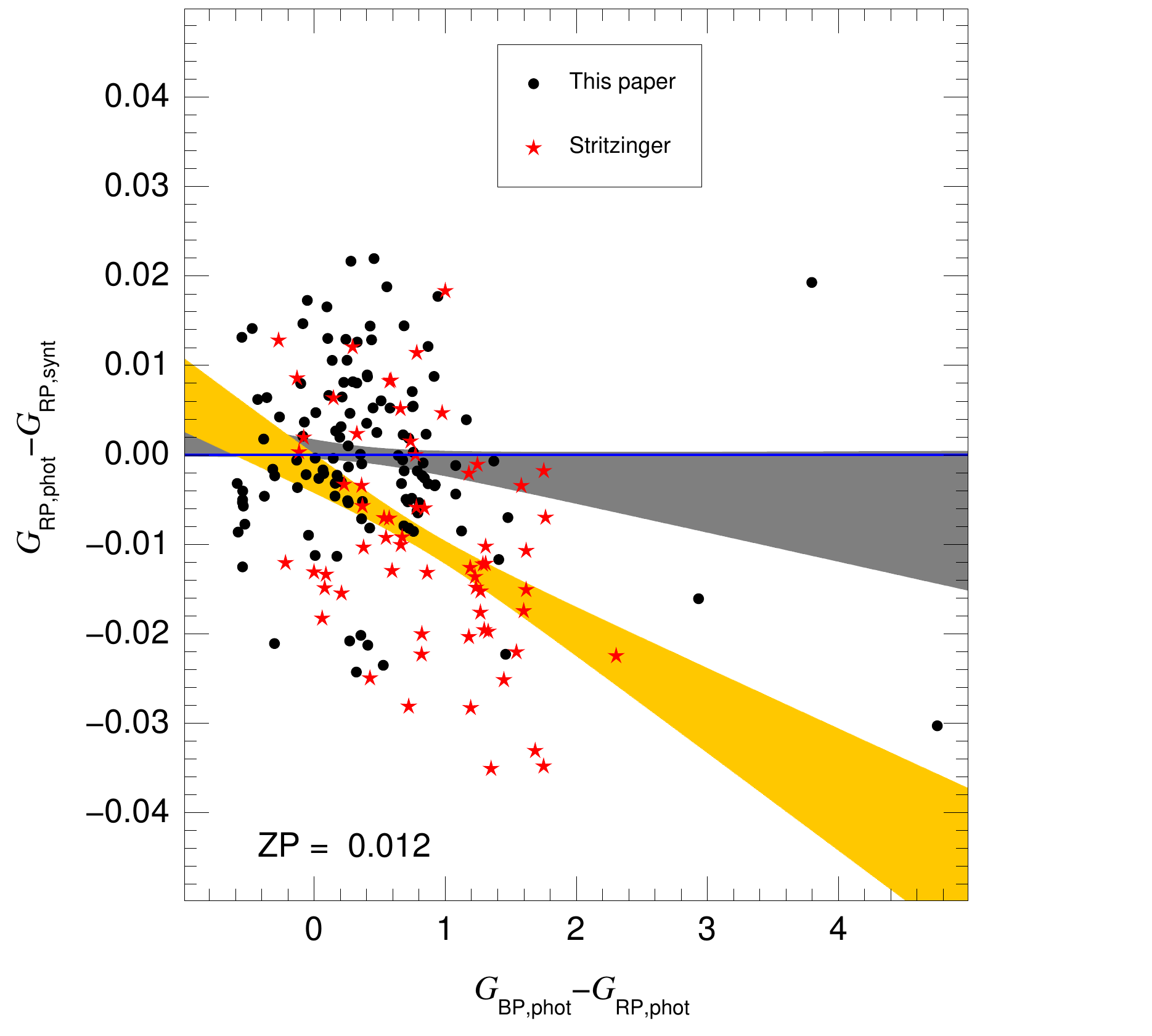} \
            \includegraphics[width=0.33\linewidth,  bb=87 90 507 512]{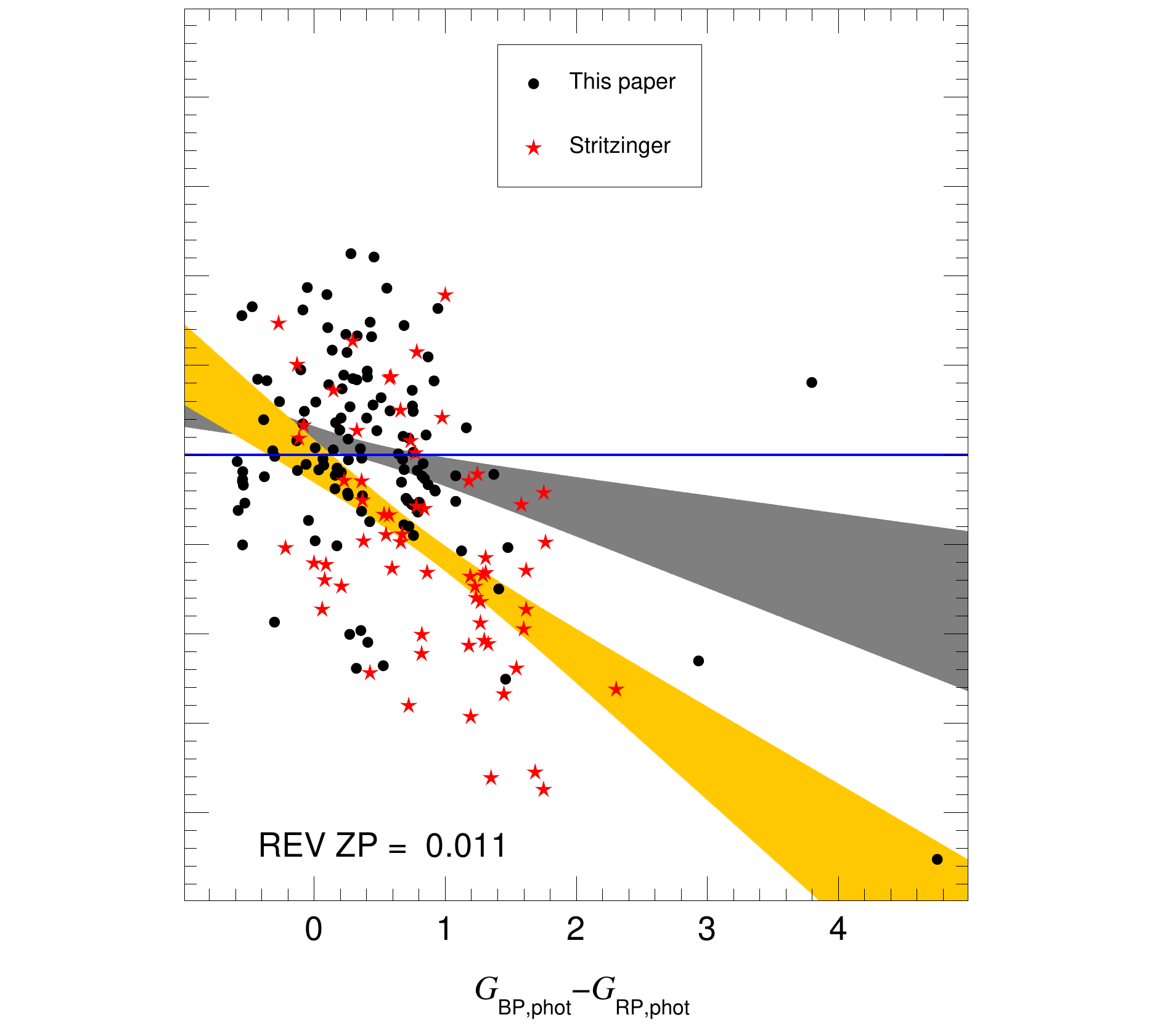} \
            \includegraphics[width=0.33\linewidth, bb=146 90 566 512]{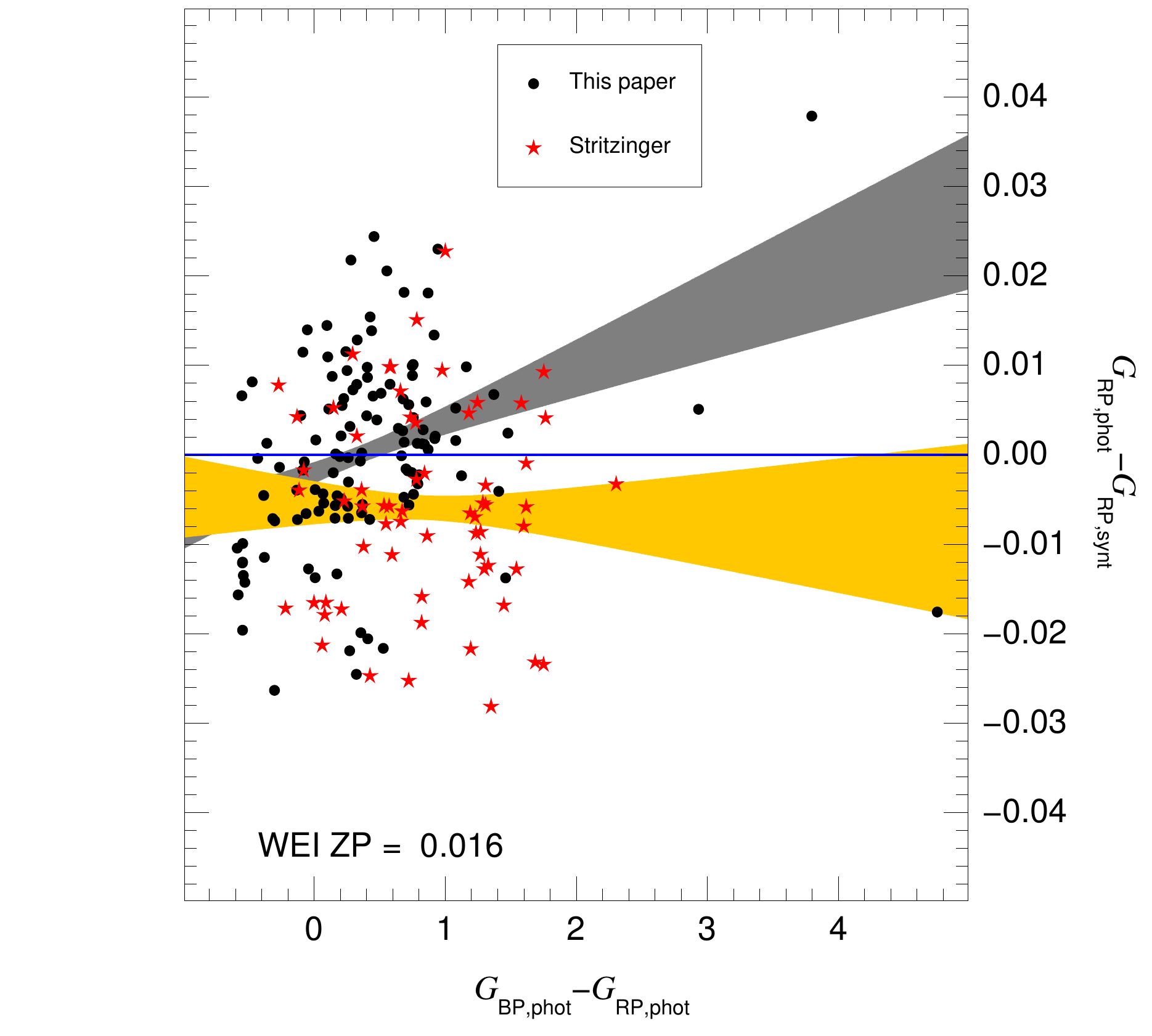}}
\centerline{\vspace{12mm}}
\caption{Comparison between the (corrected in the \GG\ case) observed magnitudes and the synthetic \GG\ magnitudes as a function of $\GBPp-\GRPp$ for
         the sample in this paper (black circles) and the Stritzinger sample (red stars). The top. middle, and bottom rows show the results for \GG, 
         \GBP, and \GRP, respectively. The left, center, and right columns show the results for MAW, REV, and WEI, respectively. The region shaded in 
         gray shows the 1~$\sigma$ confidence range for the unrestricted fit for the sample in this paper. The region shaded in light orange shows the 
         equivalent for the Stritzinger sample.}
\label{Striplots}
\end{figure*}

$\,\!$\indent In order to test the sensitivity curves derived in this work, we compute the synthetic photometry for the spectrophotometric standard stars 
from \citet{Strietal05} and compare it with the corresponding {\it Gaia} DR2 photometry. We omit a comparison with the NGSL because of the large 
level of uncertainty in these spectra and with the SPSS data set because it has not been published yet. Figure~\ref{Striplots} shows the resulting 
residuals as a function of $\GBP-\GRP$ color for the MAW, REV, and WEI sensitivity curves and for the three {\it Gaia} passbands, respectively.

In all three {\it Gaia} passbands, an offset in the residuals from \citet{Strietal05} with respect to our calibration data set is visible, 
indicating a difference in zero point. For the \GG\ passband, all three sets of sensitivity curves result in a similar color trend in the residuals for 
\citet{Strietal05}, with MAW providing the least color dependency. For \GBP, the REV passband results in a color dependency for 
\citet{Strietal05} residuals, which is related to the break in the \GBP\ photometry. All blue \citet{Strietal05} stars with 
$\GBP - \GRP < 0.6$ belong to the bright magnitude regime, which is not well represented by the REV sensitivity curve. The WEI sensitivity 
curves for \GBP\ remove the color dependency for the \citet{Strietal05} residuals. However, WEI does not fully remove the effects of 
the break in photometry in the set of calibration spectra used in this work, which may also affect the color dependency of the \citet{Strietal05} 
residuals. The MAW sensitivity curves describe the calibration sources of this work better than WEI, but at the same time introduce a color dependency in 
the residuals for the \citet{Strietal05} spectra.

For \GRP, the color dependency in the residuals for \citet{Strietal05} spectra is strongest. For the REV sensitivity curves, a clear color
dependency of the \citet{Strietal05} residuals is visible. A similar color dependency of the residuals was also observed for the SPSS set of 
calibration spectra by \citet{Weil18}, and the WEI sensitivity curve for \GRP\ was constructed to remove this color term from the SPSS 
residuals. As seen in Fig.~\ref{Striplots}, the WEI curve also entirely removes the color dependency from the \citet{Strietal05} residuals. 
The MAW sensitivity curve removes that color term from the calibration spectra of this work, becoming very similar in its overall shape to the REV 
passband, but it re-introduces the color term in the \citet{Strietal05} data set. We are thus in the situation that the WEI passband provides a better 
description of the \GRP\ photometric system if the ground-based SPSS and \citet{Strietal05} spectra are used as a standard, while the REV and 
the MAW sensitivity curves provide a better description if the calibration spectra in this work are used as reference.

The origin for the discrepancy between the different sets of calibration spectra remains unknown. It is however not related to the choice of the 
orthogonal component of the \GRP\ sensitivity curve. We may use the angle $\gamma$ as defined in \citet{Weiletal18} to describe the sensitivity
of the \citet{Strietal05} spectra to the choice of the orthogonal component of the sensitivity curve. Approximately computing this angle for all 
\citet{Strietal05} spectra results in very small values below 2\degr\ for all spectra in \GG\ and \GRP, with only one spectrum exceeding 
2\degr\ in \GBP. The synthetic photometry for the \citet{Strietal05} set of spectra is thus strongly dominated by the parallel 
component of the sensitivity curve with respect to the calibration set used in this work and the systematic difference between the ground-based 
calibration spectra (SPSS and \citet{Strietal05}) and the STIS spectra used for calibration in this work is a small but likely real effect. The 
difference in shape between the WEI and MAW sensitivity curves for the \GRP\ passband, although appearing large, eventually reflects this small 
effect.

\subsection{$\GBP-\GGc$ vs. $\GGc-\GRP$ diagrams}

\begin{figure*}
\centerline{\includegraphics[width=\linewidth]{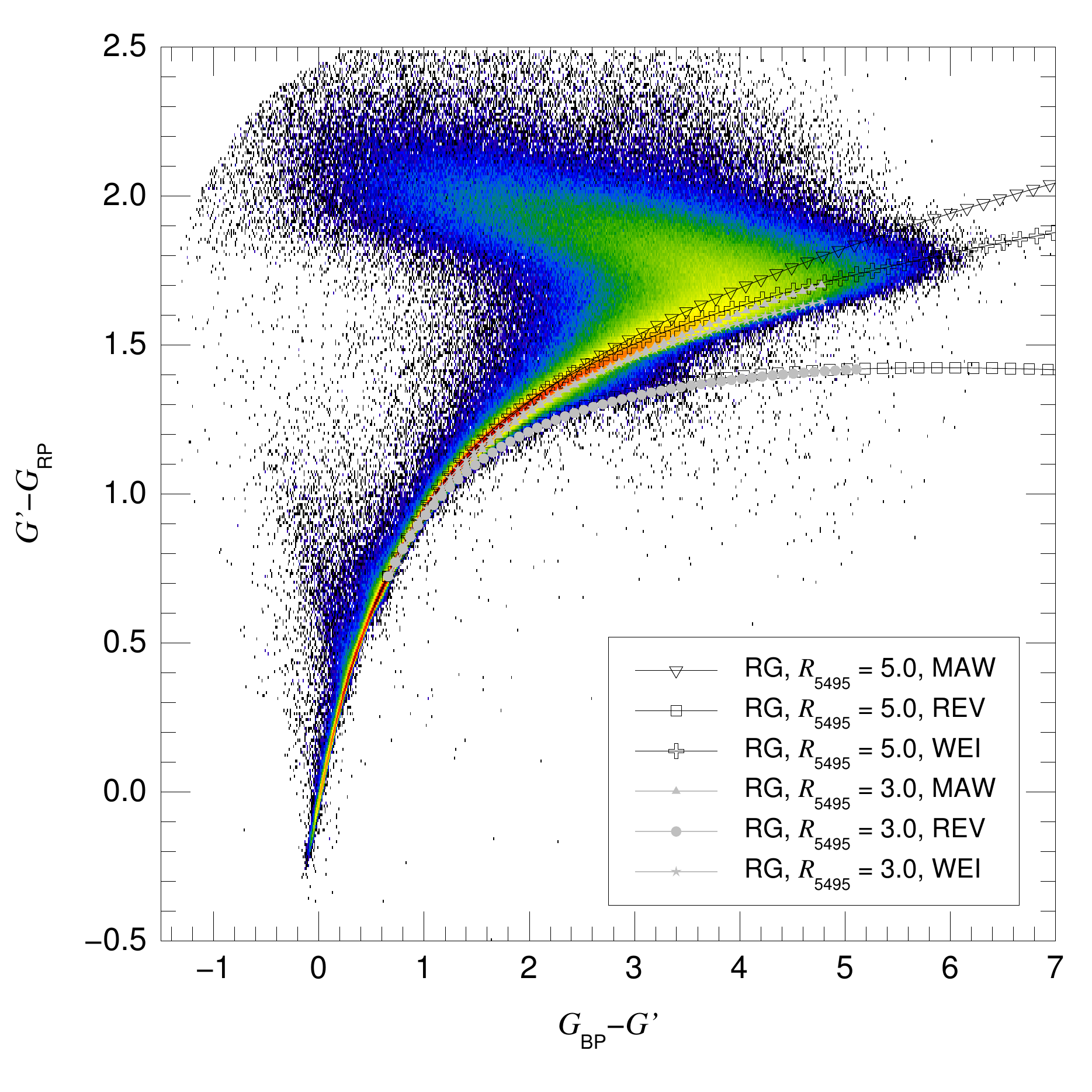}}
\caption{$\GBP-\GGc$ vs. $\GGc-\GRP$ diagram that includes all stars with 2MASS counterparts, good-quality photometry, and $K < 9$~mag. The intensity 
 scale is logarithmic. The lines with symbols mark the extinction trajectories of a red giant with $\GGp > 10.87$~mag using the family of extinction laws 
 of \citet{Maizetal14a} (symbols are spaced by $\Delta E(4405-5495) = 0.1$~mag and reach to $E(4405-5495) = 5.0$~mag) combined in six ways by selecting 
 (a) $R_{5495} = 3$ (normal extinction) or $R_{5495} = 5$ (H\,{\sc ii} region extinction) and (b) MAW, REV, or WEI sensitivity curves. See the text 
 for more details.}
\label{BG_GR_1}
\end{figure*}

\begin{figure*}
\centerline{\includegraphics[width=0.49\linewidth]{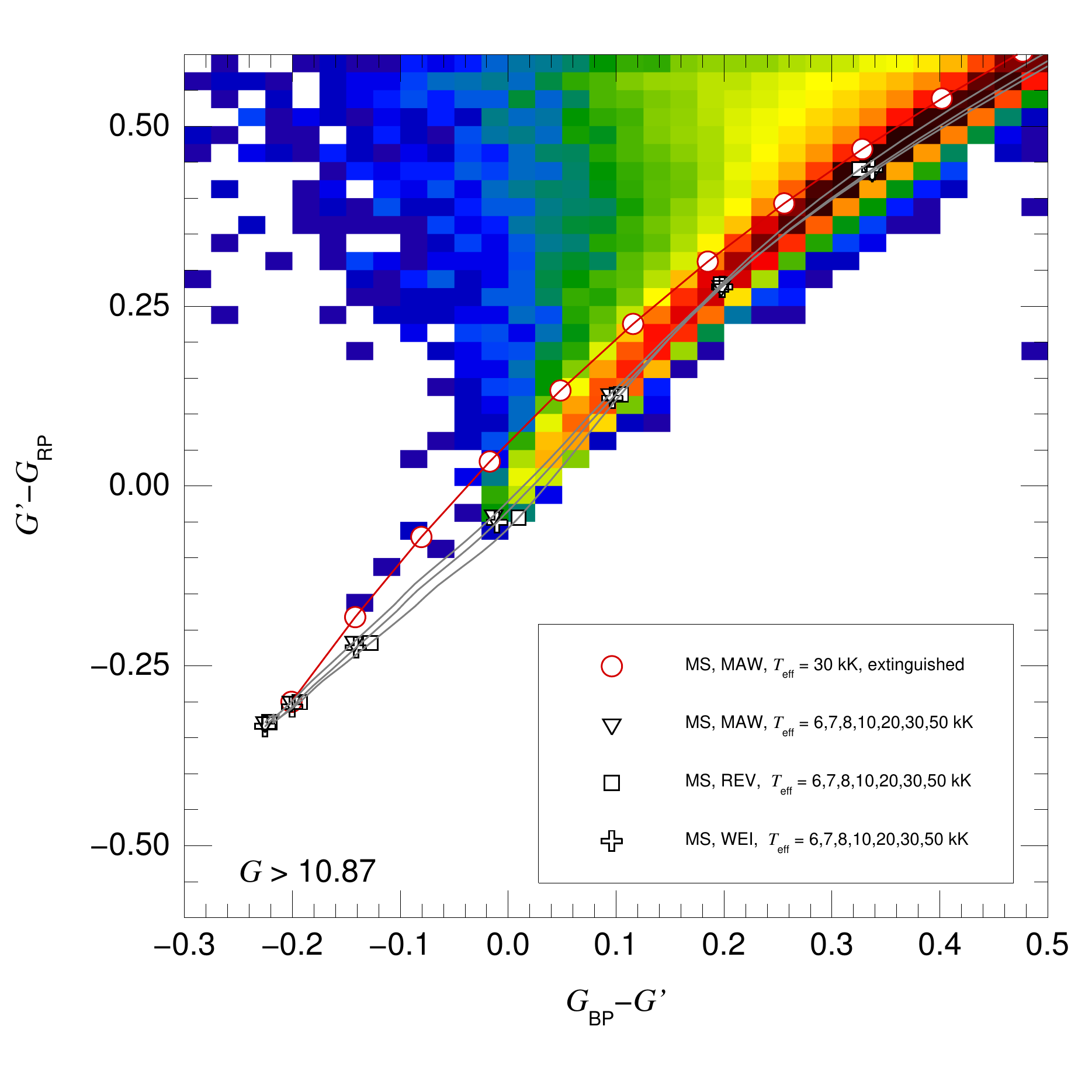}
            \includegraphics[width=0.49\linewidth]{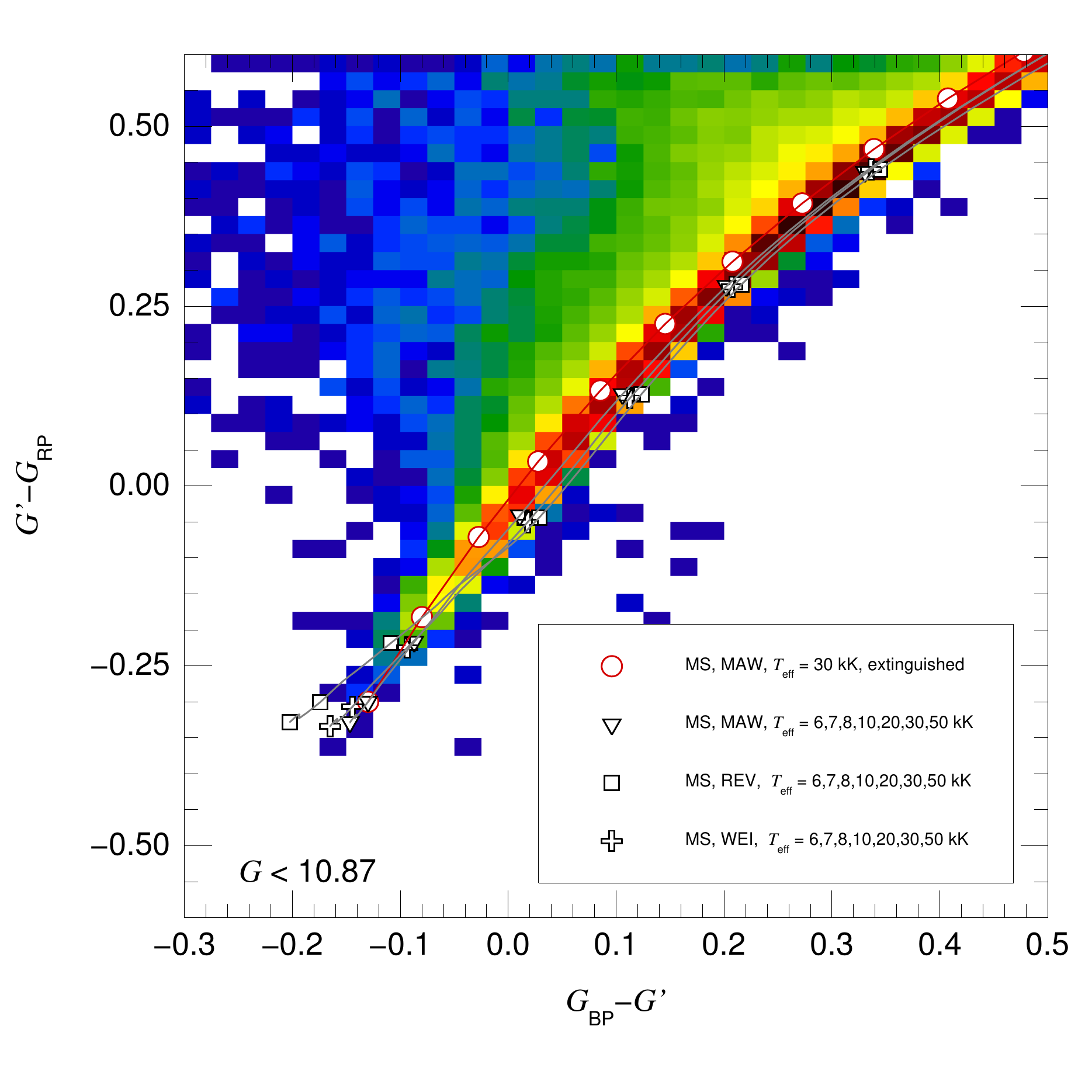}}
\caption{Lower-right region of the $\GBP-\GGc$ vs. $\GGc-\GRP$ diagram that includes all stars with 2MASS counterparts, good-quality photometry, and 
 $K < 11$. The left panel shows the faint stars ($\GGp > 10.87$~mag) and the right panel the bright ones ($\GGp < 10.87$~mag). The intensity scale is 
 logarithmic. The black and white symbols mark the location of the main sequence using the MAW, REV, or WEI sensitivity curves. The red and white symbols 
 mark the extinction trajectory of a 30~kK main-sequence star using the family of extinction laws of \citet{Maizetal14a} (symbols are spaced by 
 $\Delta E(4405-5495) = 0.1$~mag).}
\label{BG_GR_2}
\end{figure*}

$\,\!$\indent A final test of the validity of the different sensitivity curves can be done by comparing the stellar locus in the $\GBP-\GGc$ vs. 
$\GGc-\GRP$ observed color-color diagram with the synthetic photometry from stellar models. For this purpose, we cross-matched the {\it Gaia} DR2 catalog 
with 2MASS \citep{Skruetal06}
and selected the stars with good-quality photometry and $K$ magnitudes less than 9~or~11, respectively. 
That selection reduces the dispersion in the color-color diagram and 
preferentially selects luminous stars, as the bright population in $K$ selects mostly high luminosity and nearby stars for blue colors and mostly red 
giants for red colors.  For the synthetic photometry we use the \citet{Maiz13a} solar metallicity grid and the \citet{Maizetal14a} family of extinction 
laws\footnote{We have repeated the analysis below with the \citet{Cardetal89} family of extinction laws and the results are very similar.}.

Figure~\ref{BG_GR_1} shows the $\GBP-\GGc$ vs. $\GGc-\GRP$ for the sample described above with $K < 9$~mag. The sample follows a tight correlation in the
color-color diagram as one moves diagonally from blue colors to red ones, with an extension that goes from $\GBP-\GGc \sim 4$, $\GGc-\GRP \sim 1.6$ 
towards
$\GBP-\GGc \sim 0$, $\GGc-\GRP \sim 2.1$. As described by \citet{Evanetal18}, the tight correlation that extends from the lower left to the upper right
is the real stellar locus while the extension towards the upper left corner is caused by objects with ``flux excess'' i.e. objects where crowding, 
nebulosity, or background subtraction introduce contamination in \GBP\ and/or \GRP. The lower left part of the stellar locus is populated mostly by
low-extinction stars (with some intermediate-extinction O+B stars) with ``normal'' colors while the central and upper right parts are populated mostly
by red giants of increasing extinctions as one moves from center to right. The intrinsic {\it Gaia} colors of red giants are relatively well 
characterized (most of them are bluer than the the three M dwarfs in our calibration set) but the extinction trajectories depend on the type
of extinction (i.e. the $R_{5495}$ value for the \citealt{Maizetal14a} family of extinction laws) and the sensitivity curves of the three {\it Gaia}
passbands. Figure~\ref{BG_GR_1} shows the extinction trajectories for a solar-metallicity red giant using two different assumptions for $R_{5495}$ and the
three sets of sensitivity curves described in this paper. The extinction trajectories of the REV sensitivity curves follow very similar paths
independently of $R_{5495}$ (but note that the position along the trajectory is not the same for a fixed $E(4405-5495)$ if $R_{5495}$ changes) but the
trajectories are well below the stellar locus by up to several tenths of a magnitude. This indicates that one or more of the REV sensitivity curves does
not correctly describe its passband. Our previous analysis suggests that all three bands have systematic errors for very red objects, with \GG\ being the
worst of the three. On the other hand, the extinction trajectories for the MAW and WEI passbands show a small dependence with $R_{5495}$, with the 
$R_{5495}=5$ case predicting a higher value of $\GG-\GRP$ for a given $\GBP-\GG$ than the $R_{5495}=3$ case. Both MAW and WEI yield trajectories that
are consistent with the stellar locus but MAW has the advantage that the center of the stellar locus lies between the $R_{5495}=3$ and $R_{5495}=5$
trajectories, which is the expected result \citep{MaizBarb18}. Therefore, the {\it Gaia} color-color diagram for high-extinction stars indicates that 
the MAW sensitivity curves are slightly better than the WEI ones and significantly better than the REV ones. Another interesting consequence of this
analysis is that the extinction trajectories for red giants in the {\it Gaia} color-color plane depend more strongly on the definition of the passbands
than on the extinction law.

Figure~\ref{BG_GR_2} shows the same color-color diagram but for the equivalent sample with $K < 11$~mag divided into the two subsamples limited by the
$\GGp = 10.87$ magnitude that separates the two \GBP\ sensitivity curves. The left panel shows that for faint stars there are few objects hotter that 
10~kK, which is expected because normal stars with $K < 11$~mag and $\GGp > 10.87$~mag can be only slightly bluer than Vega in any color formed by 
filters with most of their sensitivity to the right of the Balmer jump. That panel is dominated by low-extinction AFG stars and the three sets of 
sensitivity curves describe that stellar locus correctly assuming zero extinction. 
Note that for faint stars it is possible to use the {\it Gaia} colors to differentiate between a zero-extinction A0 
main-sequence star and an O star with $E(4405-5495) \sim 0.3$, as there is a significant separation between the zero-extinction main-sequence stellar
locus and the extinction trajectory for hot stars. However, at higher values of $E(4405-5495)$ the zero-extinction stellar locus and the extinction 
trajectory get closer and such a distinction is no longer possible. That is the reason why the stellar locus at high extinctions is so narrow: temperature
and extinction become degenerate. The distinction is possible at lower extinctions because of the extra sensitivity of \GBP\ to the left of the Balmer
jump.

The right panel of Fig.~\ref{BG_GR_2} shows differences with the left panel. In the first place, the diagram is populated up to the extreme of the
zero-extinction stellar locus i.e. there are a few low-extinction O and early B stars with $K < 11$~mag and $\GGp < 10.87$~mag. That leads to an important 
difference: the position of the tip agrees with the prediction of the MAW and WEI sensitivity curves but not with that of the REV sensitivity curves. This
is another consequence of the reality of the existence of two \GBP\ sensitivity curves. Another difference takes place for redder colors, as the stellar
locus is located between the zero-extinction prediction (for any set of sensitivity curves) and the extinction trajectory for hot stars. This indicates
that for brighter stars there is a significant fraction of either (a) OBA stars with non-negligible extinction or (b) AF supergiants\footnote{We decided 
not to represent in Fig.~\ref{BG_GR_2} the zero-extinction stellar locus for supergiants in order to reduce confusion but for AF stars it is also 
displaced in the same direction as the extinction trajectory for hot stars. Note, however, that the AF supergiant phase is a short evolutionary phase, so
it has few members.}. Finally, the switch of \GBP\ sensitivity curve at $\GGp = 10.87$~mag 
has the consequence of decreasing the spacing between A0 main-sequence stars with no extinction and O stars with $E(4405-5495) \sim 0.3$,
as the \GBP\ sensitivity to the left of the Balmer jump is lower for bright stars. This is unfortunate, as it makes the use of {\it Gaia} photometry by
itself to distinguish between populations less useful. In a future paper we will analyze how the combination with 2MASS photometry helps in this issue.

In summary, the $\GBP-\GGc$ vs. $\GGc-\GRP$ color-color diagrams provide additional evidence that the MAW sensitivity curves are clearly better than
the REV ones and that they are slightly better than the WEI ones.

\section{Summary and future work}

$\,\!$\indent In this work we produced an extension of the CALSPEC set of spectrophotometric standard stars by compiling and re-calibrating suitable 
HST/STIS observations. The resulting set of calibration spectra was used to derive new sensitivity curves for the three passbands of the {\it Gaia} DR2. 
We used the functional analytic framework by \citet{Weiletal18} for the passband computations.

For the \GG\ passband we confirm a systematic magnitude-dependent trend in the photometric system. For the range of magnitudes between 6 and 16, we 
derive a linear correction of 3.2 mmag/mag. This correction needs to be applied for an accurate comparison of {\it Gaia} DR2 photometry with synthetic 
photometry. The sensitivity curve for \GG\ derived in this work results in a small improvement compared to the WEI sensitivity curve, and a large 
improvement over the REV one, which shows a strong color term for very red sources.

For the \GBP\ passband we confirm the existence of a color-dependent break in the DR2 photometry that was described by \citet{Weil18}. This break can be 
accurately modelled by two different sensitivity curves for \GBP, valid for bright and faint sources, respectively. From an analysis of the 
uncertainties of the \GBP\ fluxes, we can constrain the position of the break better than it was done in \citet{Weil18} and we confirm that the break is 
a result of the \GG\ magnitude rather than the \GBP\ magnitude. We therefore present two \GBP\ sensitivity curves, valid for $\GGp > 10.87$~mag and 
$\GGp <10.87$~mag, respectively. These new sensitivity curves result in a strong improvement as compared to REV, and still a clear improvement as 
compared to WEI.

For the \GRP\ sensitivity curve we obtained a solution that is similar in its shape to the REV curve, but resulting in an improvement by removing a color
dependency in the residuals. However, this solution re-introduces a color dependency in ground-based calibration spectra that was removed by the WEI 
sensitivity curves. The difference between WEI and the sensitivity curve of this work is thus related to the different sets of calibration spectra used. 
It is not a lack of constraints on the sensitivity curves in one of the sets of calibration spectra that is causing the difference in shape, but 
rather a small but systematic difference in spectral shapes.


To verify the consistency of the sensitivity curves for the different passbands, we compared synthetic color-color relationships derived from stellar 
models with observed color-color relationships. We extend this approach, described in \citet{Weil18}, by also including high-extinction sources 
into the comparison. The set of sensitivity curves presented in this work result in a very good agreement with {\it Gaia} DR2 color-color relationships 
over a wide range of colors.

We have demonstrated that HST/STIS observations provide excellent means for calibrating {\it Gaia} photometry, allowing for an accuracy better than any 
other set of spectrophotometric standards. However, there is still room for improvement in the calibration of {\it Gaia} DR2 photometry, as the
definition of the sensitivity curves for \GG\ and \GRP\ depend strongly on three M dwarfs. It is urgent that additional HST spectrophotometry of several 
tens of very red sources is obtained to solve this deficiency. Once we have seen the diffuclties in producing an accurate
calibration for the photometry in {\it Gaia} DR2, it is necessary to plan ahead for the calibration of the spectrophotometry in {\it Gaia} DR3.

\hyphenation{pu-mi-lio-ni-bus}
\begin{acknowledgements}
We thank J.A.C. Escurialensis for his help cum pumilionibus rubris.
This work has made use of data from the European Space Agency (ESA) mission {\it Gaia} ({\tt https://www.cosmos.esa.int/gaia}), processed by the 
{\it Gaia} Data Processing and Analysis Consortium (DPAC, {\tt https://www.cosmos.esa.int/web/gaia/dpac/consortium}). Funding for the DPAC has been 
provided by national institutions, in particular the institutions participating in the {\it Gaia} Multilateral Agreement. 
J.M.A. acknowledges support from the Spanish Government Ministerio de Ciencia, Innovaci\'on y Universidades through grant AYA2016-75\,931-C2-2-P.
M.W. acknowledges support from the Spanish Government Ministerio de Ciencia, Innovaci\'on y Universidades through grants ESP2016-80\,079-C2-1-R 
(MICINN/FEDER, UE), ESP2014-55\,996-C2-1-R (MICINN/FEDER, UE), and MDM-2014-0369 of ICCUB (Unidad de Excelencia ``Mar{\'\i}a de Maeztu'').

\end{acknowledgements}

\bibliographystyle{aa}
\bibliography{general}

\begin{appendix}
\section{Zero points and conversions between magnitude systems}

$\,\!$\indent We follow the notation of \citet{Maiz07a} to write the synthetic magnitudes based on the Vega system for a filter $p$ as:

\begin{equation}
m_{{\rm Vega},p} = 
 -2.5\log_{10}\left(\frac{\int P_p(\lambda)f_{\lambda,s}(\lambda)\lambda\,d\lambda}
                         {\int P_p(\lambda)f_{\lambda,{\rm Vega}}(\lambda)\lambda\,d\lambda}\right)
                       + {\rm ZP}_{{\rm Vega},p},
\label{Vega}   
\end{equation}

\noindent where $P_p(\lambda)$ is the total-system dimensionless sensitivity curve; $f_{\lambda,s}$ and $f_{\lambda,{\rm Vega}}$ are the star 
and Vega SEDs, respectively; and ZP$_{{\rm Vega},p}$ is the filter zero point. One usually tries to define a system with values of zero for the 
ZPs but, in practice, the ZPs are small but non-zero and have to be calculated from external sources to ensure photometric compatibility across 
surveys (that is what we have done in this paper, see \citealt{Maiz07a} for other examples). For consistency with our previous work, we use the
Vega spectrum provided by \citet{Bohl07}\footnote{Available from \url{ftp://ftp.stsci.edu/cdbs/calspec/alpha_lyr_stis_003.fits}.}.

Vega-based magnitude systems have been commonly used in astronomy for decades but they have been critized because they depend on an assumed Vega 
SED and different authors provide different ones (which makes sense as our knowledge improves over time, see e.g \citealt{Bohl14}). That criticism 
is valid only as a consistency issue because if one indicates which Vega SED is being used (and makes it available) and defines the ZPs 
consistently using Eqn.~\ref{Vega}, the magnitudes are correctly defined. If, at one point in the future, somebody comes up with a better Vega 
SED, the ZPs will change accordingly:

\begin{eqnarray}
2.5\log_{10}\left(\int P_p(\lambda)f_{\lambda,{\rm Vega,new}}(\lambda)\lambda\,d\lambda\right) + {\rm ZP}_{{\rm Vega,new},p} & = & \nonumber \\
2.5\log_{10}\left(\int P_p(\lambda)f_{\lambda,{\rm Vega,old}}(\lambda)\lambda\,d\lambda\right) + {\rm ZP}_{{\rm Vega,old},p} & . &
\label{Vega2}    
\end{eqnarray} 

\noindent and the resulting $m_{{\rm Vega},p}$ will remain unchanged. Therefore, it is possible to work with the Vega-based definitions on this 
paper without having to resort to systems based on other reference SEDs.

Nevertheless, some readers may prefer to use the alternative AB system defined as:

\begin{equation}
m_{{\rm AB},p} = 
 -2.5\log_{10}\left(\frac{\int P_p(\lambda)f_{\lambda,s}(\lambda)\lambda\,d\lambda}
                         {\int P_p(\lambda)f_{\lambda,{\rm AB}}(\lambda)\lambda\,d\lambda}\right)
                       + {\rm ZP}_{{\rm AB},p},
\label{AB}     
\end{equation}

\noindent where $f_{\nu,{\rm AB}} = 3.63079\cdot 10^{-20}$ erg s$^{-1}$ cm$^{-2}$ Hz$^{-1}$ (constant), leading to:

\begin{equation}
m_{{\rm AB},p} = 
 -2.5\log_{10}\left(\frac{\int P_p(\lambda)f_{\lambda,s}(\lambda)\lambda\,d\lambda}
                         {\int P_p(\lambda)cf_{\nu,{\rm AB}}(\lambda)/\lambda\,d\lambda}\right)
                       + {\rm ZP}_{{\rm AB},p}.
\label{AB2}     
\end{equation}

On a survey where AB magnitudes are used by default, the values of ZP$_{{\rm AB,p}}$ will be close to zero (as it is also done for surveys where
Vega magnitudes are used, see above) but not exactly so\footnote{See \url{https://www.sdss.org/dr12/algorithms/fluxcal/} for the SDSS case.}. 
However, Gaia uses Vega magnitudes by default so if one imposes the condition $m_{{\rm Vega},p} = m_{{\rm AB},p}$ necessary to compare the 
synthetic magnitudes calculated this way with the observed ones we find:

\begin{eqnarray}
2.5\log_{10}\left(\int P_p(\lambda)f_{\lambda,{\rm Vega}}(\lambda)\lambda\,d\lambda\right) + {\rm ZP}_{{\rm Vega},p} & = & \nonumber \\
2.5\log_{10}\left(\int P_p(\lambda)cf_{\nu,{\rm AB}}(\lambda)/\lambda\,d\lambda\right) + {\rm ZP}_{{\rm AB},p} & , &
\label{VegaAB}    
\end{eqnarray} 

\noindent from where we get:

\begin{eqnarray}
{\rm ZP}_{{\rm AB},p} & = & -2.5\log_{10}\left(\frac{\int P_p(\lambda)f_{\lambda,{\rm Vega}}(\lambda)\lambda\,d\lambda}
                                                    {\int P_p(\lambda)cf_{\nu,{\rm AB}}(\lambda)/\lambda\,d\lambda}\right)
                            + {\rm ZP}_{{\rm Vega},p} \nonumber \\
                      & = & m_{{\rm AB,ZP=0},p}({\rm Vega}) + {\rm ZP}_{{\rm Vega},p}, 
\label{VegaAB2}     
\end{eqnarray}

\noindent where $m_{{\rm AB,ZP=0},p}({\rm Vega})$ is the magnitude of Vega in the filter $p$ using the default AB system (the one where ZP = 0).
The values of those quantities for the three filters in this paper using the \citet{Bohl07} Vega spectral energy distribution are 
$m_{{\rm AB,ZP=0},G}({\rm Vega}) = 0.125$~mag, $m_{{\rm AB,ZP=0},G_{\rm BP}}({\rm Vega}) = 0.044$~mag, and 
$m_{{\rm AB,ZP=0},G_{\rm RP}}({\rm Vega}) = 0.369$~mag.

We can do the same analysis for the ST system, whose magnitudes are defined as:

\begin{equation}
m_{{\rm ST},p} = 
 -2.5\log_{10}\left(\frac{\int P_p(\lambda)f_{\lambda,s}(\lambda)\lambda\,d\lambda}
                         {\int P_p(\lambda)f_{\lambda,{\rm ST}}(\lambda)\lambda\,d\lambda}\right)
                       + {\rm ZP}_{{\rm ST},p},
\label{ST}     
\end{equation}

\noindent where $f_{\lambda,{\rm ST}} = 3.63079\cdot 10^{-9}$ erg s$^{-1}$ cm$^{-2}$ \AA$^{-1}$ (constant), to reach:

\begin{eqnarray}
{\rm ZP}_{{\rm ST},p} & = & -2.5\log_{10}\left(\frac{\int P_p(\lambda)f_{\lambda,{\rm Vega}}(\lambda)\lambda\,d\lambda}
                                                    {\int P_p(\lambda)f_{\lambda,{\rm ST}}(\lambda)\lambda\,d\lambda}\right)
                            + {\rm ZP}_{{\rm Vega},p} \nonumber \\
                      & = & m_{{\rm ST,ZP=0},p}({\rm Vega}) + {\rm ZP}_{{\rm Vega},p}, 
\label{VegaST}     
\end{eqnarray}

\noindent where $m_{{\rm ST,ZP=0},p}({\rm Vega})$ is the magnitude of Vega in the filter $p$ using the default ST system (the one where ZP = 0).
The values of those quantities for the three filters in this paper using the \citet{Bohl07} Vega spectral energy distribution are 
$m_{{\rm ST,ZP=0},G}({\rm Vega}) = 0.405$~mag, $m_{{\rm ST,ZP=0},G_{\rm BP}}({\rm Vega}) = -0.137$~mag, and 
$m_{{\rm ST,ZP=0},G_{\rm RP}}({\rm Vega}) = 1.124$~mag. Note that these values and the equivalent ones for the AB system are not close to zero, as
expected, as we have forced the AB and ST magnitudes to be the same as the Vega ones. In other words, these AB (or ST) magnitudes are far from the
exact or default AB (or ST) system (defined as the one with ZP=0).

\end{appendix}

\end{document}